\documentclass[preprint]{aastex631}
\shorttitle{Searching for neutrinos from solar flares in Super-Kamiokande}
\shortauthors{Okamoto~et al.}
\graphicspath{{./}{figures/}}

\begin{document}

\title{Searching for neutrinos from solar flares across solar cycles 23 and 24 \\ with the Super-Kamiokande detector}

\newcommand{\AFFicrr}{\affiliation{Kamioka Observatory, Institute for Cosmic Ray Research, University of Tokyo, Kamioka, Gifu 506-1205, Japan}}
\newcommand{\AFFkashiwa}{\affiliation{Research Center for Cosmic Neutrinos, Institute for Cosmic Ray Research, University of Tokyo, Kashiwa, Chiba 277-8582, Japan}}
\newcommand{\AFFicrronly}{\affiliation{Institute for Cosmic Ray Research, University of Tokyo, Kashiwa, Chiba 277-8582, Japan}}
\newcommand{\AFFipmu}{\affiliation{Kavli Institute for the Physics and
Mathematics of the Universe (WPI), The University of Tokyo Institutes for Advanced Study,
University of Tokyo, Kashiwa, Chiba 277-8583, Japan }}
\newcommand{\AFFmad}{\affiliation{Department of Theoretical Physics, University Autonoma Madrid, 28049 Madrid, Spain}}
\newcommand{\AFFubc}{\affiliation{Department of Physics and Astronomy, University of British Columbia, Vancouver, BC, V6T1Z4, Canada}}
\newcommand{\AFFbu}{\affiliation{Department of Physics, Boston University, Boston, MA 02215, USA}}
\newcommand{\AFFuci}{\affiliation{Department of Physics and Astronomy, University of California, Irvine, Irvine, CA 92697-4575, USA }}
\newcommand{\AFFcsu}{\affiliation{Department of Physics, California State University, Dominguez Hills, Carson, CA 90747, USA}}
\newcommand{\AFFcnm}{\affiliation{Institute for Universe and Elementary Particles, Chonnam National University, Gwangju 61186, Korea}}
\newcommand{\AFFduke}{\affiliation{Department of Physics, Duke University, Durham NC 27708, USA}}
\newcommand{\AFFfukuoka}{\affiliation{Junior College, Fukuoka Institute of Technology, Fukuoka, Fukuoka 811-0295, Japan}}
\newcommand{\AFFgifu}{\affiliation{Department of Physics, Gifu University, Gifu, Gifu 501-1193, Japan}}
\newcommand{\AFFgist}{\affiliation{GIST College, Gwangju Institute of Science and Technology, Gwangju 500-712, Korea}}
\newcommand{\AFFuh}{\affiliation{Department of Physics and Astronomy, University of Hawaii, Honolulu, HI 96822, USA}}
\newcommand{\AFFicl}{\affiliation{Department of Physics, Imperial College London , London, SW7 2AZ, United Kingdom }}
\newcommand{\AFFkek}{\affiliation{High Energy Accelerator Research Organization (KEK), Tsukuba, Ibaraki 305-0801, Japan }}
\newcommand{\AFFkobe}{\affiliation{Department of Physics, Kobe University, Kobe, Hyogo 657-8501, Japan}}
\newcommand{\AFFkyoto}{\affiliation{Department of Physics, Kyoto University, Kyoto, Kyoto 606-8502, Japan}}
\newcommand{\AFFliv}{\affiliation{Department of Physics, University of Liverpool, Liverpool, L69 7ZE, United Kingdom}}
\newcommand{\AFFmiyagi}{\affiliation{Department of Physics, Miyagi University of Education, Sendai, Miyagi 980-0845, Japan}}
\newcommand{\AFFnagoya}{\affiliation{Institute for Space-Earth Environmental Research, Nagoya University, Nagoya, Aichi 464-8602, Japan}}
\newcommand{\AFFkmi}{\affiliation{Kobayashi-Maskawa Institute for the Origin of Particles and the Universe, Nagoya University, Nagoya, Aichi 464-8602, Japan}}
\newcommand{\AFFpol}{\affiliation{National Centre For Nuclear Research, 02-093 Warsaw, Poland}}
\newcommand{\AFFsuny}{\affiliation{Department of Physics and Astronomy, State University of New York at Stony Brook, NY 11794-3800, USA}}
\newcommand{\AFFokayama}{\affiliation{Department of Physics, Okayama University, Okayama, Okayama 700-8530, Japan }}
\newcommand{\AFFosaka}{\affiliation{Department of Physics, Osaka University, Toyonaka, Osaka 560-0043, Japan}}
\newcommand{\AFFox}{\affiliation{Department of Physics, Oxford University, Oxford, OX1 3PU, United Kingdom}}
\newcommand{\AFFqmul}{\affiliation{School of Physics and Astronomy, Queen Mary University of London, London, E1 4NS, United Kingdom}}
\newcommand{\AFFregina}{\affiliation{Department of Physics, University of Regina, 3737 Wascana Parkway, Regina, SK, S4SOA2, Canada}}
\newcommand{\AFFseoul}{\affiliation{Department of Physics, Seoul National University, Seoul 151-742, Korea}}
\newcommand{\AFFsheff}{\affiliation{Department of Physics and Astronomy, University of Sheffield, S3 7RH, Sheffield, United Kingdom}}
\newcommand{\AFFshizuokasc}{\affiliation{Department of Informatics in
Social Welfare, Shizuoka University of Welfare, Yaizu, Shizuoka, 425-8611, Japan}}
\newcommand{\AFFstfc}{\affiliation{STFC, Rutherford Appleton Laboratory, Harwell Oxford, and Daresbury Laboratory, Warrington, OX11 0QX, United Kingdom}}
\newcommand{\AFFskk}{\affiliation{Department of Physics, Sungkyunkwan University, Suwon 440-746, Korea}}
\newcommand{\AFFtokyo}{\affiliation{The University of Tokyo, Bunkyo, Tokyo 113-0033, Japan }}
\newcommand{\AFFtodai}{\affiliation{Department of Physics, University of Tokyo, Bunkyo, Tokyo 113-0033, Japan }}
\newcommand{\AFFtit}{\affiliation{Department of Physics,Tokyo Institute of Technology, Meguro, Tokyo 152-8551, Japan }}
\newcommand{\AFFtus}{\affiliation{Department of Physics, Faculty of Science and Technology, Tokyo University of Science, Noda, Chiba 278-8510, Japan }}
\newcommand{\AFFtoronto}{\affiliation{Department of Physics, University of Toronto, ON, M5S 1A7, Canada }}
\newcommand{\AFFtriumf}{\affiliation{TRIUMF, 4004 Wesbrook Mall, Vancouver, BC, V6T2A3, Canada }}
\newcommand{\AFFtokai}{\affiliation{Department of Physics, Tokai University, Hiratsuka, Kanagawa 259-1292, Japan}}
\newcommand{\AFFtsinghua}{\affiliation{Department of Engineering Physics, Tsinghua University, Beijing, 100084, China}}
\newcommand{\AFFynu}{\affiliation{Department of Physics, Yokohama National University, Yokohama, Kanagawa, 240-8501, Japan}}
\newcommand{\AFFllr}{\affiliation{Ecole Polytechnique, IN2P3-CNRS, Laboratoire Leprince-Ringuet, F-91120 Palaiseau, France }}
\newcommand{\AFFbari}{\affiliation{ Dipartimento Interuniversitario di Fisica, INFN Sezione di Bari and Universit\`a e Politecnico di Bari, I-70125, Bari, Italy}}
\newcommand{\AFFnapoli}{\affiliation{Dipartimento di Fisica, INFN Sezione di Napoli and Universit\`a di Napoli, I-80126, Napoli, Italy}}
\newcommand{\AFFroma}{\affiliation{INFN Sezione di Roma and Universit\`a di Roma ``La Sapienza'', I-00185, Roma, Italy}}
\newcommand{\AFFpadova}{\affiliation{Dipartimento di Fisica, INFN Sezione di Padova and Universit\`a di Padova, I-35131, Padova, Italy}}
\newcommand{\AFFkeio}{\affiliation{Department of Physics, Keio University, Yokohama, Kanagawa, 223-8522, Japan}}
\newcommand{\AFFwinnipeg}{\affiliation{Department of Physics, University of Winnipeg, MB R3J 3L8, Canada }}
\newcommand{\AFFkcl}{\affiliation{Department of Physics, King's College London, London, WC2R 2LS, UK }}
\newcommand{\AFFwarwick}{\affiliation{Department of Physics, University of Warwick, Coventry, CV4 7AL, UK }}
\newcommand{\AFFral}{\affiliation{Rutherford Appleton Laboratory, Harwell, Oxford, OX11 0QX, UK }}
\newcommand{\AFFwu}{\affiliation{Faculty of Physics, University of Warsaw, Warsaw, 02-093, Poland }}
\newcommand{\AFFbcit}{\affiliation{Department of Physics, British Columbia Institute of Technology, Burnaby, BC, V5G 3H2, Canada }}
\newcommand{\AFFtohoku}{\affiliation{Department of Physics, Faculty of Science, Tohoku University, Sendai, Miyagi, 980-8578, Japan }}
\newcommand{\AFFicise}{\affiliation{Institute For Interdisciplinary Research in Science and Education, ICISE, Quy Nhon, 55121, Vietnam }}
\newcommand{\AFFilance}{\affiliation{ILANCE, CNRS, University of Tokyo International Research Laboratory, Kashiwa, Chiba 277-8582, Japan}}
\newcommand{\AFFibs}{\affiliation{Institute for Basic Science (IBS), Daejeon, 34126, Korea}}

\AFFicrr
\AFFkashiwa
\AFFicrronly
\AFFmad
\AFFbu
\AFFbcit
\AFFuci
\AFFcsu
\AFFcnm
\AFFduke
\AFFllr
\AFFfukuoka
\AFFgifu
\AFFgist
\AFFuh
\AFFibs
\AFFicise
\AFFicl
\AFFbari
\AFFnapoli
\AFFpadova
\AFFroma
\AFFilance
\AFFkeio
\AFFkek
\AFFkcl
\AFFkobe
\AFFkyoto
\AFFliv
\AFFmiyagi
\AFFnagoya
\AFFkmi
\AFFpol
\AFFsuny
\AFFokayama
\AFFox
\AFFral
\AFFseoul
\AFFsheff
\AFFshizuokasc
\AFFstfc
\AFFskk
\AFFtohoku
\AFFtokai
\AFFtokyo
\AFFtodai
\AFFipmu
\AFFtit
\AFFtus
\AFFtoronto
\AFFtriumf
\AFFtsinghua
\AFFwu
\AFFwarwick
\AFFwinnipeg
\AFFynu
\author{K.~Okamoto}
\AFFicrr
\author{K.~Abe}
\author{Y.~Hayato}
\AFFicrr
\AFFipmu
\author{K.~Hiraide}
\AFFicrr
\AFFipmu
\author{K.~Hosokawa}
\author{K.~Ieki}
\AFFicrr
\author{M.~Ikeda}
\AFFicrr
\author{J.~Kameda}
\AFFicrr
\AFFipmu
\author{Y.~Kanemura}
\author{Y.~Kaneshima}
\author{Y.~Kataoka}
\author{Y.~Kashiwagi}
\author{S.~Miki}
\AFFicrr
% ...


\collaboration{245}{The Super-Kamiokande Collaboration}



\begin{abstract}
Neutrinos associated with solar flares~(solar-flare neutrinos) provide information on particle acceleration mechanisms during the impulsive phase of solar flares. We searched using the Super-Kamiokande detector for neutrinos from solar flares that occurred during solar cycles $23$ and $24$, including the largest solar flare~(X28.0) on November 4th, 2003. In order to minimize the background rate we searched for neutrino interactions within narrow time windows coincident with $\gamma$-rays and soft X-rays recorded by satellites. In addition, we performed the first attempt to search for solar-flare neutrinos from solar flares on the invisible side of the Sun by using the emission time of coronal mass ejections~(CMEs). By selecting twenty powerful solar flares above X5.0 on the visible side and eight CMEs whose emission speed exceeds $2000~\mathrm{km \, s^{-1}}$ on the invisible side from 1996 to 2018, we found two~(six) neutrino events coincident with solar flares occurring on the visible~(invisible) side of the Sun, with a typical background rate of $0.10$~($0.62$)~events per flare in the MeV--GeV energy range. No significant solar-flare neutrino signal above the estimated background rate was observed. As a result we set the following upper limit on neutrino fluence at the Earth $\mathit{\Phi}<1.1\times10^{6}~\mathrm{cm^{-2}}$ at the $90\%$~confidence level for the largest solar flare. The resulting fluence limits allow us to constrain some of the theoretical models for solar-flare neutrino emission.

\end{abstract}



\section{Introduction} \label{sec:intro}

Solar flares are the largest explosive events that occur around the surface of the Sun. This phenomenon is caused by the reconnection of magnetic field lines above sun spots and produces electromagnetic radiation from radio to $\gamma$-rays~\citep{1974IAUS...57..105K}. Solar flares sometimes occur associated with coronal mass ejections~(CMEs), which are eruptions of the atmospheric plasma into interplanetary space~\citep{1974JGR....79.1799C, 1975SoPh...40..439G}.
The frequency of these explosive events is strongly correlated with the activity of the Sun.

For both types of energetic events, the typical energy released by the explosion is estimated in the range of $10^{27}$--$10^{32}$~erg~\citep{1963QJRAS...4...62E,1994ESASP.373..409H, 2010ApJ...722.1522V}. When a solar flare occurs, high energy particles that do not normally exist in the solar atmosphere are generated~\citep{1973Natur.241..333C, 1974SoPh...35..193D}. Solar imaging methods can partially identify the location at which these particles are generated, such as the magnetic loop-top~\citep{1994Natur.371..495M, 2010ApJ...714.1108K}, the loop-foot~\citep{2008ApJ...675.1645F}, and the reconnection point~\citep{2014ApJ...787..125N}. This indicates that these particles are accelerated by solar flares.  Although the acceleration mechanisms remain poorly understood, several theoretical models of solar flares are proposed~\citep{1998ApJ...495L..67T, 2004A&A...419.1159K, 2008ApJ...676..704L}.

Neutral particles associated with solar flares, such as $\gamma$-rays and neutrinos, are important to test theoretical aspects of particle acceleration in the magnetic reconnection because they can escape from the acceleration site\footnote{Some of line $\gamma$-rays cannot escape from the photosphere due to Compton scattering.}. Their observation reveals both the spatial and time profile of primary particle acceleration while primary and secondary charged particles are trapped by the magnetic field.

Neutrinos are only produced by accelerated protons above $300$~MeV~\citep{1975SSRv...18..341R, 1995ARA&A..33..239H}, which can generate pions~($\pi^{\pm}$ and $\pi^{0}$) by interacting with dense plasma in the lower solar atmosphere during solar flares. The generated $\pi^{\pm}$ produce neutrinos in their decay chain. 
\begin{figure*}[]
\centering\includegraphics[width=0.8\textwidth]{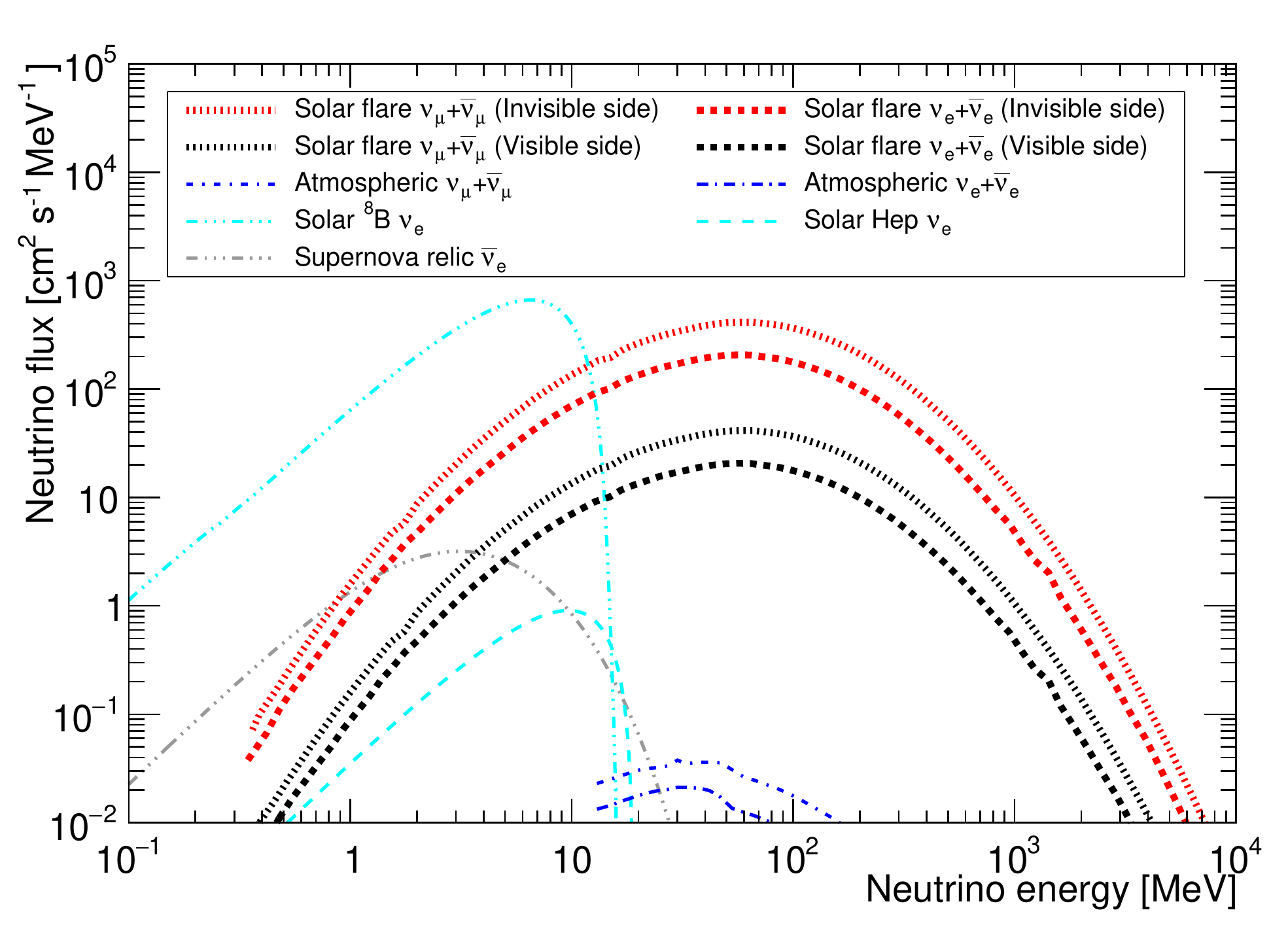}
\caption{Typical solar-flare neutrino fluxes from a powerful solar flare~(\cite{2003ChJAS...3...75F}, red thick and black thick lines) together with other neutrino fluxes, such as atmospheric neutrinos~(\cite{2005APh....23..526B}, blue lines), solar neutrinos~(\cite{2001ApJ...555..990B}, light blue lines), and relic neutrinos~(\cite{2009PhRvD..79h3013H}, light gray line).\label{fig:flare-flux}}
\end{figure*}
Figure~\ref{fig:flare-flux} shows the typical neutrino fluxes from a powerful solar flare together with other neutrino fluxes, such as atmospheric neutrinos, solar neutrinos, and supernova relic neutrinos. In this article, we refer to such neutrinos from solar-flares as solar-flare neutrinos.

Solar-flare neutrinos have been searched for by neutrino detectors since the 1980s. However, no clear signal has been found in spite of setting a timing gate coincident with soft X-rays from visible solar flares. Setting a narrow search window, which covers only the period of neutrino production, \cite{2016arXiv160600681D} and \cite{2020SoPh..295..133O} proposed to open search windows coincident with $\gamma$-rays originating from $\pi^{0}$ decays and nuclear interactions. These methods are helpful to minimize the background rate in the neutrino searches.

During the solar cycles 23~($1996$--$2008$) and 24~($2008$--$2019$), the Super-Kamiokande detector~(hereafter SK) had been operating since April 1st, 1996~\citep{2003NIMPA.501..418F} with five distinct periods, from SK-I to SK-V with ultra-pure water. Although several neutrino telescopes are running during those solar cycles, SK is unique in its search for solar-flare neutrinos because its data set covers almost two cycles of solar activity. The largest solar flare, whose class is X$28.0$ and which occurred on November 4th 2003~\citep{2005AA...433.1133K}, is also included. 

This paper is organized as follows. In Section~\ref{sec:review} we provide a brief overview of neutrinos associated with solar flares and the determination of the search windows to find solar-flare neutrinos. In Section~\ref{sec:analysis} we describe the performance of the SK detector and the analysis methods to search for solar-flare neutrinos within the selected search windows. In Section~\ref{sec:result} and Section~\ref{sec:discuss} we present analysis results and make comparisons to results from other neutrino experiments. In the final section we conclude this study and give future prospects.

\section{solar-flare neutrinos} \label{sec:review}

\subsection{Particle acceleration and neutrino production in solar flare} \label{sec:time-intro}

In many solar flares, the hard X-ray, (line)~$\gamma$-ray, and microwave emissions are observed almost simultaneously~\citep{1981ApJ...244L.171C, 1983Natur.305..292N, 1984JPSJ...53.4499Y}. Those observations suggest that electrons, protons, and ions are accelerated over a short period of time. 

The $\gamma$-ray emissions from solar flares provide information on the processes of proton acceleration and the subsequent reactions of the protons in the chromosphere. For example, line $\gamma$-rays from neutron capture on hydrogen~\citep{1973Natur.241..333C} and de-excitation $\gamma$-rays from $\mathrm{^{12}C}$ and $\mathrm{^{16}O}$~\citep{2003ApJ...595L..81S} imply the acceleration and nuclear reactions of protons. The time profile of $\gamma$-rays from $\pi^{0}$ decays~($\pi^{0}\to2\gamma$) also provides information on the time scale of neutrino production since charged pions can be generated at the same time as neutral pions. Observations of $\gamma$-rays from $\pi^{0}$ decay have been performed by several instruments on-board satellites in geostationary and polar orbits~\citep{1986AdSpR...6f.115F, 1987ApJ...318..913C,1993A&AS...97..345L,1997ApJ...479..997D,1993A&AS...97..349K,2010CosRe..48...70K,2017ApJ...835..219A}. Those $\gamma$-ray observations indirectly demonstrate that protons are accelerated up to relativistic energies and subsequently neutrinos should be produced during solar flares.

On the theoretical side, several simulations of neutrino emission from solar flares have been developed in order to estimate the expected event rate in neutrino detectors. Table~\ref{tb:neutrino-model} summarizes the features of three theoretical models for solar-flare neutrinos by~\cite{1991NCimC..14..417K}, \cite{2003ChJAS...3...75F}, and \cite{2013ICRC...33.3656T}.

\begin{table*}[]
    \begin{center}
    \caption{The summary of theoretical models for solar-flare neutrinos. In each theoretical model, the number of expected interactions in the SK detector are calculated. For the expected number of events in the SK detector from~\cite{2003ChJAS...3...75F}, the conversion factor~($\eta$ defined in~\cite{2003ChJAS...3...75F}) is assumed to be ${\sim}0.10$ for the visible side and ${\sim}1.0$ for the invisible side.}
        \label{tb:neutrino-model}
        \begin{tabular}{cccccc}
        \hline
        Theoretical model & Side of the Sun & Power index & Directional feature of  & Number of expected  \\
        (Reference) & &  of proton & generated neutrino & events in SK~[$\mathrm{flare^{-1}}$] \\ \hline
        \cite{1991NCimC..14..417K} & Visible side & $3$--$4$ & Isotropic &$1.36 \times 10^{-4}$ \\ 
        \cite{1991NCimC..14..417K} & Invisible side & $1$ & Beam like & $0.85$ \\ 
        \cite{2004JHEP...06..045F} & Visible side & -- & Isotropic & $0.75$  \\ 
        \cite{2004JHEP...06..045F} & Invisible side & -- & Beam like & $7.5$ \\ 
        \cite{2013ICRC...33.3656T} & Invisible side & $3$ & Isotropic & $9.0 \times 10^{-5}$ \\ 
        \cite{2013ICRC...33.3656T} & Invisible side & $1$ & Beam like &$3.8 \times 10^{-6}$ \\ \hline
        \end{tabular}
    \end{center}
\end{table*}

These theoretical models describe neutrino emission from the most powerful solar flares whose energy is larger than $10^{31}$~erg. The neutrino fluxes change depending on the assumptions in the models, such as the proton spectral index, the interaction cross sections between accelerated protons and nuclei in the chromosphere, the angular distribution of neutrinos with respect to the proton direction, and the location of the solar flare as summarized in Table~\ref{tb:neutrino-model}. Although the absolute fluxes are quite different, the predicted energy spectrum of solar-flare neutrinos is almost identical to the atmospheric neutrino energy spectrum.  This is because in both cases neutrinos are created through the same production process. 

\cite{2004JHEP...06..045F} estimated the number of neutrino interactions between neutrinos and the free protons in the water of the SK detector;
\begin{equation}
    n_{\mathrm{int}} = 
     7.5 \, \eta \left(\frac{E_{\mathrm{FL}}}{10^{31}~\mathrm{erg}} \right), \label{eq_eta}
\end{equation}
\noindent where $n_{\mathrm{int}}$ is the number of interactions in the SK detector~(fiducial volume $22.5$~kton),  $E_{\mathrm{FL}}$ is the total energy of the solar flare, and $\eta$ is an energy conversion factor from the solar flare~$E_{\mathrm{FL}}$ to neutrino energy, as defined in~\cite{2003ChJAS...3...75F}. 
According to~\cite{2004JHEP...06..045F}, several neutrino interactions are expected in the SK detector when a solar flare classified as the largest explosion~($\geq 10^{32}$~erg) occurs on the visible~(invisible) side  of the Sun and $\eta {\sim} 0.10~(1.0)$. On the other hand, \cite{2013ICRC...33.3656T} argues that the assumed value of $\eta$ is questionable and it is typically of order $10^{-6}$. Therefore, experimental searches for neutrinos from powerful solar flares can test theoretical aspects of neutrino production during the impulsive phase of the solar flares.

\subsection{Searches for solar-flare neutrinos using neutrino detectors} \label{sec:neutrino}

The possibility of detecting solar-flare neutrinos with neutrino experiments has been discussed since the 1980's~\citep{1982JETPL..35..341B, 1983ICRC....7..104E}. In 1988, the Homestake experiment reported an excess of neutrino events when energetic solar flares occurred~\citep{1994PrPNP..32...13D}. This observation suggested a possible correlation between solar flares and the neutrino capture rate on $\mathrm{^{37}Cl}$~\citep{1988PhRvL..61.2650B,1987ApJ...320L..69B}. Soon after the Mont Blanc Neutrino Detector searched for solar-flare neutrinos but no significant signal was found in time coincidence with any solar flares that occurred between 1988 to 1991, including the largest solar flare in 1989~\citep{1991ApJ...382..344A}. Since then various neutrino detectors have searched for solar-flare neutrinos by analyzing different solar flare samples~\citep{1988PhRvL..61.2653H, 1990ApJ...359..574H, 2012ApJ...745..193G, 2014APh....55....1A, Agostini:2019yuq, 2021PhRvD.103j2001A, 2022ApJ...924..103A}. However, no significant signal for solar-flare neutrinos has been found by any of these experiments. 

\subsection{Search window for solar flares on the visible side of the Sun} \label{sec:window}

Atmospheric neutrinos are continuously produced by collisions between primary cosmic rays and nuclei in the Earth's atmosphere~\citep{2016PhRvD..94e2001R}. In neutrino experiments the separation between atmospheric neutrinos and solar-flare neutrinos is technically difficult since their energy ranges overlap with each other due to their identical production process. While atmospheric neutrinos are generated constantly, solar-flare neutrinos are released only during the period of particle acceleration during the flare. Therefore, a search window that is appropriately narrow in time allows neutrino detectors to substantially reduce the atmospheric neutrino background rate.

The first proposal to use search windows when searching for solar-flare neutrinos was published by~\cite{2016arXiv160600681D}, which analyzed the detection time of $\gamma$-rays from $\pi^{0}$ decays using the Fermi-Large Area Telescope~(LAT) satellite~\citep{2009ApJ...697.1071A}. Following this proposal, the IceCube collaboration searched for neutrinos from solar flares in the energy range from $500$~MeV to $5$~GeV~\citep{2021PhRvD.103j2001A} and constrained the integrated neutrino flux emitted during the considered time window according to the catalog of $\gamma$-ray flares recorded by Fermi-LAT~\citep{2021ApJS..252...13A}. However, this catalog covers the period of solar cycle~$24$~($2008$--$2019$) after the launch of Fermi-LAT in 2008. Hence, a different method must be used to identify search windows for solar flares that occurred before 2008.

\cite{2020SoPh..295..133O} proposed to determine the search window by analyzing $2.2$~MeV line $\gamma$-rays and the derivative of soft X-rays to improve the signal-to-noise ratio to find solar-flare neutrinos. The former channel selected three solar flares across solar cycles 23 and 24 with the observation of line $\gamma$-rays recorded by the RHESSI satellite\footnote{The Reuven Ramaty High-Energy Solar Spectroscopic Imager~(RHESSI)}~\citep{2002SoPh..210....3L} on July 23rd 2002~(X5.1), November 2nd 2003~(X9.2), and January 20th 2005~(X7.1). The latter channel selected twenty-three solar flares~(above X5.0) recorded by the GOES\footnote{Geostationary Operational Environmental Satellite~(GOES)}~\citep{2004SPIE.5171...65L} across solar cycles 23 and 24. Note that this selection set the search window for the largest flare~(X28.0) on November 4th, 2003. Although the derivative of soft X-rays extracts the time scale of non-thermal electron acceleration in general, this channel is still appropriate to improve the signal-to-noise ratio for finding solar-flare neutrinos because the recent study by Fermi-LAT concluded ions and electrons are accelerated, transported, and interact with the ambient medium at the same time~\citep{2021ApJS..252...13A}. In the latter section, we separately searched in the SK detector for neutrinos from selected solar flares occurred on the visible side of the Sun within these two different search windows.

\subsection{Search windows for solar flares on the invisible side of the Sun} \label{sec:invisible}

Energetic proton flux directed back to the Sun generates a nuclear cascade in the solar atmosphere. Such flux results in narrow beam of relativistic protons with a rather hard spectrum from solar flares on the invisible side of the Sun. Hence the searches for neutrinos associated from solar flares occurring at the invisible side provide information about acceleration mechanism of downward going proton flux. \cite{2003ChJAS...3...75F} argue that the probability of solar-flare neutrino detection increases when solar flares occur on the invisible side of the Sun due to efficient collisions between the accelerated protons and the dense plasma at the surface of the Sun. Searching for neutrinos from solar flares on the invisible side of the Sun allows us to test these proposed neutrino production models. However, a selection of solar flares that occur on the invisible side of the Sun has never been performed for this purpose. 

To select solar flares that occur on the invisible side of the Sun, the time of CME emission allows one to infer the occurrence time of solar flares because large energetic solar flares are usually accompanied by CMEs~\citep{2003SoPh..218..261A}. The observation of CMEs occurring on the invisible side of the Sun has been performed by the LASCO\footnote{The Large Angle Spectroscopic Coronagraph~(LASCO) on the Solar and Heliospheric Observatory~(SOHO)} coronagraph~\citep{1995SoPh..162....1D, 1995SoPh..162..357B} and the CME emission times are listed in a catalog maintained by NASA~\citep{2004JGRA..109.7105Y}.

From the catalog we selected energetic CMEs whose emission speed is more than $2000~\mathrm{km{\,}s^{-1}}$, which roughly corresponds to class X2.0 solar flares. This criteria allowed us to select ten~CMEs from 1996 April to 2018 May. The search window for solar flares occurring on the invisible side is set to $7238$~s as explained in Appendix~\ref{app:cme}. The date of the selected CMEs is summarized in Table~\ref{tb:time-invisible} in Appendix~\ref{app:cme}.

\section{Detector and analysis} \label{sec:analysis}

\subsection{The Super-Kamiokande detector}

Super-Kamiokande is a water Cherenkov detector in a cavern beneath Ikeno-yama mountain, Japan~\citep{2003NIMPA.501..418F}. It is a cylindrical stainless tank structure and contains $50$~kiloton~(ktons) of ultra-pure water. The detector is divided into two regions by the tank structure, separated optically by Tyvek sheets: one is the inner detector~(ID) and the other is the outer detector~(OD). The ID serves as the target volume for neutrino interactions and the OD is used to veto external cosmic-ray muons as well as $\gamma$-rays from the surrounding rock. In the ID, the diameter~(height) of the cylindrical tank is $33.8$~m~($36.2$~m). It contains $32$~kton of water and holds $11,129$~inward-facing $20$-inch photomultipliers~(PMTs)\footnote{The SK-I detector used $11,149$~PMTs while the other phases use $11,129$~PMTs except for SK-II, which used $5,182$.} to observe the Cherenkov light produced by charged particles. The diameter~(height) of the OD tank is $39.3$~m~($41.4$~m). The detector simulation has been developed using the \textsc{Geant3} toolkit~\citep{Brun:1994aa} and tuned to calibration data. The details of the detector configuration, the calibration, and the performance can be found elsewhere~\citep{2014NIMPA.737..253A}. In this article, we analyzed the data taken in SK-I through SK-IV~(from April 1996 to May 2018) to cover solar cycles 23 and 24\footnote{During SK-V~(from January 2019 to July 2020), no solar flare above X5.0 occurred due to low solar activity.}.

In order to determine the initial neutrino interaction vertex and the trajectories and momenta of any subsequent charged particles, event reconstruction is performed by analyzing the timing and the ring pattern of the observed Cherenkov light in the SK detector. Using this water Cherenkov technique the SK detector has sensitivity to a wide range of neutrino energies, from a few MeV to tens of GeV. The neutrino events are categorized into two samples depending on the energy of the reconstructed charged particles after the initial neutrino interaction. A neutrino event reconstructed with less than $100$~MeV of visible energy is categorized as part of the ``low energy sample'' and is mainly used for studies of solar neutrinos~\citep{2016PhRvD..94e2010A} and supernova neutrinos~\citep{2021PhRvD.104l2002A}. In this energy region the reconstruction tool searches for an interaction point because the track length of the charged particle is at most $30$~cm, which is small compared to the reconstructed vertex resolution~(typically more than $50$~cm). On the other hand, a neutrino event reconstructed with more than $100$~MeV of visible energy is categorized as part of the ``high energy sample'' and is mainly used for the study of atmospheric neutrinos~\citep{2016PhRvD..94e2001R} and to search for proton decay~\citep{2020PhRvD.102k2011T}. In this energy region the majority of neutrino interactions occur on nuclei and can produce a number of charged particles. The event reconstruction algorithm then determines the number of Cherenkov rings in the event, identifies the particle type that created each ring, locates the interaction vertex and predicts the energy of each charged particle.

\subsection{Low energy sample}

The SK detector can potentially reconstruct the energies of charged particles down to a few MeV. In the energy range of the low energy sample, the dominant reaction is the inverse beta decay~(IBD) of electron anti-neutrinos because of its relatively large cross section. Other sub-dominant reactions are elastic scattering between electrons and electron neutrinos, and the charged current and neutral current interactions with oxygen~\citep{2002PhRvD..66a3007K}. Even though we set an appropriate search window for solar-flare neutrinos, the signal-to-noise ratio is still poor below $16$~MeV due to solar neutrinos and background events originated from radioactive isotopes dissolved in the SK water~\citep{2020NIMPA.97764297N} and produced by penetrating muons~\citep{Super-Kamiokande:2015xra}. Hence, we set the energy threshold to $16$~MeV in this analysis, where the total energy of a positron produced by the IBD is considered. For the energy range above $16$~MeV, the possible backgrounds are atmospheric neutrino interactions, decay electrons originated from invisible muons, and low energy pions. For selecting positrons from IBD reactions we applied the selection cuts used for supernova relic neutrino searches since these cut criteria are optimized to maximize the event selection efficiency of the positron. The detailed analysis method is described in~\cite{2012PhRvD..85e2007B} and \cite{2021PhRvD.104l2002A}.

\begin{figure*}[]
    \begin{minipage}{0.5\hsize}
        \centering\includegraphics[width=1.0\linewidth]{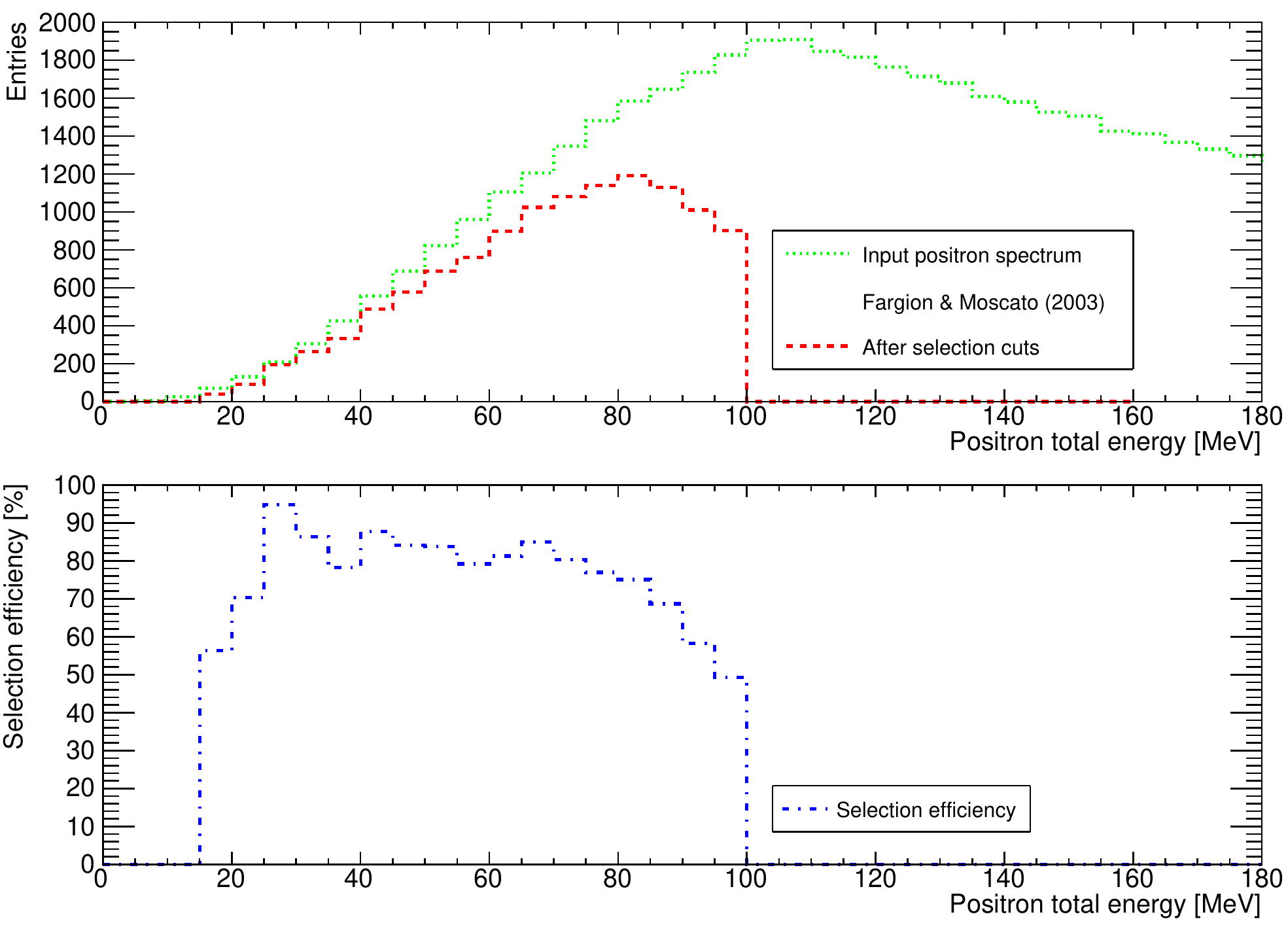}
    \end{minipage}
    \begin{minipage}{0.5\hsize}
        \centering\includegraphics[width=1.0\linewidth]{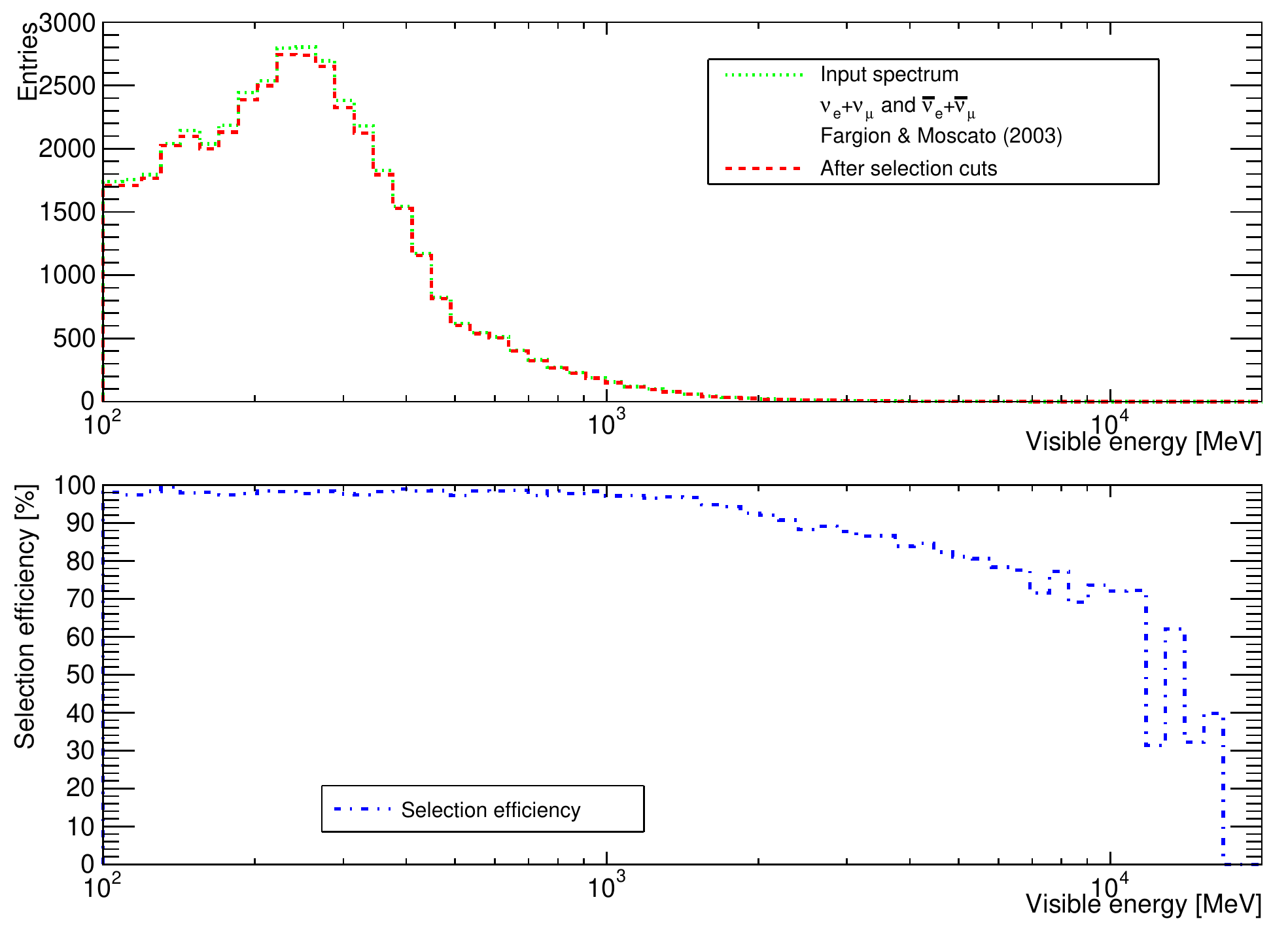}
    \end{minipage}
    \caption{Energy spectra and selection efficiencies for the low~(left) and high~(right) energy samples. The distributions of the reconstructed positron energy and the visible energy calculated using the MC simulation which uses the energy spectra of solar-flare neutrinos from~\cite{2003ChJAS...3...75F}. The horizontal axis is reconstructed positron kinetic energy in the left panel~(reconstructed visible energy in the right panel), and the vertical axis shows the number of events. The light-green histograms represent the energy spectra of the generated events and the red histograms represent the energy spectrum after the selection cuts are applied. The blue histograms represent the selection efficiencies. \label{fig:e-dist-input}}
\end{figure*}

For evaluating the event selection efficiency in the low energy sample, we first simulated positrons from IBD reactions. In this simulation, we use the neutrino energy spectrum from~\cite{2003ChJAS...3...75F}. Then, the positron energy is calculated by considering the cross section of the IBD reaction from~\cite{2003PhLB..564...42S}. Note that  only the IBD reaction is considered in this simulation because of its large cross section in this energy range. Figure~\ref{fig:e-dist-input}~(left-top) shows the input energy distribution of positrons from the neutrino interactions and the reconstructed total positron energy after selection cuts.

Assuming the electron anti-neutrino energy spectrum of~\cite{2003ChJAS...3...75F}, the selection efficiency, defined as $\varepsilon_\mathrm{low}^{\mathrm{Fargion}}$, in the low energy sample is about $27\%$, since the energy range covered by the low energy sample is relatively narrow. We also evaluated the selection efficiency for neutrinos between $16$ and $100$~MeV by generating a flat neutrino energy distribution. This produced the model independent analysis detailed in Section~\ref{sec:ind}. That selection efficiency, defined as $\varepsilon^{\mathrm{Ind}}_{\mathrm{low}}$, is about $75\%$. Table~\ref{tb:bg_rate} summarizes the livetime, the selection efficiencies, and the background rate after all reduction cuts in the low energy sample using all of the SK data sets.

\begin{table*}[]
    \begin{center}
    \caption{The summary of the dates, the livetimes, the selection efficiencies~($\varepsilon^{\mathrm{Fargion}}_{\mathrm{low}}$ and $\varepsilon^{\mathrm{Fargion}}_{\mathrm{high}}$) for the energy spectrum from~\cite{2003ChJAS...3...75F}, the model-independent selection efficiencies~($\varepsilon^{\mathrm{Ind}}_{\mathrm{low}}$), and the background rates of the low energy sample and the high energy sample.  The difference in their livetimes comes from differences in the SK detector data quality between the low and high energy analyses.}
        \label{tb:bg_rate}
        \begin{tabular}{cccccc}
            \hline
            Category& SK phase & SK-I & SK-II & SK-III & SK-IV \\ \hline
            Date & Start & Apr. 1996 & Oct. 2002 & Jul. 2006 & Sep. 2008 \\  
            & End & Jul. 2001 & Oct. 2005 & Aug. 2008 & May 2018 \\
            \hline
            & Livetime~[day] & 1497 & 794 & 562 & 2970 \\
            Low energy & Selection efficiency~($\varepsilon_{\mathrm{low}}^{\mathrm{Fargion}}$)~[$\%$] & $26.2$ & $27.1$ & $27.8$ & $28.3$ \\
            & Selection efficiency~($\varepsilon^{\mathrm{Ind}}_{\mathrm{low}}$)~[$\%$] & $72.3$ & $74.8$ & $76.6$ & $78.1$ \\
            & Background rate~[$\mathrm{event \, day^{-1}}$] & $0.20\pm0.01$ & $0.19\pm0.02$ & $0.20\pm0.01$ & $0.19\pm0.01$ \\ \hline
            & Livetime~[day] & 1489 & 825 & 522 & 3235 \\
            High energy & Selection efficiency~($\varepsilon_{\mathrm{high}}^{\mathrm{Fargion}}$)~[$\%$] & $61.9$ & $62.1$ & $61.8$ & $61.6$ \\
            & Background rate~[$\mathrm{event \, day^{-1}}$] & $7.45\pm0.07$ & $7.33\pm0.09$ & $7.53\pm0.12$ & $7.48\pm0.05$ \\
            \hline
        \end{tabular}
    \end{center}
\end{table*}

After the installation of new front-end electronics at SK-IV~\citep{Super-Kamiokande:2010kjr}, the SK trigger system allows the detector to tag neutron signals after IBD reactions using the delayed coincidence technique. However,  we did not require the neutron signal to identify electron anti-neutrinos in this analysis since the trigger system did not allow us to record the neutron signals in SK-I, -II, and -III.

\subsection{High energy sample}

This sample is further divided into three sub-samples based on the event topology: a fully contained~(FC) sample, a partially contained~(PC) sample, and an upward going muon~(UPMU) sample. In this study, we only analyzed the FC sample because the energy of solar-flare neutrinos is less than $10$~GeV and this results in the tracks of the all charged particles being essentially contained in the inner tank. 
In the energy region of the high energy sample different interactions occur depending on the neutrino energy. For simulating neutrino interactions with hydrogen and oxygen in the detector, we used the NEUT generator~\citep{2021EPJST.230.4469H}. Figure~\ref{fig:e-dist-input}~(right-top) shows the input energy distribution of charged particles from the neutrino interactions and the reconstructed energies after selection cuts. The event selection criteria for the high energy sample are detailed in~\cite{2005PhRvD..71k2005A}. The selection efficiency for the neutrino spectrum from~\cite{2003ChJAS...3...75F} is typically $62\%$ after all reduction cuts. Table~\ref{tb:bg_rate} also summarizes the livetime, the selection efficiency, and the background rate of the high energy sample for each SK phase.

\begin{figure*}[]
\centering\includegraphics[width=1.0\textwidth]{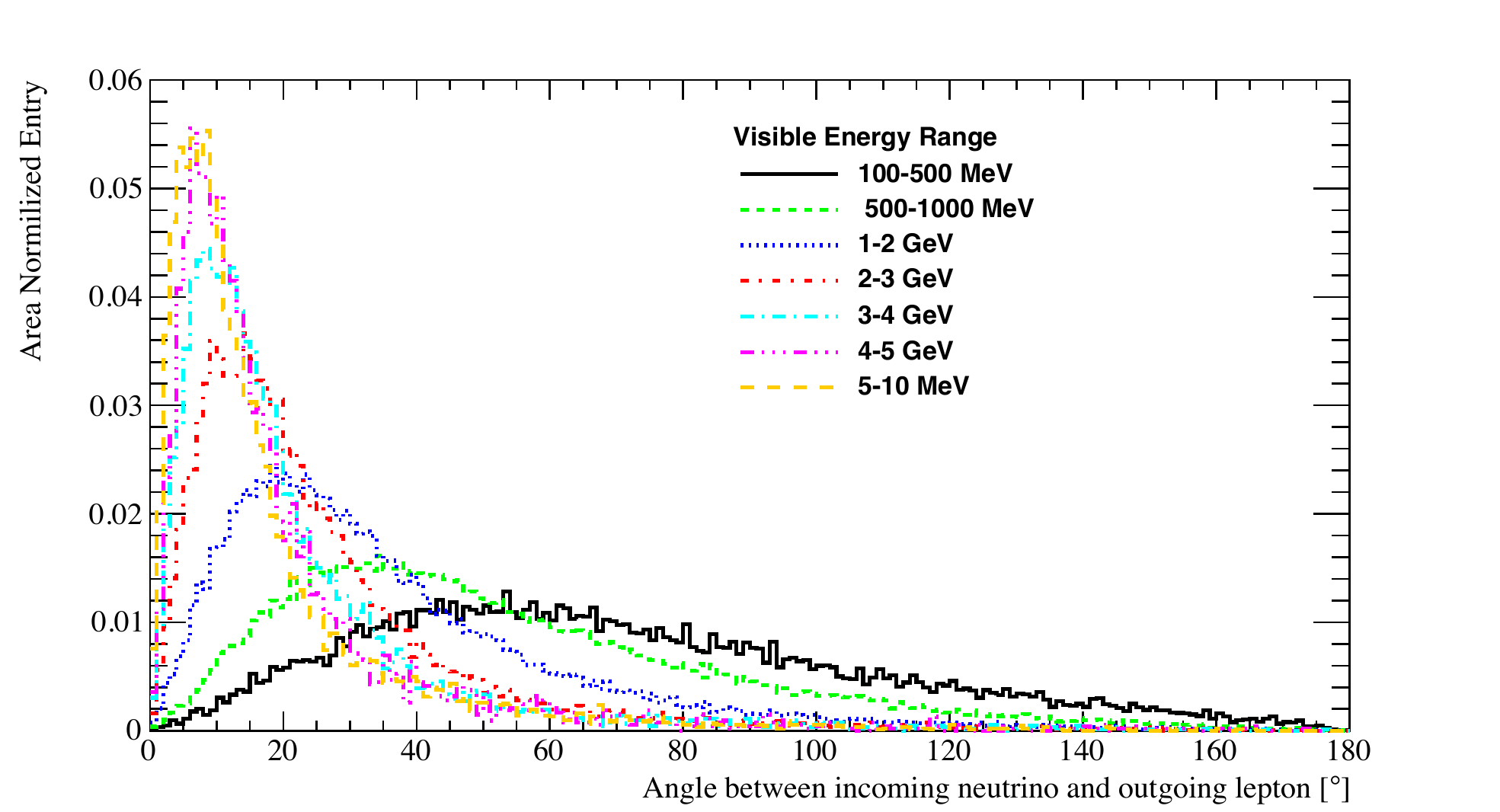}
\caption{Angular distribution between the direction of the incident neutrino and the reconstructed directions of produced charged particles~($e^{\pm}$, $\mu^{\pm}$, and $\pi^{\pm}$) from the MC simulation of the high energy sample. For multi-ring events, the direction of the neutrino is reconstructed as the momentum weighted sum of the directions of all the identified rings. 
\label{fig:angular-resolution}}
\end{figure*}

Directional information can be used to test whether neutrino signals come from a specific astrophysical source or not. Figure~\ref{fig:angular-resolution} shows the typical distribution of angles between the incoming neutrino and the direction of the final state charged particles based on the MC simulation. In the energy region above $1$~GeV the direction of the final state charged particles is highly correlated with the direction of the incoming neutrino and this tendency clearly depends on the neutrino energy.

\section{Results} \label{sec:result}
\subsection{Results for solar-flare neutrinos coincident with line $\gamma$-ray observations} \label{result_line}

As explained in Section~\ref{sec:time-intro}, $2.2$~MeV line $\gamma$-rays are produced by the acceleration of hadrons, their interactions with nuclei in the chromosphere, and the production of neutrinos. Hence, the signal-to-noise ratio of solar-flare neutrinos is high when in coincidence with line $\gamma$-rays. For this reason, we searched for neutrino candidate events within the search windows determined by the light curve of line $\gamma$-rays by \cite{2020SoPh..295..133O}. As explained in Section~\ref{sec:window}, three solar flares on the visible side of the Sun are selected by~\cite{2020SoPh..295..133O}. The SK data does not cover the period of the solar flare that occurred on July 23rd 2002 due to the re-instrumentation of the SK detector following the implosion accident in 2001. Within the remaining two search windows no signal was found in either the low or high energy samples. 

\subsection{Results for solar flares on the visible side of the Sun} \label{result_soft}

We searched for solar-flare neutrinos from the visible side of the Sun within the search windows determined by the time derivative of soft X-rays recorded by the GOES satellite, as described in Section~\ref{sec:window}. \cite{2020SoPh..295..133O} selected twenty-three solar flares using this channel, with SK missing three of these (on August 25th 2001, December 13th 2001 and July 23rd 2002) because of the re-instrumentation work discussed previously. Hence, we searched for neutrinos from twenty solar flares across solar cycles 23 and 24. No signal was found in the low energy sample while two events are found in the high energy sample. The first event was observed on November 4th 2003 and the second event on September 6th 2017, as summarized in Table~\ref{tb:summary-event}.

\begin{table*}[]
    \begin{center}
    \caption{Summary of two events observed within the search window for neutrinos associated with a solar flare on the visible side of the Sun. The estimated background rate is normalized by the duration of the corresponding search window determined by~\cite{2020SoPh..295..133O}.}
        \label{tb:summary-event}
        \begin{tabular}{ccc}
            \hline
            Date~(UTC) & November 4th, 2003 & September 6th, 2017 \\
            Solar flare class & X$28.0$ & X$9.4$ \\
            \hline
            SK phase & SK-II & SK-IV \\
            Observed time~(UTC) & 19:42:26 & 12:03:05 \\
            Duration of window~[s] & 1144 & 521 \\
            Event topology & $2$-ring  $e$-like & $1$-ring $\mu$-like \\
            Reconstructed energy & $178.3$~MeV & $1.2$~GeV \\
            $\theta_{\mathrm{Sun}}$ & $67.1^{\circ}$ & $39.6^{\circ}$ \\
            Estimated background rate~[$\mathrm{event{\,}flare^{-1}}$] & $0.20$ & $0.12$ \\
            $p$-value of the null hypothesis & $18.1\%$ & $11.3\%$ \\
            \hline
        \end{tabular}
    \end{center}
\end{table*}

\begin{figure*}[]
    \begin{minipage}{0.5\hsize}
        \centering\includegraphics[width=1.0\textwidth]{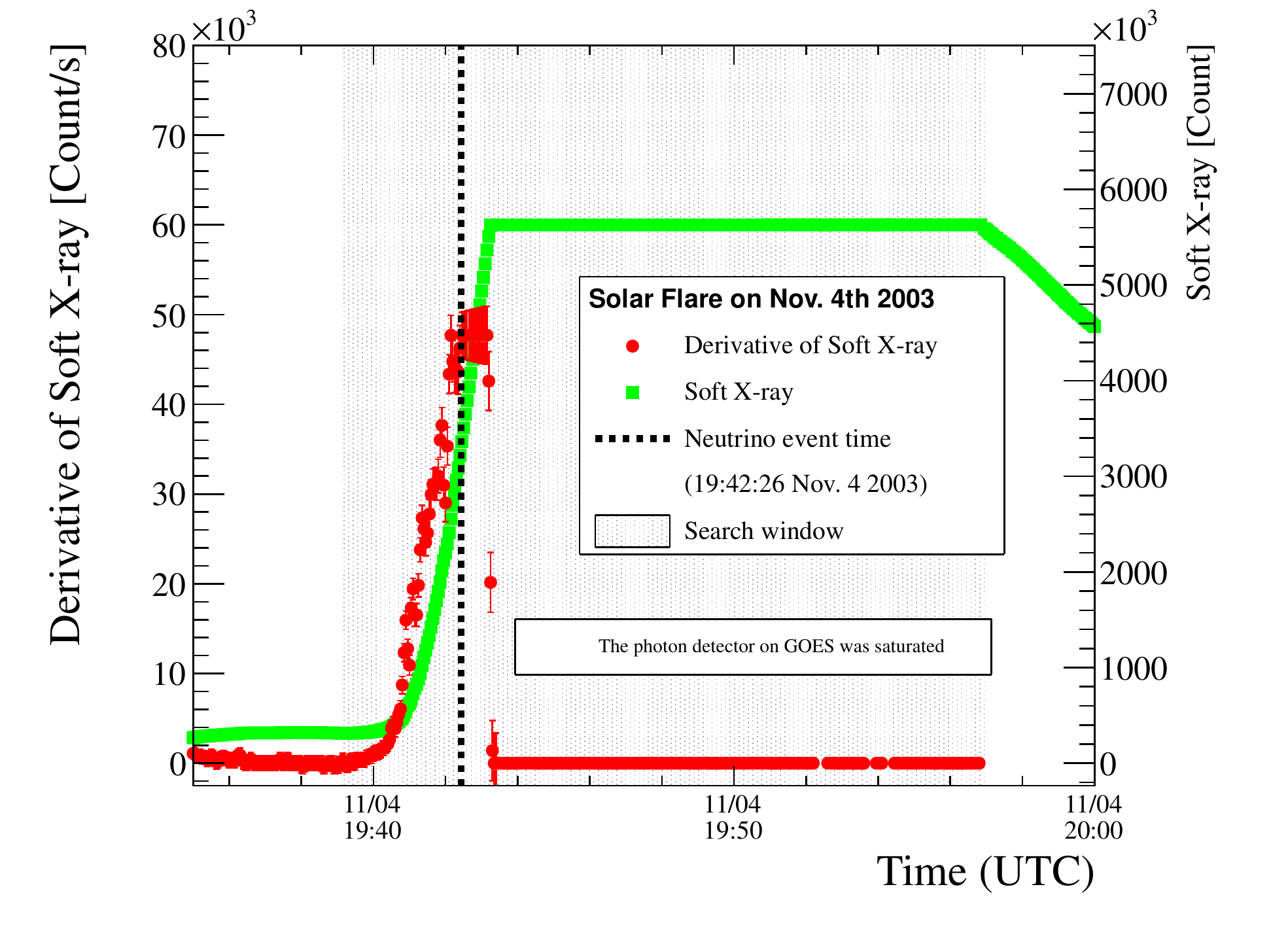}
    \end{minipage}
    \begin{minipage}{0.5\hsize}
        \centering\includegraphics[width=1.0\textwidth]{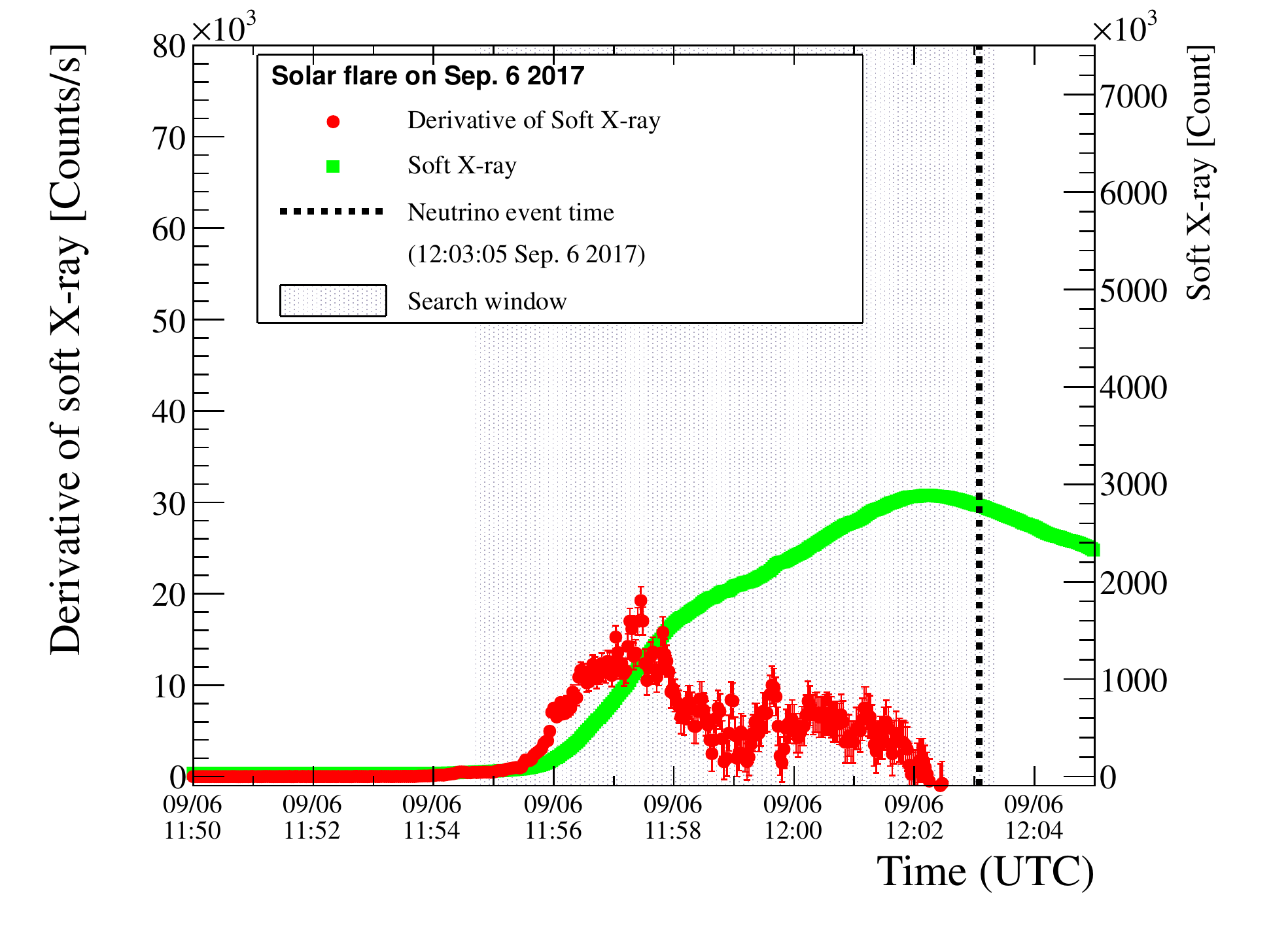}
    \end{minipage}
    \caption{The time of observed neutrino events for the solar flare on November 4th, 2003~(left), and July 6th, 2017~(right). The black vertical line shows the time of neutrino event in the SK detector. The red~(green) plot shows the derivative of the light curves~(original light curve) recorded by the GOES satellite. The shaded region shows the search windows determined by using the derivative of soft X-ray according to the method developed by \cite{2020SoPh..295..133O}. In the case of the solar flare on November 4th, 2003~(left), the instrument on the GOES satellite saturated due to the high intensity of soft X-rays. That resulted in the satellite not recording  data for more than $15$~minutes from 19:45 to 20:00. \label{fig:light-curve-visible}}
\end{figure*}

Figure~\ref{fig:light-curve-visible} shows the time of the observed neutrino events together with the light curves recorded by the GOES satellite. The event on November 4th 2003 was observed during the impulsive phase of the solar flare, where particle acceleration is expected to be active. Furthermore, \cite{Watanabe_2006} reported that relativistic neutrons associated with this solar flare were observed by the neutron monitors on the ground at 19:45~(UTC), which is about $3$~minutes after the detection of the neutrino candidate in SK. This simultaneous observation also indicates that hadrons~(ions) were accelerated to more than $1$~GeV during this solar flare. On the other hand, the event on September 6th 2017 was observed during the dimming phase after the peak of the soft X-ray light curve, when all processes of particle acceleration are likely to have been completed. The event displays and sky-maps for the two observed candidates are shown in Figure~\ref{fig:skymap-visible1} and Figure~\ref{fig:skymap-visible2} in Appendix~\ref{app:visible}.

\begin{figure*}[]
    \begin{center}
        \includegraphics[width=0.8\linewidth]{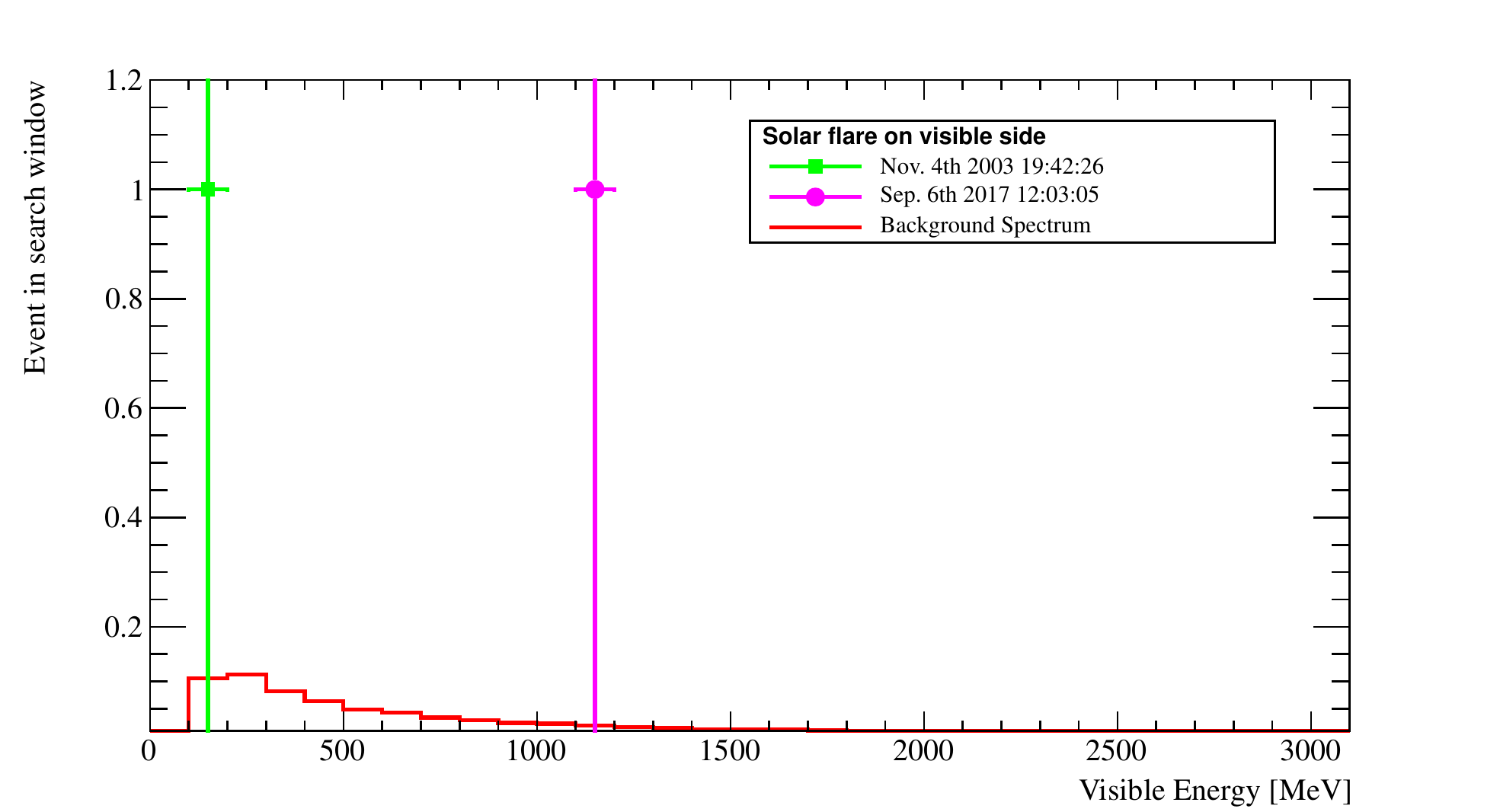}
    \end{center}
\caption{The energy distribution of the simulated neutrino events from solar flares occurring on the visible side of the Sun compared to the background sample. The first event~(in the green square) was observed as a $2$-ring $e$-like event on November 4th, 2003 while the other~(in the magenta circle) was observed as a $1$-ring $\mu$-like event on September 6th, 2017. The background spectrum~(red histogram) is normalized such that it corresponds to the expectation for a search window with an average duration of $700$~s, determined from the derivative of the soft X-ray light curve by~\cite{2020SoPh..295..133O}. \label{fig:rate}}
\end{figure*}

Figure~\ref{fig:rate} shows the energies of the two observed events compared to the expected background energy spectrum. Here, the background spectrum is from events accumulated outside the search windows and the main component is the interaction of atmospheric neutrinos. The expected number of background events in the high energy sample in the search window is $0.20$~events~($0.12$~events) for the solar flare event on November 4th 2003~(September 6th 2017).  The $p$-value of the null hypothesis is $18.1\%$~($11.3\%$).

\begin{figure*}[]
    \begin{minipage}{0.5\hsize}
        \centering\includegraphics[width=1.0\linewidth]{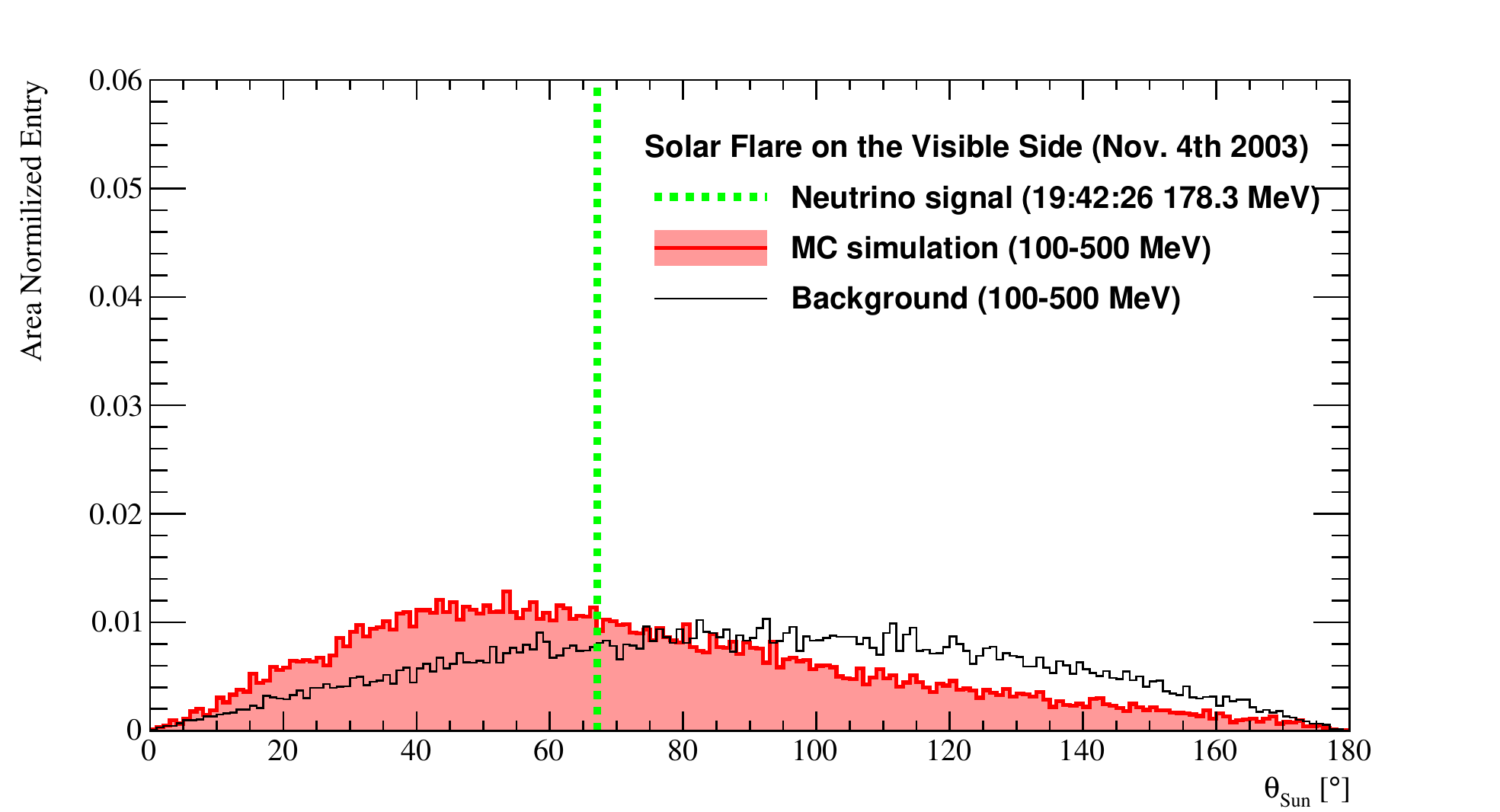}
    \end{minipage}
    \begin{minipage}{0.5\hsize}
        \centering\includegraphics[width=1.0\linewidth]{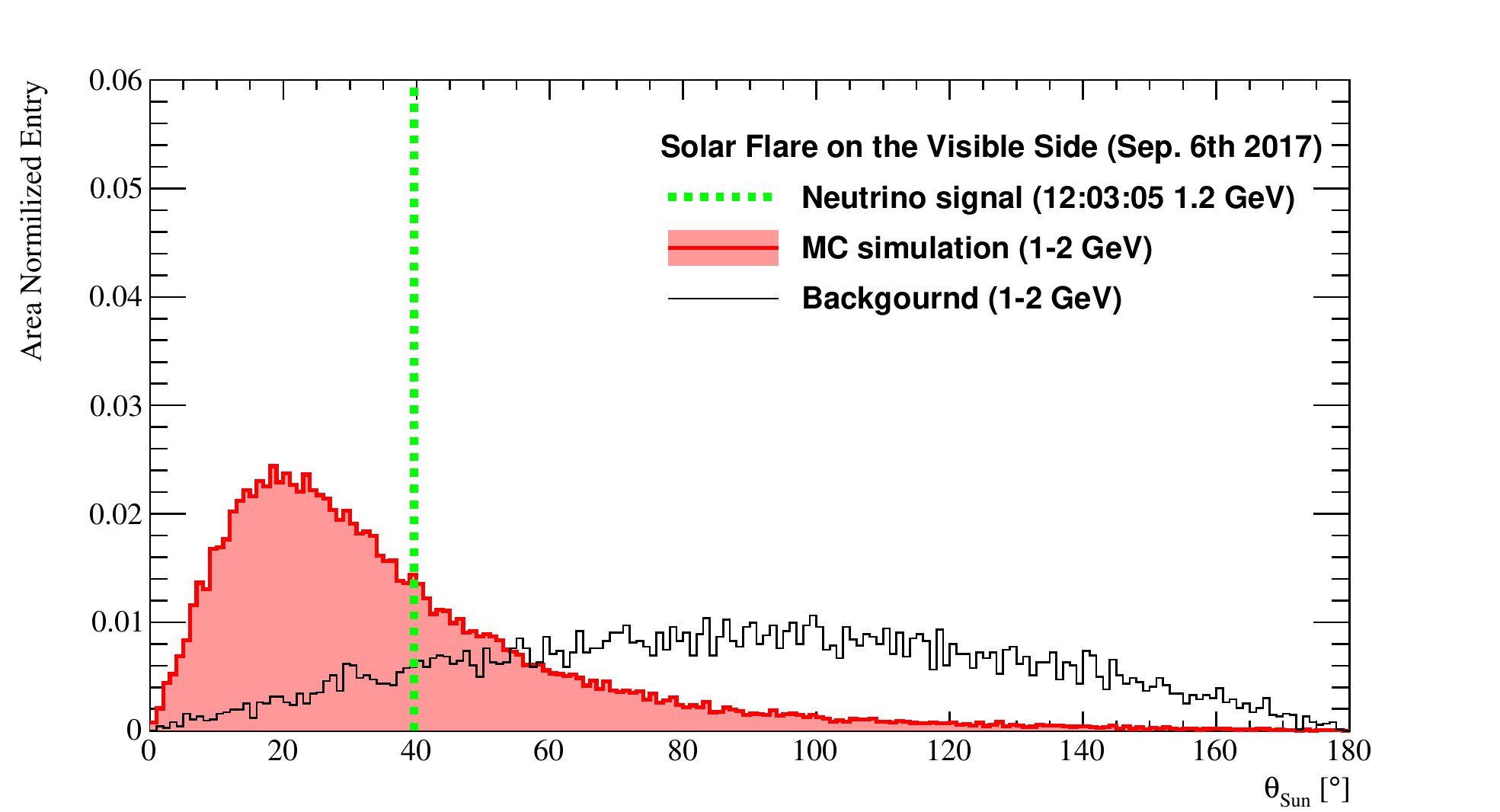}
    \end{minipage}
\caption{Reconstructed angles of the observed neutrino events in coincidence with solar flares occurring on November 4th, 2003~(left panel) and September 6th, 2017~(right panel) together with the angular distributions from MC sample for signal and background. The light green dashed line shows the angles between the reconstructed event direction and the direction of the Sun, $\theta_{\rm Sun}$, at the time when the neutrino event was observed. The red~(black) histograms show the angular distribution of the MC~(background) sample in the given energy range. \label{fig:angle_front}}
\end{figure*}

In order to investigate whether neutrino candidate events come from the direction of the Sun or not, we examined the angular distribution of $\theta_{\mathrm{Sun}}$, which is defined as the angle between the reconstructed direction of the charged particles and the direction pointing to the Sun. In the case of a multi-ring event, the direction of the neutrino is reconstructed as the momentum weighted sum of the directions of all the identified rings. The value of $\theta_{\mathrm{Sun}}$ of the candidate event on November 4th, 2003 (September 6th, 2017) is $67.1^{\circ}$~($39.6^{\circ}$). Figure~\ref{fig:angle_front} shows the $\theta_{\mathrm{Sun}}$ of the two observed events together with the angular distribution derived from the MC simulation.

\subsection{Results for solar flares on the invisible side of the Sun} \label{result_soft_invisible}

As explained in Section~\ref{sec:invisible}, we selected ten large CMEs that occurred on the invisible side of the Sun by setting criteria on their emission speed. However, SK did not take data for the two CMEs that occurred on July 18th and 19th, 2002 due to the detector re-instrumentation work. Hence, we searched for solar-flare neutrinos from the remaining eight CMEs that occurred on the invisible side of the Sun. 

\begin{figure*}[]
    \begin{center}
        \includegraphics[width=0.8\linewidth]{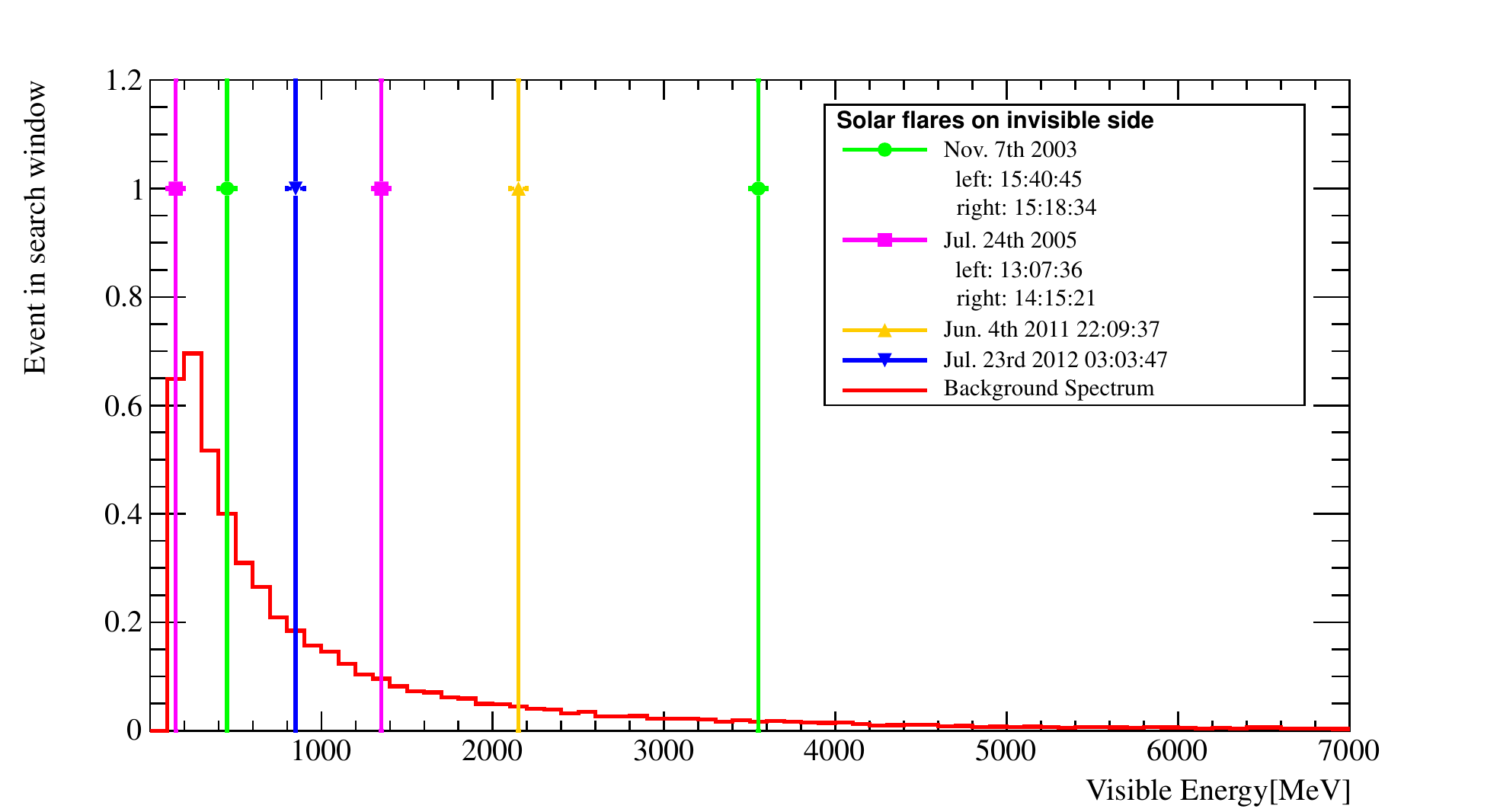}
    \end{center}
\caption{The reconstructed energies of the neutrino events from solar flares that occurred on the invisible side of the Sun and the typical energy distribution of events in the background sample. The green circles, magenta squares, yellow upward triangles, and blue downward triangles are the energy of observed events on November 7th, 2003, July 24th, 2005, June 4th, 2011, and July 23rd, 2012, respectively. The background spectrum is normalized by the time duration of $7238$~s according to the method described in Appendix~\ref{app:cme}. \label{fig:e-dist-invisible}}
\end{figure*}

There was no signal in the low-energy sample while six events were found in the high-energy sample as summarized in Table~\ref{tb:summary-event-invisible}. Two neutrino events were identified for the solar flares on November 7th, 2003 and July 24th, 2005 while one event was observed for those on June 4th, 2011 and July 23rd, 2012. The expected number of background events in the high-energy sample is $0.62~\mathrm{event \, flare^{-1}}$.  The $p$-value for the null hypothesis finding one~(two) events for these solar flares is $10.2\%$~($33.5\%$). Figure~\ref{fig:e-dist-invisible} shows the energies of the observed events together with the background energy spectrum. 
The event displays and sky-maps for the two observed candidates are shown from Figure~\ref{fig:skymap-invisible1} to Figure~\ref{fig:skymap-invisible4} in Appendix~\ref{app:invisible}.

\begin{longrotatetable}
\begin{deluxetable}{ccccccc}
    \tablecaption{The summary of events observed within the search window for neutrinos associated with solar flares that occurred on the invisible side of the Sun. The duration of the search window is $7238$~s as detailed in Appendix~\ref{app:cme}. The number of expected background events in the search window is $0.62 \pm 0.01$.}
        \label{tb:summary-event-invisible}
        \startdata
    \tablehead{
    \colhead{Date~(UTC)} & \twocolhead{November 7th, 2003} & \twocolhead{July 24th, 2005}  & \colhead{June 4th, 2011} & \colhead{July 23rd, 2012}
    } 
            SK phase & \multicolumn{2}{c}{SK-II} & \multicolumn{2}{c}{SK-II} & SK-IV & SK-IV \\
            Observed time~(UTC) & 15:18:34 & 15:40:45 & 13:07:36 &  14:15:21 & 21:05:07 & 03:03:47 \\
            Time difference between two events & \multicolumn{2}{c}{$1131$~s} & \multicolumn{2}{c}{$4065$~s} & -- & -- \\
            Event topology & $1$-ring  $e$-like & $2$-ring $e$-like & $1$-ring $\mu$-like & $1$-ring $\mu$-like & $4$-ring  $e$-like & $3$-ring  two $\mu$-like, $e$-like \\
            Reconstructed energy & $3.58$~GeV & $493$~MeV & $126$~MeV & $1.35$~GeV & $2.14$~GeV & $834$~MeV \\
            $\theta_{\mathrm{Sun}}$ & $20.0^{\circ}$ & $71.4^{\circ}$ & $100.4^{\circ}$ & $94.0^{\circ}$ & $101.0^{\circ}$ & $76.7^{\circ}$ \\
            $p$-value of the null hypothesis & \multicolumn{2}{c}{$10.2\%$} & \multicolumn{2}{c}{$10.2\%$}  & $33.5\%$ & $33.5\%$ \\
            Probability of background event  & \multicolumn{2}{c}{$10.2\%$} & \multicolumn{2}{c}{$34.5\%$}  & -- & --\\
            from timing distribution &  & & &  &  & \\
        \enddata            
\end{deluxetable}
\end{longrotatetable}

\begin{figure*}[]
    \begin{minipage}{0.5\hsize}
        \centering\includegraphics[width=1.0\linewidth]{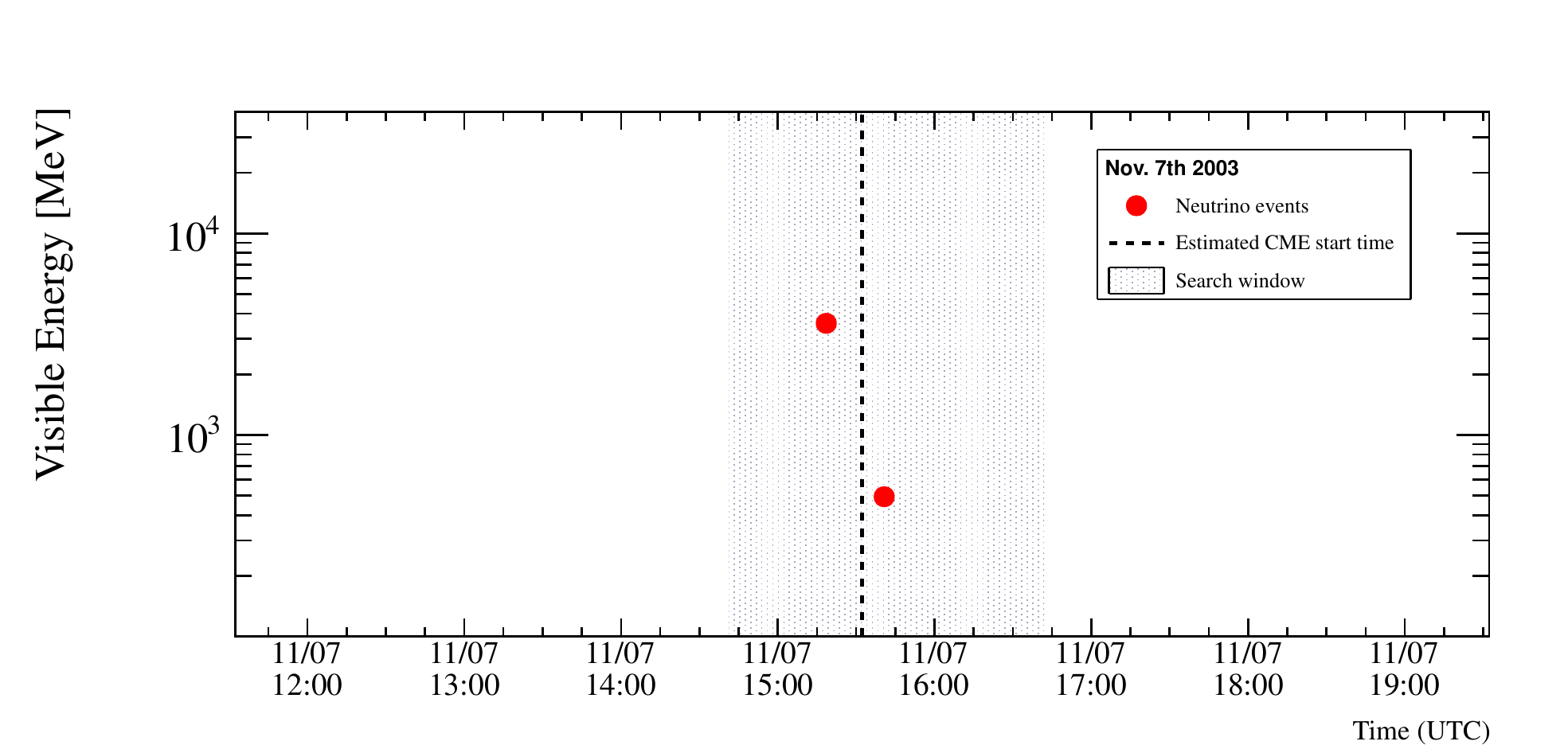}
    \end{minipage}
    \begin{minipage}{0.5\hsize}
        \centering\includegraphics[width=1.0\linewidth]{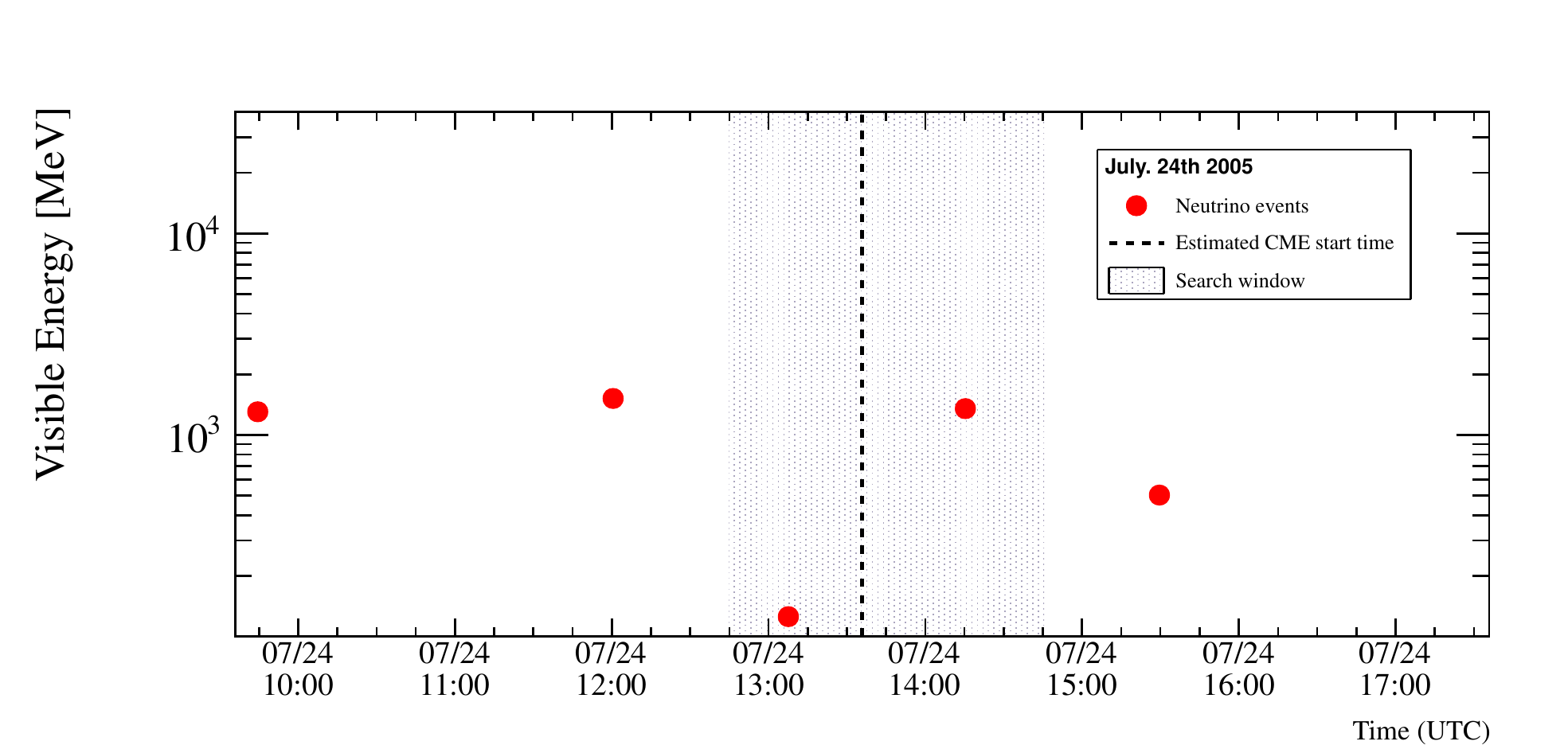}
    \end{minipage} \\
    \begin{minipage}{0.5\hsize}
        \centering\includegraphics[width=1.0\linewidth]{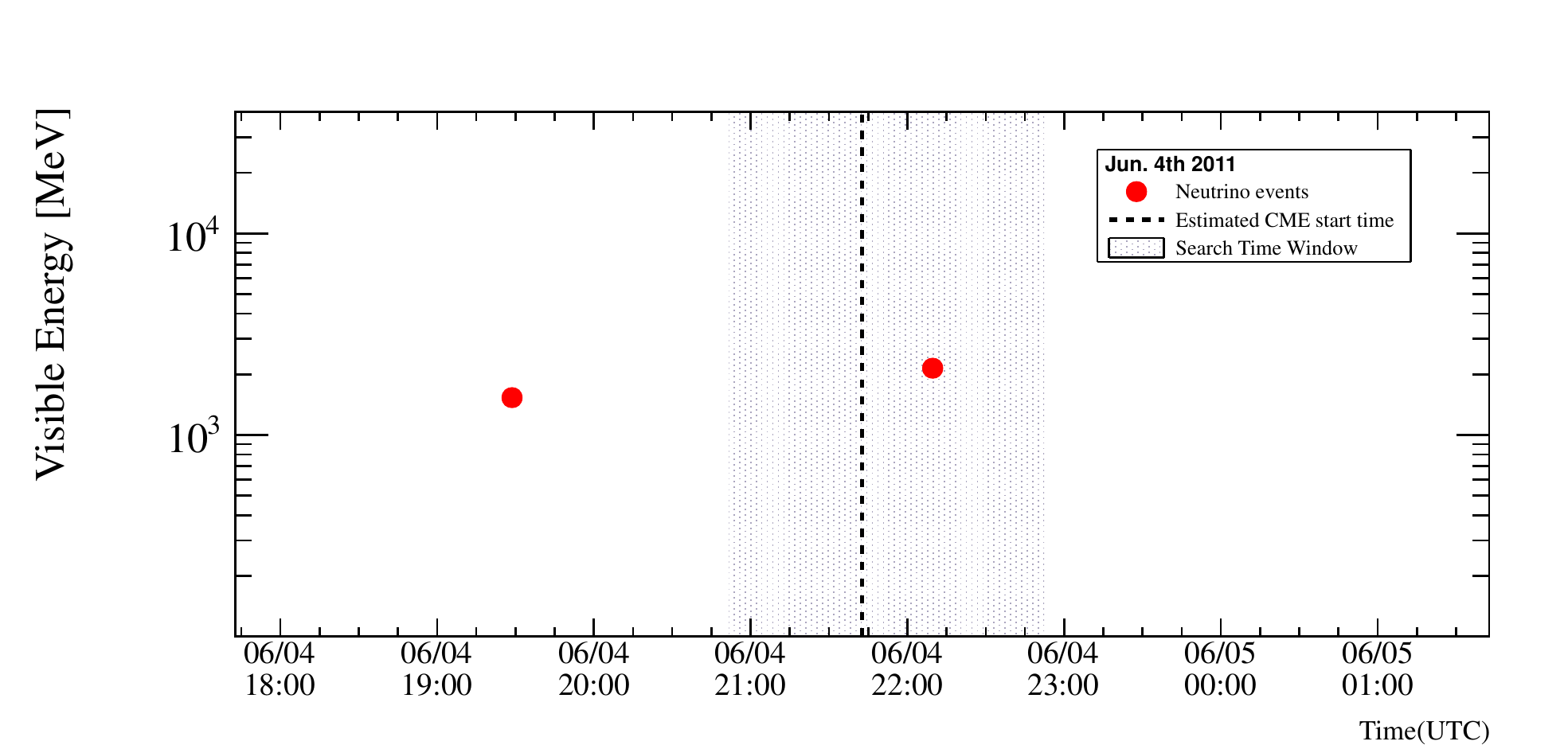}
    \end{minipage}
    \begin{minipage}{0.5\hsize}
        \centering\includegraphics[width=1.0\linewidth]{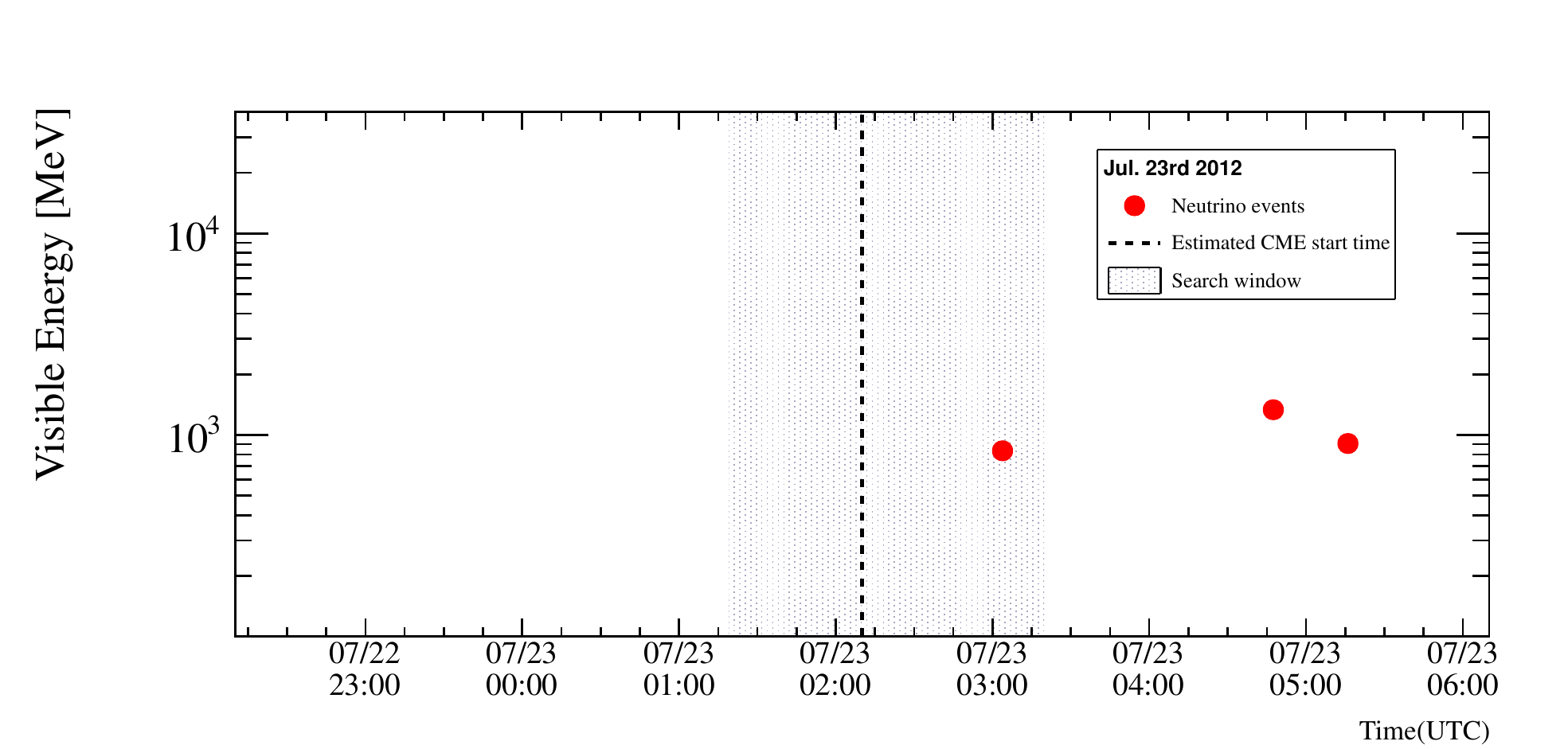}
    \end{minipage}
    \caption{The time distributions of neutrino events around the solar flares that occurred on the invisible side of the Sun on November 7th, 2003~(top-left), July 24th, 2005~(top-right), June 4th, 2011~(bottom-left), and  July 23rd, 2012~(bottom-right). The red points show the times of the neutrino events, which are summarized in Table~\ref{tb:summary-event-invisible}. The dashed vertical lines show the estimated start time of the particular CME emission and the shaded regions show the search windows~($7238$~s) according to the method described in Section~\ref{sec:invisible}.  \label{fig:time-dist-invisible}}
\end{figure*}

Figure~\ref{fig:time-dist-invisible} shows the observed neutrino events around the time of the solar flare. Note that the duration of the search window for solar flares on the invisible side of the Sun is uniform~($7238$~s) as detailed in Appendix~\ref{app:cme}.

\begin{figure}[]
    \begin{center}
        \includegraphics[width=0.8\linewidth]{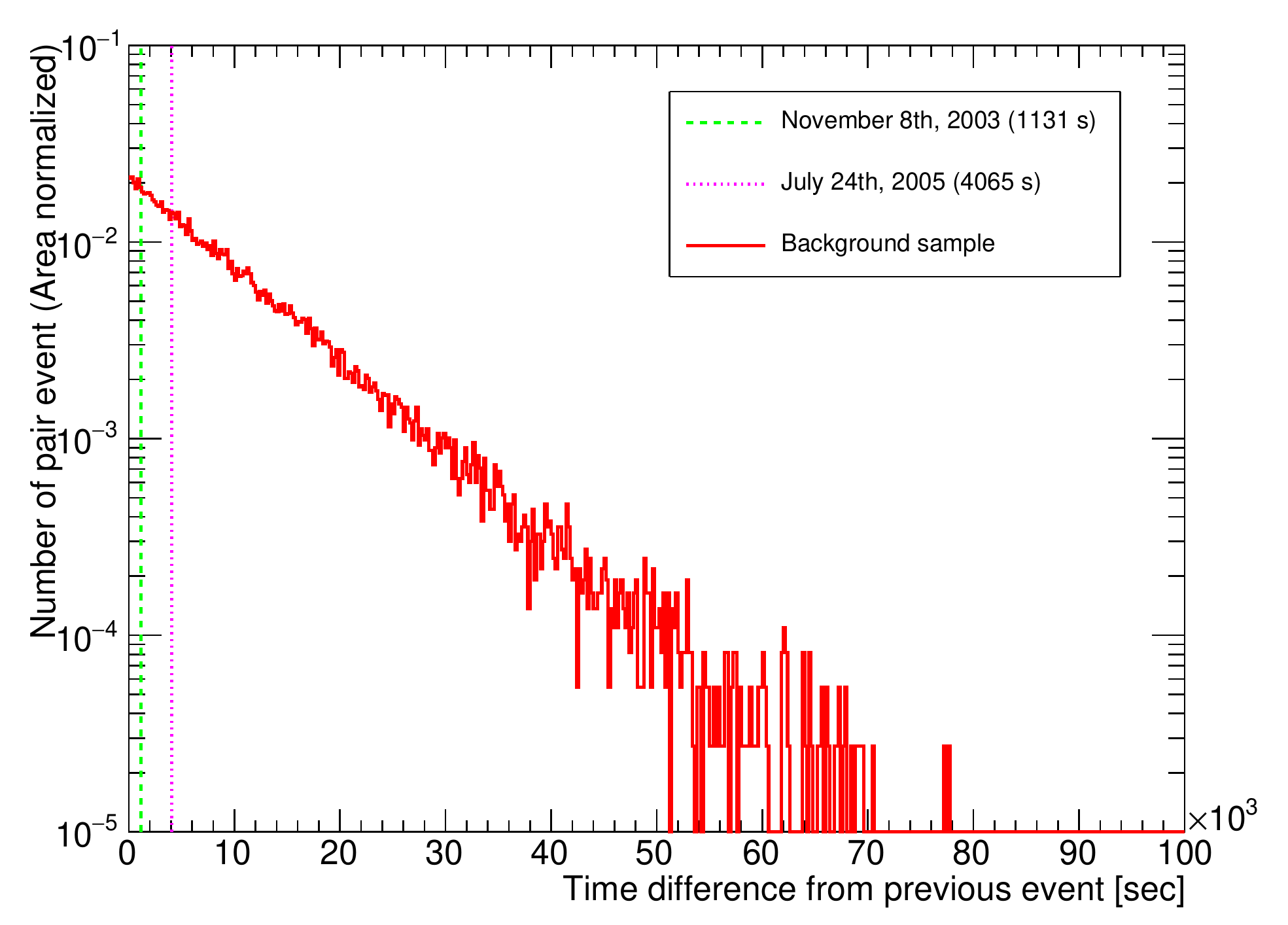}
    \end{center}
\caption{The time difference between the two events observed within the search windows for solar flares on November 7th, 2003~($1131$~s, dashed green line) and July 24th, 2005~($4065$~s, dotted pink line). The red histogram shows the distribution of the time difference between consecutive events in the background sample using the combined data from SK-I to SK-IV. \label{fig:tdiff-dist-invisible}}
\end{figure}

In the case of the two solar flares on November 7th, 2003 and on July 24th, 2005, we found two consecutive neutrino events within their search windows. Their time differences are $1131$~s and $4065$~s, respectively. We analyzed the time difference distribution between consecutive events in the background sample in order to verify whether their time differences are likely or not. Figure~\ref{fig:tdiff-dist-invisible} shows the time difference of the two consecutive events observed within their search windows together with the time difference distribution of the background sample.
Comparing the time difference distribution of the background sample, we estimated the occurrence probabilities for each solar flare, which correspond to $10.2\%$ and $34.5\%$, respectively. 

\begin{figure*}[]
    \begin{minipage}{0.5\hsize}
        \centering\includegraphics[width=1.0\linewidth]{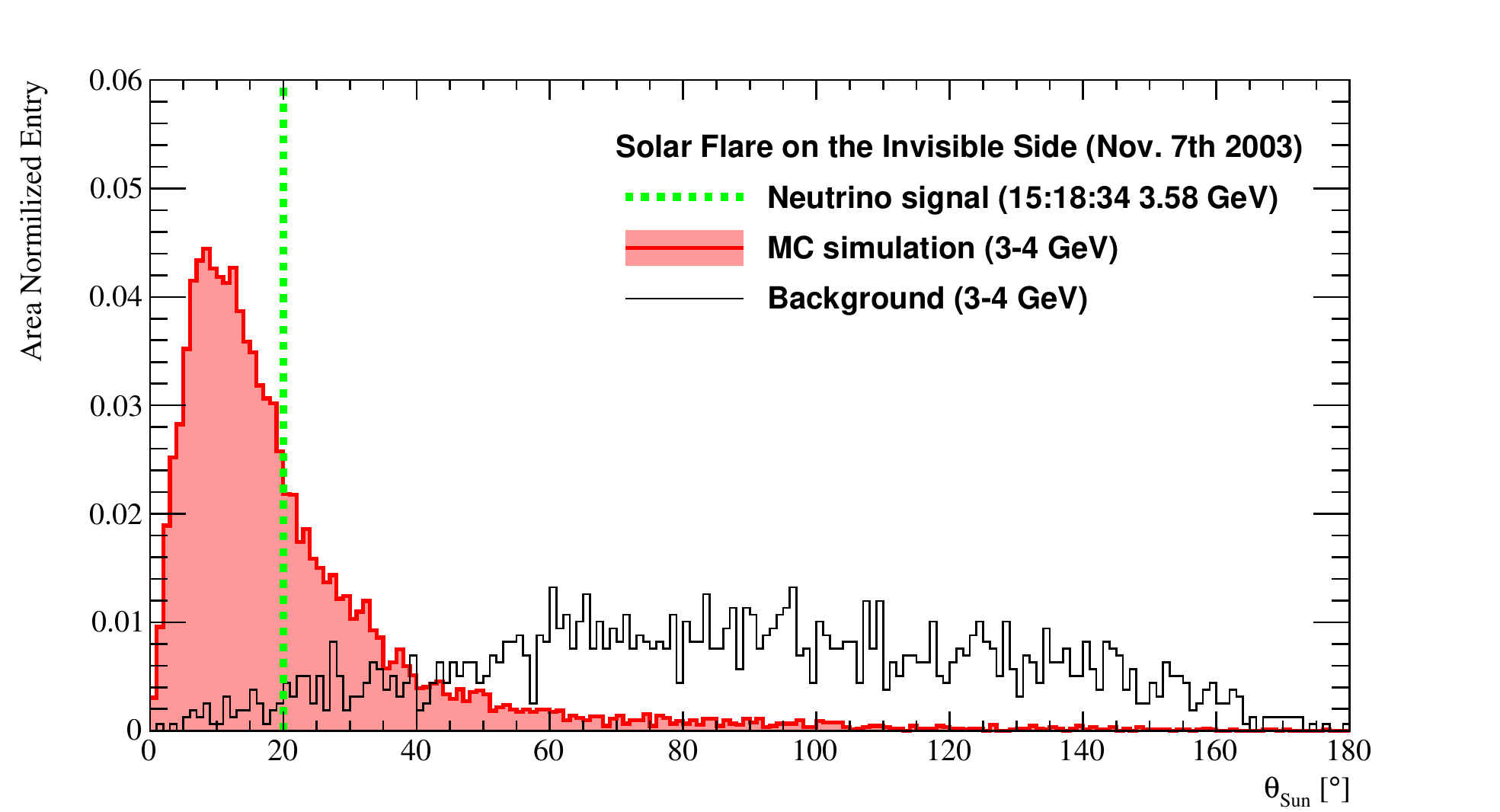}
    \end{minipage}
    \begin{minipage}{0.5\hsize}
        \centering\includegraphics[width=1.0\linewidth]{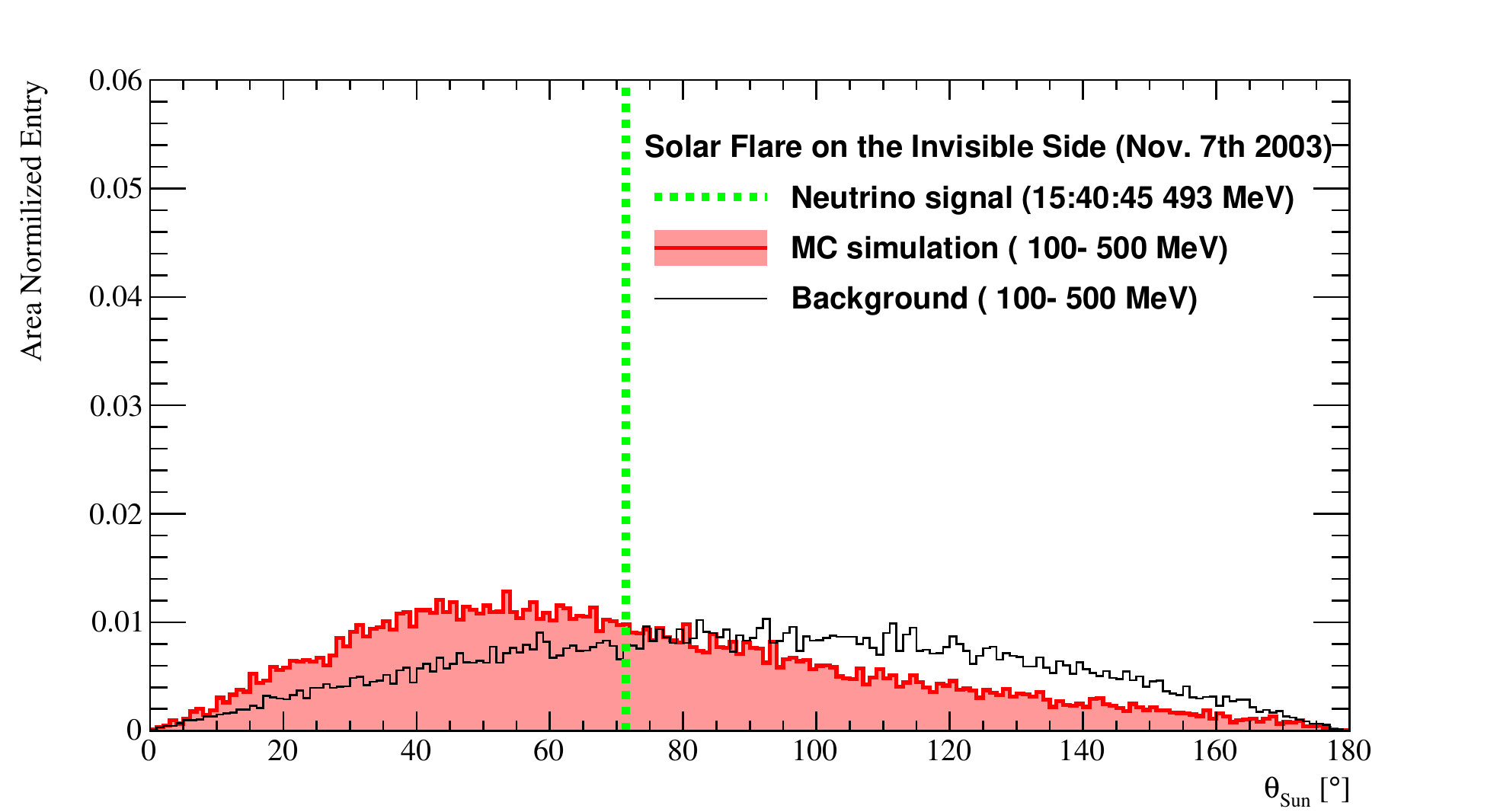}
    \end{minipage} \\
    \begin{minipage}{0.5\hsize}
        \centering\includegraphics[width=1.0\linewidth]{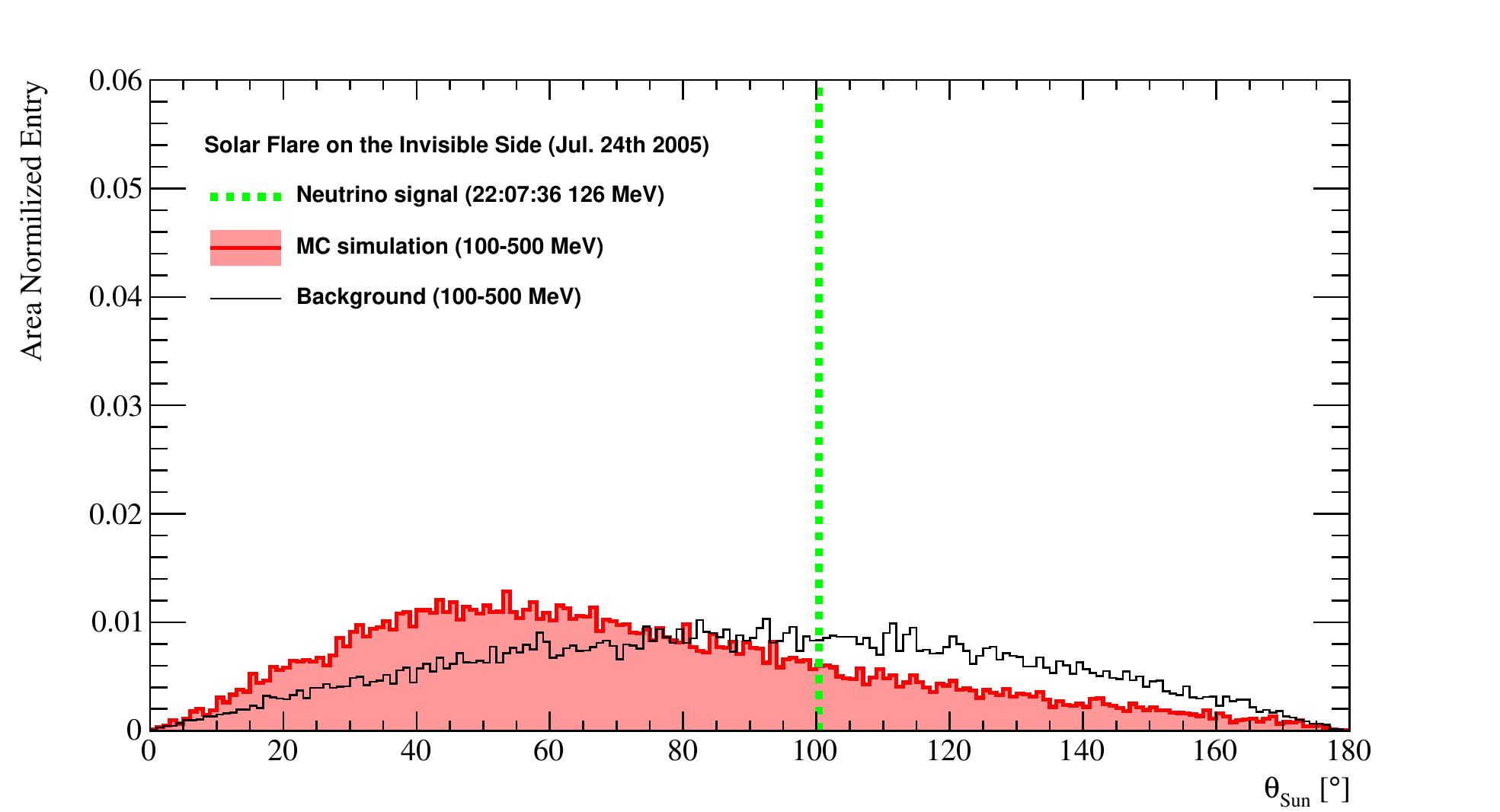}
    \end{minipage}
    \begin{minipage}{0.5\hsize}
        \centering\includegraphics[width=1.0\linewidth]{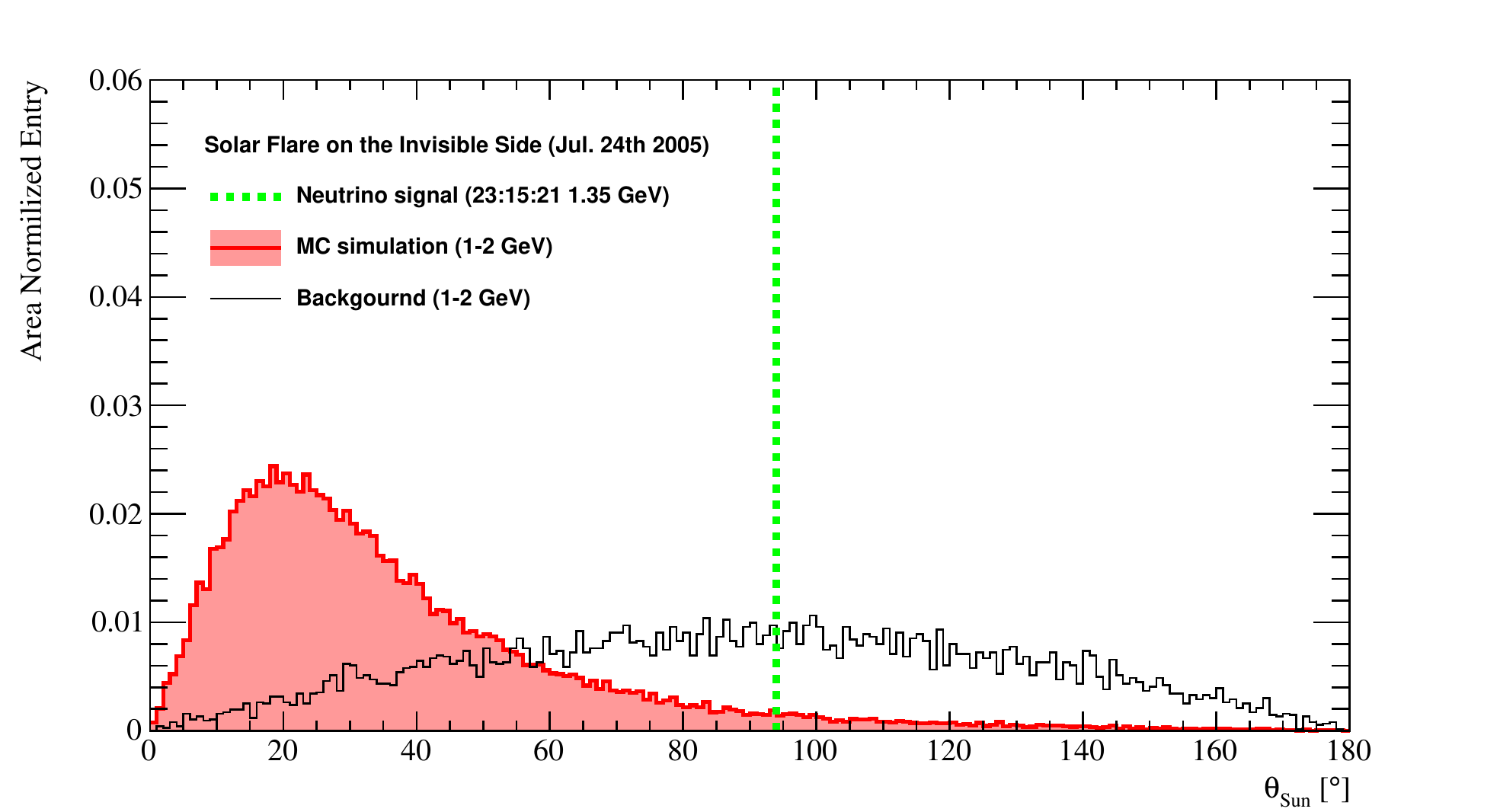}
    \end{minipage} \\
    \begin{minipage}{0.5\hsize}
        \centering\includegraphics[width=1.0\linewidth]{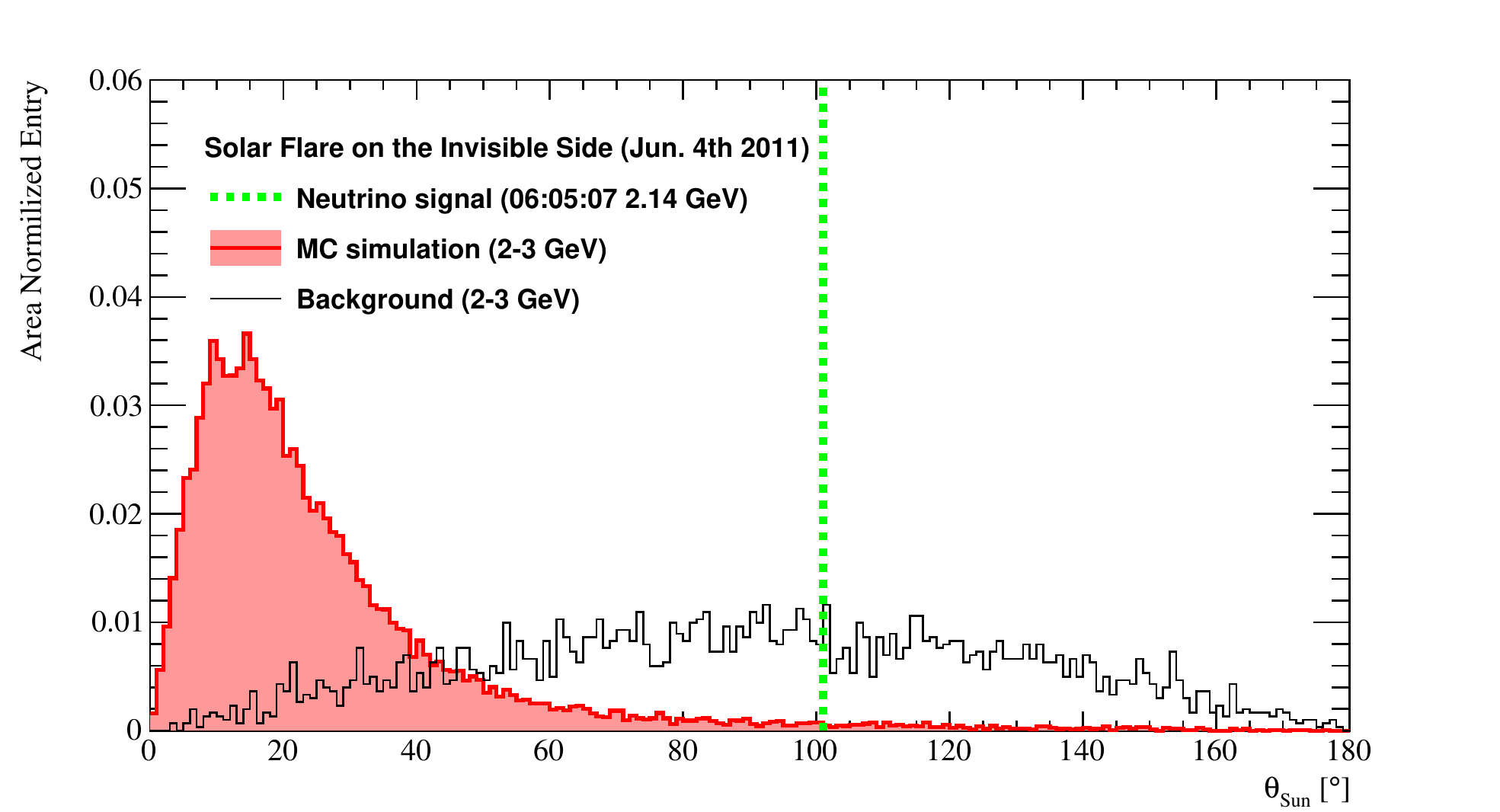}
    \end{minipage}
    \begin{minipage}{0.5\hsize}
        \centering\includegraphics[width=1.0\linewidth]{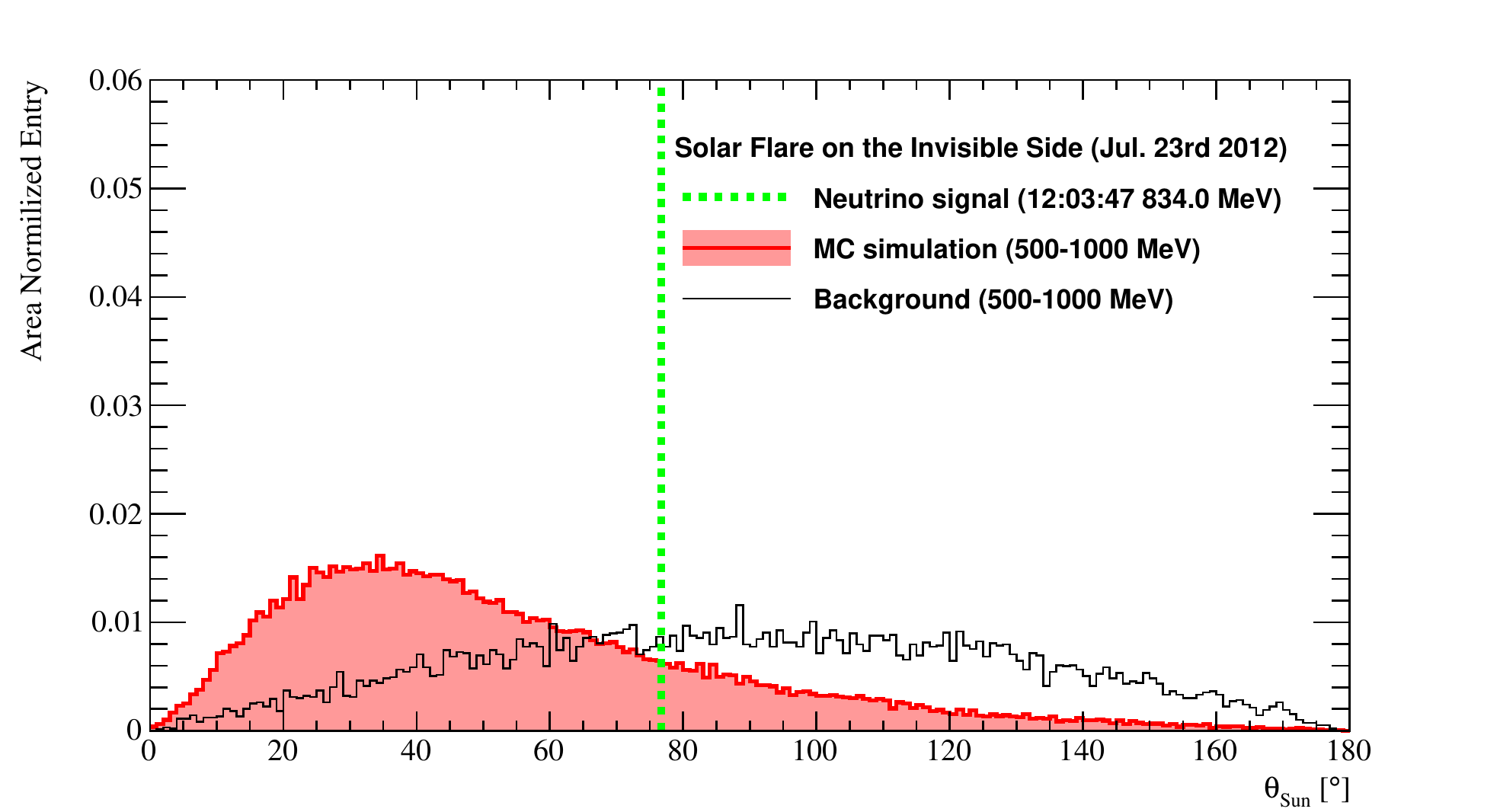}
    \end{minipage}
    \caption{The reconstructed angle of the neutrino events in coincidence with solar flares that occurred on the invisible side of the Sun together with the typical angular distributions from the MC simulation for signal and background sample. The light green dashed lines show the angle between the reconstructed event direction and the direction from the Sun to SK, $\theta_{\rm Sun}$, at the time when the neutrino candidate was observed. The red~(black) histograms show the angular distributions of the MC~(background) sample in the given energy range.\label{fig:angle_rear}}
\end{figure*}

Figure~\ref{fig:angle_rear} shows the reconstructed angle $\theta_{\mathrm{Sun}}$ together with the typical distribution derived from the MC simulation. The reconstructed values of $\theta_{\mathrm{Sun}}$ are also summarized in Table~\ref{tb:summary-event-invisible}.

\section{Discussion} \label{sec:discuss}
\subsection{solar-flare neutrino fluence derived from the theoretical predictions}

We estimated the fluence of solar-flare neutrinos produced by powerful solar flares based on the number of observed events within their corresponding search window. Here, we calculate the upper limit of the neutrino fluence using a Bayesian method, in the absence of a significant excess of observed events above the expected background rate~\citep{2000PhRvD..63a3009R}. 
We separately calculate the upper limits of neutrino fluence for low and high energy samples depending on the neutrino energies. 

The neutrino fluence at the Earth~$\mathit{\Phi}$ is calculated using the neutrino flux~$F(E_{\nu})$ at the Earth in the search window,
\begin{equation}
    \mathit{\Phi} = t_{\mathrm{emit}}\int F(E_{\nu}){\rm d}E_{\nu},
\end{equation}
where $t_{\mathrm{emit}}$ is a time duration of neutrino emissions in a solar flare, which is $100$~s according to the assumption in~\cite{2003ChJAS...3...75F}, $E_{\nu}$ is the neutrino energy, and $F(E_{\nu})$ is the predicted neutrino flux without neutrino oscillations in unit of $\mathrm{cm^{-2}\,s^{-1}\,MeV^{-1}}$. We note that the duration of the search windows is sufficient to cover the duration of neutrino emission $t_{\mathrm{emit}}$.

For the low energy sample, the expected number of neutrino interactions within the search window~$S$ is calculated using the following equation:

\begin{equation}
S \equiv  N_{p}t_{\mathrm{emit}}\int \left[ F(E_{\bar{\nu}_{\mathrm{e}}} )P_{\mathrm{ee}} +F(E_{\bar{\nu}_{\mu}} )P_{\mu\mathrm{e}} \right] \sigma_{\mathrm{IBD}}(E_{\bar{\nu}_{\mathrm{e}}}) \varepsilon_{\mathrm{low}}^{\mathrm{Fargion}}(E_{{\nu}_{\mathrm{e}}}) {\mathrm{d}}E_{{\overline{\nu}_{\mathrm{e}}}}, \label{def_s}
\end{equation}

\noindent
where $N_{p}$ is the number of target protons in the SK fiducial volume relevant to the neutrino interactions, $P_{\alpha \beta}$ is the probability for a neutrino produced as flavor $\alpha$ to oscillate to flavor $\beta$ when travelling from the Sun to the Earth, $\sigma_{\mathrm{IBD}}$ is the IBD cross section as a function of electron anti-neutrino energy derived from the theoretical model from~\cite{2003PhLB..564...42S} and $\varepsilon_{\mathrm{low}}^{\mathrm{Fargion}}$ is the event selection efficiency of the low energy sample defined in Section~\ref{sec:analysis}.

Using this expected number of neutrino interactions within the search window, the probability density function for the number of observed events is defined as follows:

\begin{equation}
P_{\mathrm{low}}(S+B | n_{\mathrm{obs}})= \frac{1}{A}\int\!\!\!\int\!\!\!\int \frac{\mathrm{e}^{-(S+B)}(S+B)^{n_{\rm obs}}}{n_{\mathrm{obs}}!} P(\sigma_{\mathrm{IBD}})P \left(\varepsilon_{\mathrm{low}}^{\mathrm{Fargion}} \right)P(B)d\sigma_{\mathrm{IBD}}d\varepsilon_{\mathrm{low}}^{\mathrm{Fargion}}dB \label{eq-fluence_low},
\end{equation}

\noindent
where $B$ is the number of expected background events in the search window, $A$ is a normalization factor representing the total integral of $P(S+B|n_{\mathrm{obs}})$, and $n_{\mathrm{obs}}$ is the number of observed events in the search window. To include the effect of systematic uncertainties, $P(\sigma_{\mathrm{IBD}})$, $P \left(\varepsilon_{\mathrm{low}}^{\mathrm{Fargion}} \right)$, and $P(B)$ are introduced as the prior probabilities for fluctuations of the IBD cross section, the event selection efficiency, and the number of expected background events in the search window, respectively. The priors are assumed to follow a Gaussian distribution,
\begin{eqnarray}
G(x) = \frac{1}{\sqrt{2\pi\delta^{2}_{x}}}\exp\left[-\frac{(x-x_{0})^{2}}{2\delta^{2}_{x}}\right], \label{eq-low}
\end{eqnarray}

\noindent
where $x$ stands for the parameters $\sigma_{\mathrm{IBD}}$, $\varepsilon_{\mathrm{low}}$, and $B$, respectively, $x_{0}$ is their best estimates, and $\delta_{x}$ stands for their systematic uncertainties, expressed as $\delta_{\sigma_{\mathrm{IBD}}}$, $\delta_{\varepsilon_{\mathrm{low}}}$, and $\delta_{B}$, respectively. For the systematic uncertainty of the IBD cross section~($\sigma_{\mathrm{IBD}}$), we assigned the uncertainty estimated in~\cite{2003PhLB..564...42S}. For the systematic uncertainty of the selection efficiency~($\delta_{\varepsilon_{\mathrm{low}}}$), the variation in the number of events after all reduction cuts was evaluated by artificially changing the reduction parameters as performed in~\cite{2021PhRvD.104l2002A}. For the systematic uncertainty of the background rate~($\delta_{B}$), the deviation of the actual background rate is conservatively assigned. Table~\ref{tb:sys_error} summarises the values of these systematic uncertainties.

With these definitions, the neutrino fluence at a given confidence level~(C.L.) is calculated as
\begin{equation}
    {\rm C.L.} = \frac{\int^{\mathit{\Phi}_{\rm limit}}_{0}P_{\mathrm{low}}(S+B|n_{\rm obs}){\rm d} \mathit{\Phi}}{\int^{\infty}_{0}P_{\mathrm{low}}(S+B|n_{\rm obs}){\rm d} \mathit{\Phi}},
\end{equation}
where $\mathit{\Phi}_{\mathrm{limit}}$ is the upper limit of the neutrino fluence to be obtained.

\begin{table*}[]
    \begin{center}
    \caption{A summary of systematic uncertainties in this analysis. The systematic uncertainties for the event selection efficiency of the low and high energy samples are estimated from~\cite{2021PhRvD.104l2002A} and~\cite{2018PhRvD..97g2001A}, respectively. The systematic uncertainty of the neutrino cross section for the low energy sample simulations has been taken from~\cite{2003PhLB..564...42S}. For the high energy sample, the difference between the cross sections from~\cite{Smith:1972xh} and from~\cite{2004PhRvC..70e5503N} is assigned as the systematic uncertainty of the cross section. The deviation of the background rate which is listed in Table~\ref{tb:bg_rate} has been used as the systematic uncertainty of the background rate.}
        \label{tb:sys_error}
        \begin{tabular}{ccccccccc} \hline
                Valuable & \multicolumn{4}{c}{Low energy sample} & \multicolumn{4}{c}{High energy sample}  \\ 
                 &SK-I& SK-II & SK-III & SK-IV & SK-I& SK-II & SK-III & SK-IV \\\hline                
                 Selection efficiency ($\delta_{\varepsilon}$)  & $5.0\%$ & $5.3\%$ & $3.5\%$ & $4.1\%$ & $1.5\%$ & $0.4\%$ & $1.5\%$ & $0.1\%$  \\
                Cross section ($\delta_{\sigma}$)  &  \multicolumn{4}{c}{$\phantom{0}1.0\%$}  & \multicolumn{4}{c}{$20.0\%$}\\
                Background rate ($\delta_{B}$) & $5.0\%$ & $10.5\%$ & $5.0\%$ & $5.3\%$ & $0.9\%$ & $1.2\%$ & $1.6\%$ & $0.6\%$ \\
            \hline
        \end{tabular}
    \end{center}
\end{table*}

Table~\ref{tb:lowe-flence} summarizes the upper limits of neutrino fluence of anti-electron neutrinos for the low energy sample. Since there are no events observed within the search windows, the upper limits of the neutrino fluence are $4.0\times10^{7}~\mathrm{cm^{-2}}$~($4.1\times10^{7}~\mathrm{cm^{-2}}$) selected solar flares occurring on SK-I~(SK-II and SK-IV).

\begin{table*}[]
    \begin{center}
    \caption{A summary of the upper limits of neutrino fluence of anti-electron neutrinos for the low energy sample. We assumed the neutrino energy spectrum from \cite{2003ChJAS...3...75F} and that IBD is the dominant reaction for reconstructed positron energies below $100$~MeV. As described in \cite{2020SoPh..295..133O}, for the solar flare on September 9th, 2005, the brightening of its soft X-ray light curve was relatively slow and the time derivative of the curve was not large enough. Due to this unexpected behavior, the time window for this solar flare was determined using the soft X-ray light curve directly instead of its derivative. Owing to such treatment, the duration of the search window is much longer than others and this results in the higher background rate.}
        \label{tb:lowe-flence}
        \begin{tabular}{ccccc}
            \hline
             Side  & Date of flare & Observed event & Expected background & Neutrino fluence \\
             (Channel)  & [UTC] & within window  & within window & [$\mathrm{cm^{-2}{\,}flare^{-1}}$] \\
            \hline
            Visible & 2002 Jul. 23 & No SK data & -- & -- \\
            (Line $\gamma$-rays) & 2003 Nov. \phantom{0}2 & $0$ & $0.0029$ & $<4.0\times10^{7}$ \\
            & 2005 Jan. 20 & $0$ & $0.0035$ & $<4.0\times10^{7}$ \\ \hline
            & 1997 Nov. \phantom{0}6 & $0$ & $0.0004$ & $<4.0\times10^{7}$ \\
            & 2000 Jul. 14 & $0$ & $0.0027$ & $<4.0\times10^{7}$ \\
            & 2001 Apr. \phantom{0}2 & $0$ & $0.0025$ & $<4.0\times10^{7}$ \\
            & 2001 Apr. \phantom{0}6 & $0$ & $0.0012$ & $<4.0\times10^{7}$ \\
            & 2001 Apr. 15 & $0$ & $0.0010$ & $<4.0\times10^{7}$ \\
            & 2001 Aug. 25 & No SK data & -- & -- \\
            & 2001 Dec. 13 & No SK data & -- & -- \\
            & 2002 Jul. 23 & No SK data & -- & -- \\
            & 2003 Oct. 23 & $0$ & $0.0022$ & $<4.1\times10^{7}$ \\
            & 2003 Oct. 28 & $0$ & $0.0016$ & $<4.1\times10^{7}$ \\
            Visible& 2003 Oct. 29 & $0$ & $0.0016$ & $<4.1\times10^{7}$ \\
            (Soft X-ray & 2003 Nov. \phantom{0}2 & $0$ & $0.0019$ & $<4.1\times10^{7}$ \\
            derivative) & 2003 Nov. \phantom{0}4 & $0$ & $0.0026$ & $<4.1\times10^{7}$ \\
            & 2005 Jan. 20 & $0$ & $0.0025$ & $<4.1\times10^{7}$ \\
            & 2005 Sep. \phantom{0}7 & $0$ & $0.0023$ & $<4.1\times10^{7}$ \\
            & 2005 Sep. \phantom{0}8 & $0$ & $0.0011$ & $<4.1\times10^{7}$ \\
            & 2005 Sep. \phantom{0}9 & $0$ & $0.018$\phantom{0} & $<4.1\times10^{7}$ \\
            & 2006 Dec. \phantom{0}5 & $0$ & $0.0016$ & $<4.1\times10^{7}$ \\
            & 2006 Dec. \phantom{0}6 & $0$ & $0.0008$ & $<4.1\times10^{7}$ \\
            & 2011 Aug. \phantom{0}9 & $0$ & $0.0007$ & $<4.1\times10^{7}$ \\
            & 2012 Mar. \phantom{0}7 & $0$ & $0.0029$ & $<4.1\times10^{7}$ \\
            & 2017 Sep. \phantom{0}6 & $0$ & $0.0012$ & $<4.1\times10^{7}$ \\
            & 2017 Sep. 10 & $0$ & $0.0024$ & $<4.1\times10^{7}$ \\ \hline
            & 2001 Apr. 18 & $0$ & $0.016$ & $<4.0\times10^{7}$ \\
            & 2002 Jul. 18 & No SK data & -- & -- \\
            & 2002 Jul. 19 & No SK data & -- & -- \\
            & 2003 Nov. \phantom{0}2 & $0$ & $0.016$ & $<4.1\times10^{7}$ \\
            Invisible & 2003 Nov. \phantom{0}7 & $0$ & $0.016$ & $<4.1\times10^{7}$ \\
            & 2003 Nov. \phantom{0}9 & $0$ & $0.016$ & $<4.1\times10^{7}$ \\
            & 2005 Jul. 24 & $0$ & $0.016$ & $<4.1\times10^{7}$ \\
            & 2011 Jun. \phantom{0}4 & $0$ & $0.016$ & $<4.1\times10^{7}$ \\
            & 2012 Jul. 23 & $0$ & $0.016$ & $<4.1\times10^{7}$ \\
            & 2014 Dec. 13 & $0$ & $0.016$ & $<4.1\times10^{7}$ \\ \hline
     \end{tabular}
    \end{center}
\end{table*}

For the high energy sample, we considered interactions of all neutrino flavors because the distance between the Sun and the Earth is sufficiently long compared with the oscillation length of neutrinos whose energy is less than $100$~GeV~\citep{2006PhRvD..74i3004F}. The flavor ratio of solar-flare neutrinos at the production point is $\nu_{e}:\nu_{\mu}:\nu_{\tau} = 1:2:0$ due to their origin from $\pi^{\pm}$ and $\mu^{\pm}$ decay while the flavor ratio at the detector is approximately $\nu_{e}:\nu_{\mu}:\nu_{\tau} = 1:1:1$~\citep{2009PhRvD..80k3006C}, where we also assume that the ratio of neutrino to anti-neutrino is approximately equal, as $\nu:\bar{\nu}=1:1$. The fluence upper limit for solar-flare neutrinos using the high energy sample can be obtained using a similar procedure as for the low energy sample.
The difference in the calculation procedure between them is the definition of the probability density function for the number of observed events. In the high energy sample it is defined as follows:

\begin{eqnarray}
        P_{\mathrm{high}}(S+B | n_{\mathrm{obs}}) & = & \displaystyle \frac{1}{A'}\int\!\!\!\int\!\!\!\int \frac{\mathrm{e}^{-(S+B)}(S+B)^{n_{\mathrm{obs}}}}{n_{\mathrm{obs}}!} P(\sigma(E_{\nu}))P \left(\varepsilon_{\mathrm{high}}^{\mathrm{Fargion}} \right)P(B)d\sigma(E_{\nu})d\varepsilon_{\mathrm{high}}^{\mathrm{Fargion}}dB, \\
        \mathrm{and}~S & = & N_T \displaystyle \int dE_{\nu} \sum_{i=e,\mu,\tau,\bar{e},\bar{\mu},\bar{\tau}} \left( \frac{F(E_{\nu_{i}}) \sigma(E_{\nu_{i}})\varepsilon_{\mathrm{high}}^{\mathrm{Fargion}}(E_{\nu_{i}})}{6} \right) \label{eq-fluence_high},
\end{eqnarray}

\noindent
where $A'$ is a normalization factor representing the total integral of $P(S+B|n_{\mathrm{obs}})$, $N_{T}$ is the number of target nuclei in the detector's fiducial volume relevant to the neutrino interactions, $\sigma(E_{\nu})$ is the combined cross section for all interactions, and $\varepsilon_{\mathrm{high}}^{\mathrm{Fargion}}(E_{\nu})$ is the event selection efficiency of the high energy sample as defined in Section~\ref{sec:analysis}. The systematic uncertainty of the total cross section~($\delta_{\sigma(E_{\nu})}$) is estimated by the difference between two theoretical models from~\cite{Smith:1972xh} and from~\cite{2004PhRvC..70e5503N}. For the other systematic uncertainties~($\delta_{\varepsilon_{\mathrm{high}}}$ and $\varepsilon_{B}$), the same procedure as for the low energy sample was performed.

Table~\ref{tb:high-flence} summarizes the fluence limits of all detactable flavors for the high energy sample. The upper limits of neutrino fluence at $90\%$~C.L. for solar flares occurring on the visible side of the Sun without neutrino candidates is $7.3\times10^{5}~\mathrm{cm^{-2}}$. The upper limit for solar flares with one neutrino candidate, which occurred on November 4th, 2003 (September 6th, 2017), is $1.1\times10^{6}~\mathrm{cm^{-2}}$ ($1.2\times10^{6}~\mathrm{cm^{-2}}$). For the solar flares occurring on the invisible side of the Sun, the upper limits of neutrino fluence at $90\%$~C.L. are $7.3\times10^{5}~\mathrm{cm^{-2}}$, $1.1\times10^{6}~\mathrm{cm^{-2}}$, and $1.6\times10^{6}~\mathrm{cm^{-2}}$ for solar flares with zero, one and two neutrino candidates, respectively.

\begin{table*}[]
    \begin{center}
    \caption{A summary of fluence limits of all detectable flavors for the high energy sample. We assumed the energy spectrum from~\cite{2003ChJAS...3...75F} and the neutrino interaction model from~\cite{2021EPJST.230.4469H}. The higher background rate for the solar flare on September 9th, 2005 is described in the caption of Table~\ref{tb:lowe-flence}.}
        \label{tb:high-flence}
        \begin{tabular}{ccccc}
            \hline
             Side  & Date of flare & Observed event & Expected background & Neutrino fluence \\
             (Channel)  & [UTC] & within window  & within window & [$\mathrm{cm^{-2}{\,}flare^{-1}}$] \\
            \hline
            Visible & 2002 Jul. 23 & No SK data & -- & --   \\
            (Line $\gamma$-rays) & 2003 Nov. \phantom{0}2 & $0$ & $0.11$ & $<7.3\times10^{5}$ \\
            & 2005 Jan. 20 & $0$ & $0.14$ & $<7.3\times10^{5}$ \\ \hline
            & 1997 Nov. \phantom{0}6 & $0$ & $0.04$ & $<7.3\times10^{5}$ \\
            & 2000 Jul. 14 & $0$ & $0.20$ & $<7.3\times10^{5}$ \\
            & 2001 Apr. \phantom{0}2 & $0$ & $0.20$ & $<7.3\times10^{5}$ \\
            & 2001 Apr. \phantom{0}6 & $0$ & $0.10$ & $<7.3\times10^{5}$ \\
            & 2001 Apr. 15 & $0$ & $0.08$ & $<7.3\times10^{5}$ \\
            & 2001 Aug. 25 & No SK data & -- & -- \\
            & 2001 Dec. 13 & No SK data & -- & -- \\
            & 2002 Jul. 23 & No SK data & -- & -- \\
            & 2003 Oct. 23 & $0$ & $0.16$ & $<7.3\times10^{5}$ \\
            & 2003 Oct. 28 & $0$ & $0.12$ & $<7.3\times10^{5}$ \\
            Visible& 2003 Oct. 29 & $0$ & $0.12$ & $<7.3\times10^{5}$ \\
            (Soft X-ray & 2003 Nov. \phantom{0}2 & $0$ & $0.14$ & $<7.3\times10^{5}$ \\
            derivative) & 2003 Nov. \phantom{0}4 & $1$ & $0.20$ & $<1.1\times10^{6}$ \\
            & 2005 Jan. 20 & $0$ & $0.18$ & $<7.3\times10^{5}$ \\
            & 2005 Sep. \phantom{0}7 & $0$ & $0.18$ & $<7.3\times10^{5}$ \\
            & 2005 Sep. \phantom{0}8 & $0$ & $0.08$ & $<7.3\times10^{5}$ \\
            & 2005 Sep. \phantom{0}9 & $0$ & $0.67$ & $<7.3\times10^{5}$ \\
            & 2006 Dec. \phantom{0}5 & $0$ & $0.12$ & $<7.3\times10^{5}$  \\
            & 2006 Dec. \phantom{0}6 & $0$ & $0.06$ & $<7.3\times10^{5}$  \\
            & 2011 Aug. \phantom{0}9 & $0$ & $0.06$ & $<7.3\times10^{5}$ \\
            & 2012 Mar. \phantom{0}7 & $0$ & $0.20$ & $<7.3\times10^{5}$ \\
            & 2017 Sep. \phantom{0}6 & $1$ & $0.12$ & $<1.2\times10^{6}$  \\
            & 2017 Sep. 10 & $0$ & $0.18$ & $<7.3\times10^{5}$ \\ \hline
            & 2001 Apr. 18 & $0$ & $0.62$ & $<7.3\times10^{5}$ \\
            & 2002 Jul. 18 & No SK data & -- & -- \\
            & 2002 Jul. 19 & No SK data & -- & -- \\
            & 2003 Nov. \phantom{0}2 & $0$ & $0.62$ & $<7.3\times10^{5}$\\
            Invisible & 2003 Nov. \phantom{0}7 & $2$ & $0.62$ & $<1.6\times10^{6}$\\
            & 2003 Nov. \phantom{0}9 & $0$ & $0.62$ & $<7.3\times10^{5}$\\
            & 2005 Jul. 24 & $2$ & $0.62$ & $<1.6\times10^{6}$ \\
            & 2011 Jun. \phantom{0}4 & $1$ & $0.62$ & $<1.1\times10^{6}$ \\
            & 2012 Jul. 23 & $1$ & $0.62$ & $<1.1\times10^{6}$ \\
            & 2014 Dec. 13 & $0$ & $0.62$ & $<7.3\times10^{5}$ \\ \hline
     \end{tabular}
    \end{center}
\end{table*}

\begin{figure}[]
    \begin{center}
        \includegraphics[width=0.8\linewidth]{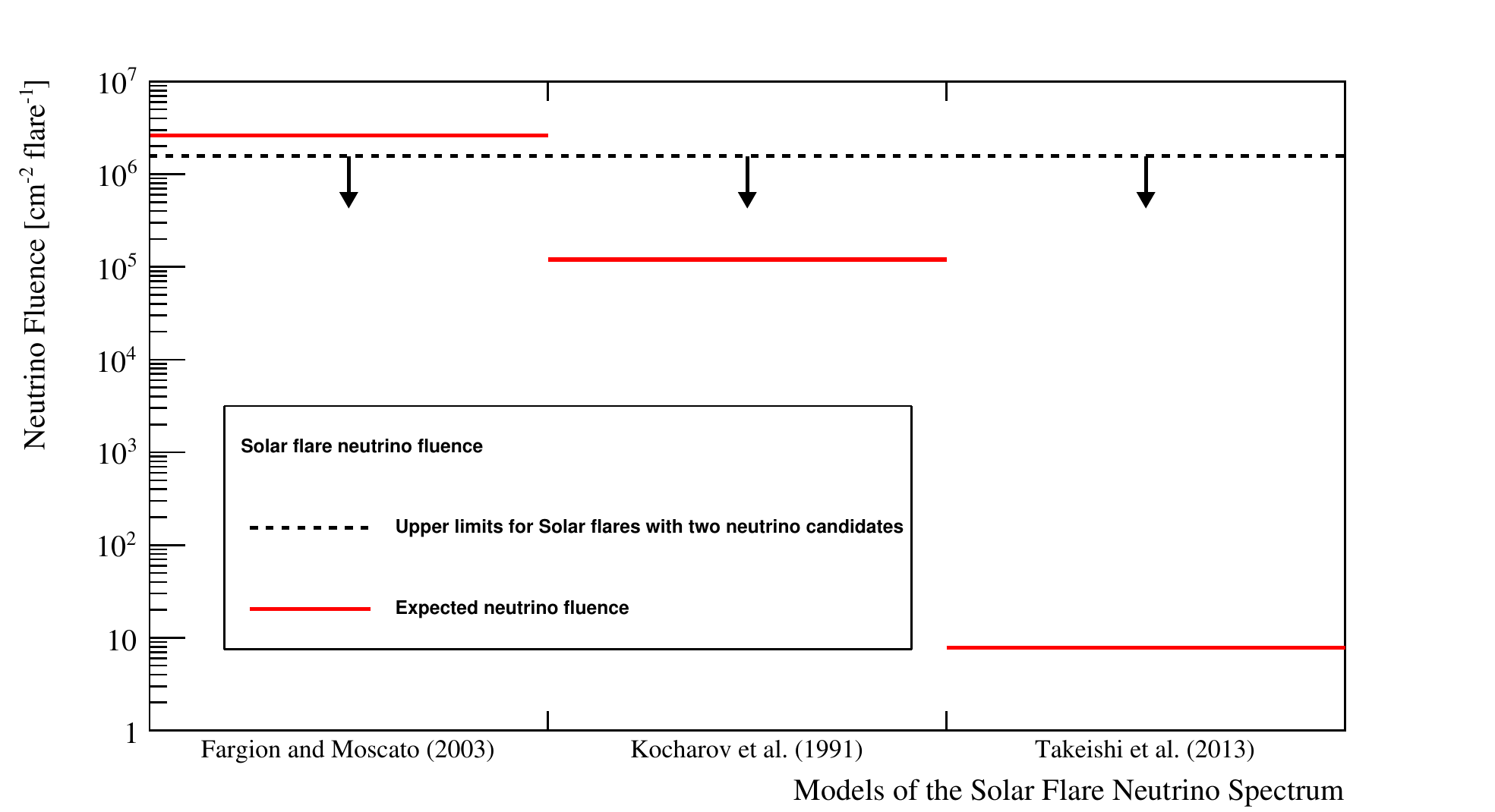}
    \end{center}
\caption{The comparison between upper limits of neutrino fluence considering neutrino spectra from the specific theoretical models of~\cite{1991NCimC..14..417K}, \cite{2004JHEP...06..045F}, and~\cite{2013ICRC...33.3656T}, and their predicted fluences. The black dashed lines with downward arrows~(red solid lines) show the upper limit of neutrino fluence~(expected neutrino fluence based on each theoretical model). The upper limits are  conservatively calculated by considering two neutrino candidates detected within the search windows. \label{fig:other_flux} }
\end{figure}

In order to calculate the upper limit with each theoretical model, $F(E_{\nu})$ in Eq.~(\ref{eq-fluence_high}) is replaced by the other flux predictions from~\cite{1991NCimC..14..417K} and~\cite{2013ICRC...33.3656T} and the selection efficiencies are also evaluated with the replaced predictions. Figure~\ref{fig:other_flux} shows the comparison between the upper limits of neutrino fluence and the predicted neutrino fluence from three theoretical models. From the results, the SK data experimentally excluded the model of~\cite{2003ChJAS...3...75F} even though this model expects several interactions in the SK detector when a energetic solar flare occurs on the invisible side of the Sun as listed in Table~\ref{tb:neutrino-model}. However, the upper limits assuming the neutrino spectra from~\cite{1991NCimC..14..417K} and  \cite{2013ICRC...33.3656T} are still higher than their predictions. As a future prospect, the model from~\cite{1991NCimC..14..417K} will be tested by the next generation of neutrino detectors with significantly larger target volumes. The model of~\cite{2013ICRC...33.3656T}, however, may be difficult to test even with the next generation of neutrino detectors.

\subsection{Model-independent solar-flare neutrino fluences} \label{sec:ind}

As explained in Section~\ref{sec:neutrino}, the excess of events reported by the Homestake experiment originally suggested the existence of solar-flare neutrinos. From this result experimental searches for solar-flare neutrinos have mainly been made by the neutrino detectors in the energy region below $100$~MeV. It should be noted that past studies by the SNO~\citep{2014APh....55....1A}, Borexino~\citep{Agostini:2019yuq}, and KamLAND~\citep{2022ApJ...924..103A} experiments searched for neutrinos from solar flares in coincidence with the soft X-ray light curves recorded by the GOES satellite,  including solar flares with smaller intensity, such as M-class flares. Due to the different assumptions and samples of selected solar flares, we cannot directly compare previous experimental results with the results presented in this article. To compare these results with those from other experiments, the upper limit of neutrino fluence without considering a specific model was also calculated using the low energy sample. In this case, the probability density function at a neutrino energy $E$ is defined as follows:

\begin{eqnarray}
P_{\mathrm{low}}(S+B | n_{\mathrm{obs}})(E) &=& \frac{1}{A}\int\!\!\!\int\!\!\!\int \frac{\mathrm{e}^{-(S(E)+B)}(S(E)+B)^{n_{\mathrm{obs}}}}{n_{\mathrm{obs}}!} P(\sigma_{\mathrm{IBD}})P \left(\varepsilon^{\mathrm{Ind}}_{\mathrm{low}} \right)P(B)d\sigma_{\mathrm{IBD}}d\varepsilon_{\mathrm{low}}^{\mathrm{Ind}}dB, \\
\mathrm{and}~S(E) &=& N_{p}t_{\mathrm{emit}}\int F(E_{\nu})\theta(E,E_\nu) \sigma_{\mathrm{IBD}}(E_{\nu}) \varepsilon_{\mathrm{low}}^{\mathrm{Ind}}{\mathrm{d}}E_{\nu}
\end{eqnarray}

\noindent
where $\varepsilon^{\mathrm{Ind}}_{\mathrm{low}}$ is the selection efficiency listed in Table~\ref{tb:bg_rate}, $\theta(E,E_{\nu})$ is a step function which is defined as, 
\begin{equation}
  \theta(E,E_\nu) = \left\{
  \begin{array}{ll}
    1 & (E-5~{\rm MeV}< E_{\nu} \le E+5~{\rm MeV}), \\
    0 & ({\rm otherwise}),
    \end{array}
    \right.
\end{equation}
and the other variables and functions are the same as those used in Eq.~(\ref{eq-fluence_low}). To convert from the reconstructed positron energy to the incoming electron anti-neutrino energy the theoretical model of~\cite{2003PhLB..564...42S} is used. To address the effect of energy resolution and the energy of the simultaneously produced neutron, the data in the neutrino energy range from $20$ to $110$~MeV was analyzed and the upper limit of neutrino fluence calculated every $10$~MeV.

\begin{figure}[]
    \begin{center}
        \includegraphics[width=0.8\linewidth]{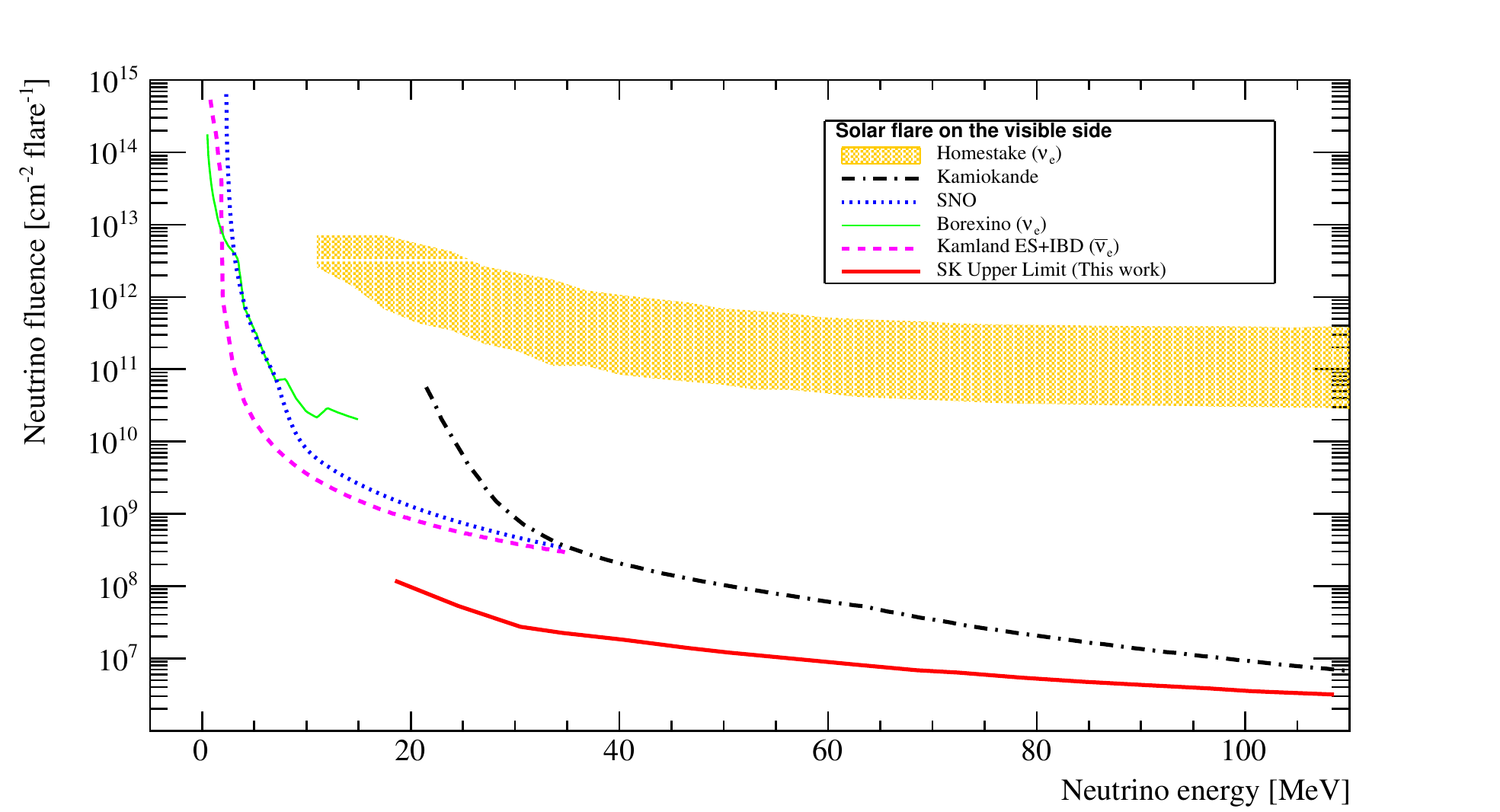}
    \end{center}
\caption{The upper limit of neutrino fluence from the data taken by SK-I, II, III, and IV~(red thick-solid) together with the other experimental results. The orange contour shows the allowed parameter region from the Homestake experiment~\citep{1994PrPNP..32...13D}. Black long-dashed-dotted, blue dotted, green thin-solid, and pink dashed lines show the upper limits from Kamiokande~\citep{1988PhRvL..61.2653H}, SNO~\citep{2014APh....55....1A}, Brexino~\citep{Agostini:2019yuq}, and KamLAND~\citep{2022ApJ...924..103A} experiments, respectively. \label{fig:fluence}}
\end{figure}

Figure~\ref{fig:fluence} shows the SK result for the upper limit of neutrino fluence without considering a specific theoretical model, together with other experimental results~\citep{1994PrPNP..32...13D, 1988PhRvL..61.2653H, 2014APh....55....1A, Agostini:2019yuq, 2022ApJ...924..103A}.
Comparing to other experimental limits of neutrino fluence, the SK limit is improved by at least an order of magnitude in the energy region from $20$ to $110$~MeV. The SK limit fully excludes the allowed parameter region which was favored by the Homestake experiment~\citep{1994PrPNP..32...13D} and gives a strong constraint for neutrino fluence from powerful solar flares.

\subsection{Energy conversion factor}
\label{sec:e_conversion}

As explained in Section~\ref{sec:time-intro}, \cite{2003ChJAS...3...75F} estimated the number of interactions in the SK detector by introducing the conversion factor $\eta$ in Eq.~(\ref{eq_eta}). The experimental search for solar-flare neutrinos by the SK detector gives a constraint on this parameter. However, estimating the total energy of a solar flare is difficult because the magnetic energy released in a solar flare is converted into a variety of different forms. Accordingly, the estimation of total energy is performed for a limited number of solar flares.

By considering the selection efficiencies of each sample and the energies of powerful solar flares, the conversion factor is calculated based on the number of observed events in each sample. For the high energy sample, we used Eq.~(\ref{eq_eta}) as the number of interactions. For the low energy sample, we used the following equation, since~\cite{2004JHEP...06..045F} also estimated the number of interactions only using the IBD reaction,
\begin{equation}
    n_{\mathrm{int}}^{\mathrm{IBD}} = \left[0.63 \left(\frac{\overline{E}_{\bar{\nu}_{e}}}{35~\mathrm{MeV}} \right) + 1.58 \right] \, \eta \left(\frac{E_{\mathrm{FL}}}{10^{31}~\mathrm{erg}} \right),
\end{equation}
\noindent where $n_{\mathrm{int}}^{\mathrm{IBD}}$ is the number of IBD interactions in the SK detector, $\overline{E}_{\bar{\nu}_{e}}$ is the average energy of the electron anti-neutrino spectrum derived from~\cite{2004JHEP...06..045F}, the first, and the second terms in the square bracket are the number of interactions in the energy range of $10$--$100$~MeV, and $100$~MeV--$1$~GeV, respectively\footnote{The expected number of interactions above $1$~GeV is estimated to be $4.9\%$ by calculating the anti-electron neutrino spectrum from~\cite{2003ChJAS...3...75F} and the IBD cross section from~\cite{2003PhLB..564...42S}. We ignored this contribution because other uncertainties, such as the total energy of the solar flare, and the fluctuation of the background event rate in the low energy sample exist in this analysis.}. Table~\ref{tb:high-flence} summarizes the $90\%$~C.L. upper limits of the conversion factors, $\eta_{\mathrm{low}}$ for the low energy sample and $\eta_{\mathrm{high}}$ for the high energy sample, from the most powerful solar flares during solar cycle 23 and 24.

In the case of the solar flare that occurred on November 4th, 2003, \cite{2005AA...433.1133K} estimated the total energy released as $1.3\times10^{34}$~ergs by analyzing electrons above $20$~keV. By analyzing the high~(low) energy sample the conversion factor~$\eta$ was found to be $<0.0006$~($<0.0025$) at $90\%$~C.L. These conversion factors are at least $10^{3}$ smaller than the assumption from~\cite{2003ChJAS...3...75F}. Hence, this result suggests that the conversion factor introduced in~\cite{2003ChJAS...3...75F} is too optimistic an assumption for the energy transfer that produces neutrinos during the solar flare. 

\begin{table*}[]
    \begin{center}
    \caption{A summary of the upper limits of the conversion factor~$\eta$. The estimated energies of selected solar flares are taken from~\cite{2004JGRA..10910104E}, \cite{2005AA...433.1133K}, \cite{2014ApJ...797...50A}, \cite{2015ApJ...802...53A}, and \cite{2020GeAe..60..929M}. Note that their flare energies are estimated using different forms of energy, i.e. magnetic energy in \cite{2004JGRA..10910104E} and \cite{2020GeAe..60..929M}, total released energy by electrons more than $20$~keV in \cite{2005AA...433.1133K}, thermal energy in \cite{2015ApJ...802...53A}, and magnetic potential energy in \cite{2014ApJ...797...50A}, respectively.}
        \label{tb:conversion}
        \begin{tabular}{ccccc}
            \hline
             Date of flare & Estimated energy & Reference &  $\eta_{\mathrm{low}}$ & $\eta_\mathrm{high}$\\
             & of solar flare~[erg]   & for estimated energy &\\ \hline
            2002 Jul. 23 & $10^{32.3}$ & \cite{2004JGRA..10910104E} & \multicolumn{2}{c}{No SK data} \\
            2003 Oct. 28 & $10^{32.3}$ & \cite{2004JGRA..10910104E} & $<0.16\phantom{00}$ & $<0.025\phantom{0}$ \\
            2003 Nov. \phantom{0}4 &  $1.3 \times 10^{34}$ & \cite{2005AA...433.1133K} & $<0.0025$ & $<0.0006$ \\
            2011 Aug. \phantom{0}9 &  $1.29 \times 10^{32}$ & \cite{2015ApJ...802...53A} & $<0.095\phantom{0}$ & $<0.038\phantom{0}$ \\
            2012 Mar. \phantom{0}7 & $1.74\times10^{33}$ & \cite{2014ApJ...797...50A} & $<0.018\phantom{0}$ & $<0.0028$ \\ 
            2017 Sep. \phantom{0}6 & $5.6\times10^{32}$ & \cite{2020GeAe..60..929M} & $<0.022\phantom{0}$ & $<0.014\phantom{0}$ \\ 
     \hline
     \end{tabular}
    \end{center}
\end{table*}

\section{Summary and future prospects}

Neutrinos from solar flares are necessary to understand the mechanisms of proton acceleration at the astrophysical site. For solar flares that occurred on the visible and invisible sides of the Sun, we first estimated the time of neutrino emission using optical light curves and CME observations by solar satellites. We then searched for neutrino events in the Super-Kamiokande detector coincident with these solar flares. Two neutrino events were observed coincident with solar flares that occurred on the visible side of the Sun while six neutrino events were observed coincident with solar flares that occurred on the invisible side of the Sun. All of them are consistent with the background rate under the usual operation of the SK detector. Based on the observed events within the search window we obtained upper limits of the neutrino fluence depending on the assumed theoretical neutrino production model. For example, the fluence limit for the largest solar flare of class X$28.0$ that occurred at the visible side of the Sun on the November 4th, 2003 is $1.1\times10^{6}~\mathrm{cm^{-2}}$. In addition, the fluence limit for the solar flare that occurred on the invisible side of the Sun on November 7th, 2003, which followed the largest solar flare, is $1.6\times10^{6}~\mathrm{cm^{-2}}$. From the obtained fluences, the upper limits on the energy conversion factor were estimated based on \cite{2003ChJAS...3...75F}. In the case of the largest solar flare on November 4th, 2003, $\eta<0.0006$ at $90\%$~C.L., which is two orders of magnitude smaller than the estimate of~\cite{2003ChJAS...3...75F}. Therefore, this experimental result suggests that the theoretical assumption of energy conversion during solar flares should be reconsidered.

In order to compare these results with other experimental searches, the fluence limit below $100$~MeV was also obtained without considering a specific theoretical model. The SK result is the most stringent constraint on the neutrino fluence from solar flares in the MeV region to date. 

In July 2020, $13$~tonnes of $\mathrm{Gd_{2}(SO_{4})_{3}\cdot8H_{2}O}$~(gadolinium sulfate octahydrate)
were dissolved into the SK water tank in order to improve its neutron detection efficiency~\citep{2022NIMPA102766248A}, followed by an additional $26$~tonnes in June 2022. The main motivation for the gadolinium loading is to increase the detector's sensitivity to diffuse supernova electron anti-neutrinos. This technique also enhances the sensitivity to solar-flare neutrinos as well. In addition to the SK phases with Gd, further searches to understand the production of solar-flare neutrinos should be performed by large scale neutrino detectors such as Hyper-Kamiokande~\citep{2018arXiv180504163H}, IceCube gen-2~\citep{2020arXiv200804323T}, and JUNO~\citep{2015arXiv150807166A} during solar cycle~$25$ that started from late 2019.

\begin{acknowledgments}

We thank D.~Fargion from the Sapienza University of Rome for providing the expected fluence of neutrinos from solar flares. We also thank S.~Masuda from Institute for Space-Earth Environmental Research, Nagoya University, T.~Terasawa from the institute for cosmic ray research, the University of Tokyo, and  S.~Yashiro from Catholic University of America, for valuable discussion related with the search windows determination at the both visible and invisible sides of the Sun. 

We gratefully acknowledge the cooperation of the Kamioka Mining and Smelting Company. The Super‐Kamiokande experiment has been built and operated from funding by the Japanese Ministry of Education, Culture, Sports, Science and Technology, the U.S. Department of Energy, and the U.S. National Science Foundation. Some of us have been supported by funds from the National Research Foundation of Korea NRF‐2009‐0083526~(KNRC) funded by the Ministry of Science, ICT, and Future Planning and the Ministry of Education~(2018R1D1A1B07049158, 2021R1I1A1A01059559), the Japan Society for the Promotion of Science, the National Natural Science Foundation of China under Grants No.11620101004, the Spanish Ministry of Science, Universities and Innovation~(grant PGC2018-099388-B-I00), the Natural Sciences and Engineering Research Council~(NSERC) of Canada, the Scinet and Westgrid consortia of Compute Canada, the National Science Centre~(UMO-2018/30/E/ST2/00441) and the Ministry of Education and Science~(DIR/WK/2017/05), Poland, the Science and Technology Facilities Council~(STFC) and GridPPP, UK, the European Union's Horizon 2020 Research and Innovation Programme under the Marie Sklodowska-Curie grant agreement no.754496, H2020-MSCA-RISE-2018 JENNIFER2 grant agreement no.822070, and H2020-MSCA-RISE-2019 SK2HK grant agreement no. 872549.

This work was carried out by the joint research program of the Institute for Space-Earth Environmental Research~(ISEE), Nagoya University. A part of this study was carried using the computational resources of the Center for Integrated Data Science, Institute for Space-Earth Environmental Research, Nagoya University, through the joint research program.
\end{acknowledgments}

\bibliography{main}{}
\bibliographystyle{aasjournal}

\appendix{}
\section{Search windows for neutrinos from invisible side of the Sun} \label{app:cme}

As discussed in Section~\ref{sec:invisible}, the searches for neutrinos associated from solar flares occurring at the invisible side provide information about acceleration mechanism of downward going proton flux. However, no satellite, except for a limited number of planet explorers, directly monitors an explosive phenomenon at the invisible side before the launch of STEREO\footnote{Solar Terrestrial Relations Observatory} satellites on 2006~\citep{2008SSRv..136....5K}.

\cite{2003SoPh..218..261A} found the association rate of CMEs with solar flares clearly increases with the flare's peak flux of soft X-rays, fluence, and time duration. Hence, an occurrence time of CMEs  is helpful to estimate the time of powerful solar flare occurring on the invisible side of the Sun instead of the data taken by limited number of planet explorers.

The catalog of CMEs observed by the SOHO spacecraft are maintained by NASA~\citep{2004JGRA..109.7105Y}\footnote{\url{https://cdaw.gsfc.nasa.gov/CME_list/}} and this catalog summarizes the time of CMEs emission and its location.

\cite{2002ApJ...581..694M}  and~\cite{2009IAUS..257..233Y} statistically examined a possible correlation between the kinetic energy of CMEs and the intensity of the solar flare monitored by GOES satellite and found the weak positive correlation between them. Based on that studies, we conservatively determined the threshold as $2000~\mathrm{km{\,}s^{-1}}$ in order to select solar flares classified to X class. This criterion roughly corresponds to solar flares with X$2.0$\footnote{Note that some flares without CMEs are reported when its class is less than X1.6~\citep{2009IAUS..257..283G}.}. By this method, ten CMEs across solar cycles 23 and 24 were selected and Table~\ref{tb:time-invisible} summarizes their dates, times, locations, and speeds of CMEs.

\cite{2009IAUS..257..233Y} also investigated the time difference between the CMEs and the solar flares. That study found the standard deviation of their time difference is $1020$~s from its distribution. For covering the time of neutrino production during solar flares, we set the start time  as  $3060$~s before the time of CMEs, which corresponds to three standard deviations of their time difference. For the end time, we conservatively set the duration of $4178$~s after the time of CMEs, where this duration is determined in~\cite{2020SoPh..295..133O} from the light curve of soft X-ray recorded by the GOES satellite since all processes, such as acceleration, energy release, and etc, occurring during a solar flare, is likely to complete within this time duration. Finally, the search window for all solar flare occurred on the invisible side of the Sun is $7238$~s in total. Table~\ref{tb:time-invisible} also summarizes the times of corresponding search window for solar-flare neutrinos from the invisible side of the Sun.

\begin{table*}[]
    \begin{center}
    \caption{The summary of the search windows estimated in this study for energetic CMEs occurred at the invisible side of the Sun. The date, time, and active region~(AR) are taken from SOHO LASCO CME CATALOG~\citep{2004JGRA..109.7105Y}.}
        \label{tb:time-invisible}
        \begin{tabular}{ccccccc}
            \hline
            Date & Time~[UTC] & $t_{\mathrm{start}}$ & $t_{\mathrm{end}}$ & AR location & Speed~[$\mathrm{km{\,}s^{-1}}$] \\
            \hline
            2001 Apr. 18 & 02:06:24 & 01:15:24 & 03:16:02 & SW90b  & $2464.2$\\
            2002 Jul. 18 & 18:58:20 & 18:07:20 & 20:07:58 & E90b & $2191.3 $\\
            2002 Jul. 19 & 16:04:34 & 15:13:34 & 17:14:12 & S15E90 & $2046.6$ \\
            2003 Nov. \phantom{0}2 & 09:00:23 & 15:13:34 & 17:14:12 & SW90b & $2036.0$ \\
            2003 Nov. \phantom{0}7 & 15:32:19 & 14:41:19 & 16:41:57 & W90b & $2237.0$ \\
            2003 Nov. \phantom{0}9 & 05:57:57 & 05:06:57 & 07:45:29 & E90b & $2008.1$ \\
            2005 Jul. 24 & 13:35:51 & 12:44:51 & 14:45:29 & E90b & $2527.8$ \\
            2011 Jun. \phantom{0}4 & 21:42:42 & 20:51:42 & 22:52:20 & N16W153 & $2425.5$  \\
            2012 Jul. 23 & 02:10:07 & 01:19:07 & 03:19:45 & S17W132 & $2003.2$ \\
            2014 Dec. 13 & 13:57:34 & 13:06:34 & 15:07:12 & S20W143 & $2221.6$ \\
            \hline
        \end{tabular}
    \end{center}
\end{table*}

\section{Event display and skymap of the observed events from solar flares on the visible side} \label{app:visible}

\begin{figure*}[]
    \begin{minipage}{0.5\hsize}
        \centering\includegraphics[width=1.0\linewidth]{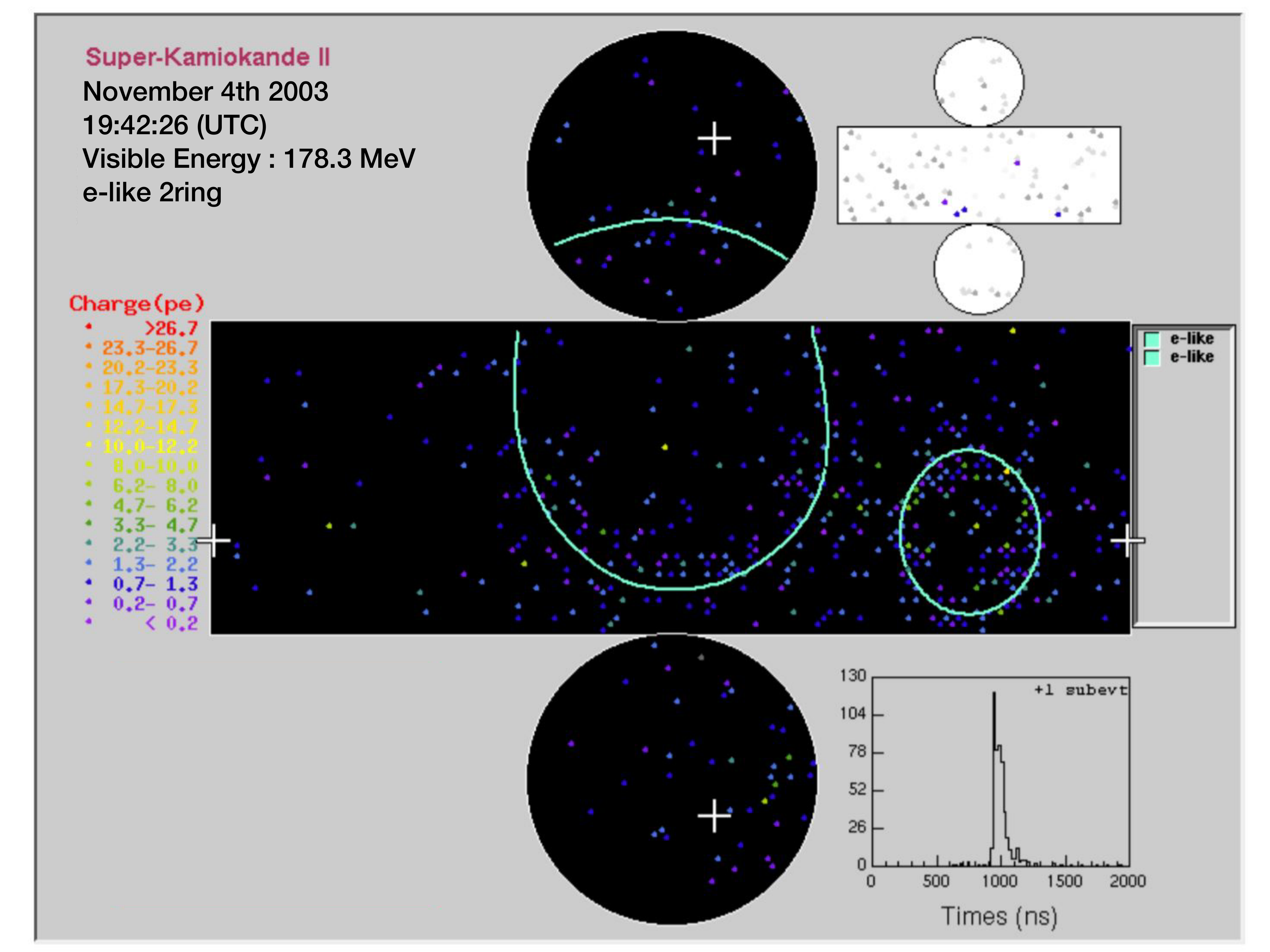}
    \end{minipage}
    \begin{minipage}{0.5\hsize}
        \centering\includegraphics[width=1.0\linewidth]{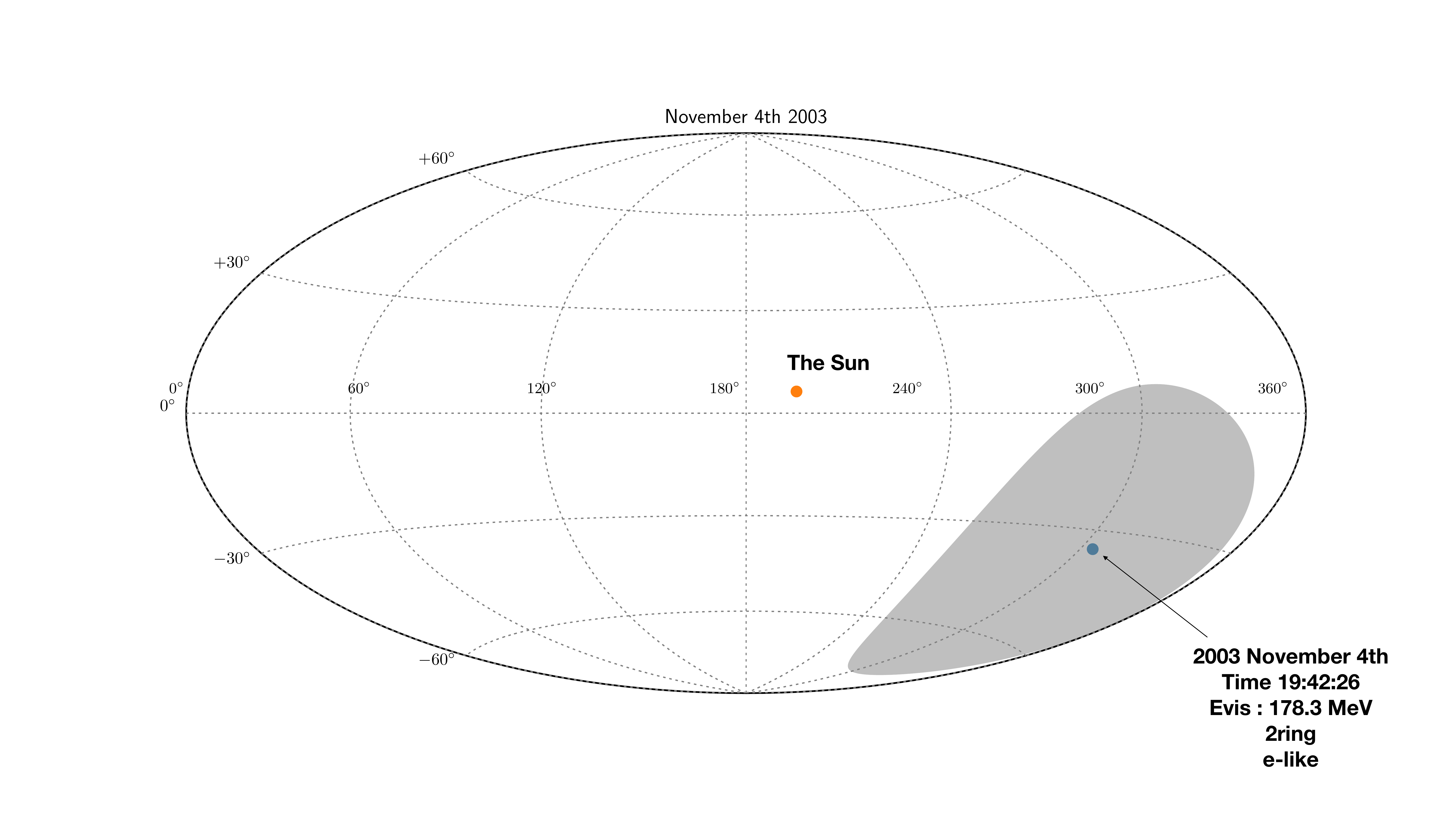}
    \end{minipage} 
   \caption{The event display and the sky-map of the observed event within the search window for the solar flare on November 4th, 2003 together with the location of the Sun at that time. The gray contour in the skymap represents the angular resolution for the observed event. \label{fig:skymap-visible1}}
\end{figure*}

\begin{figure*}[]
    \begin{minipage}{0.5\hsize}
        \centering\includegraphics[width=1.0\linewidth]{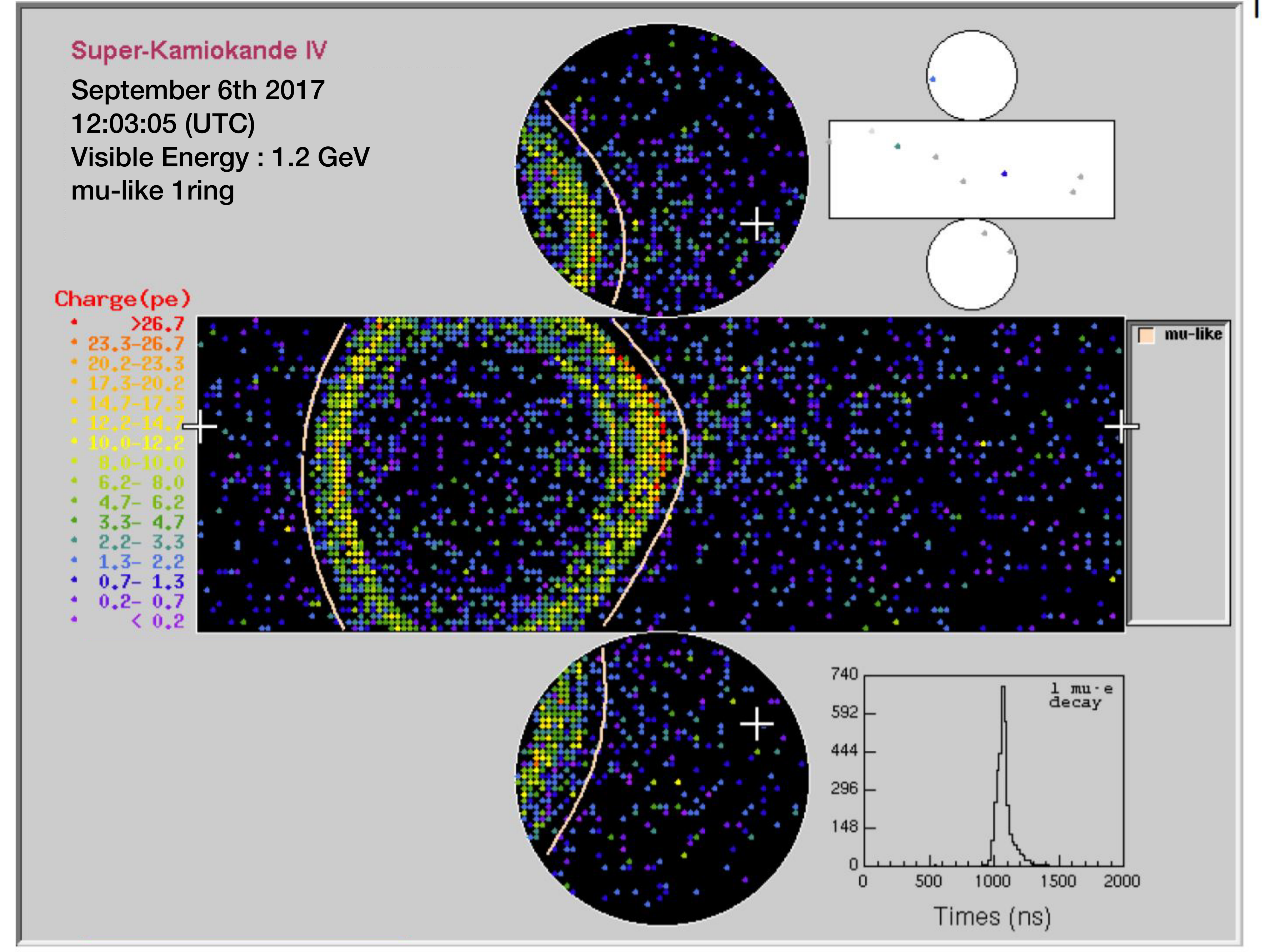}
    \end{minipage}
    \begin{minipage}{0.5\hsize}
        \centering\includegraphics[width=1.0\linewidth]{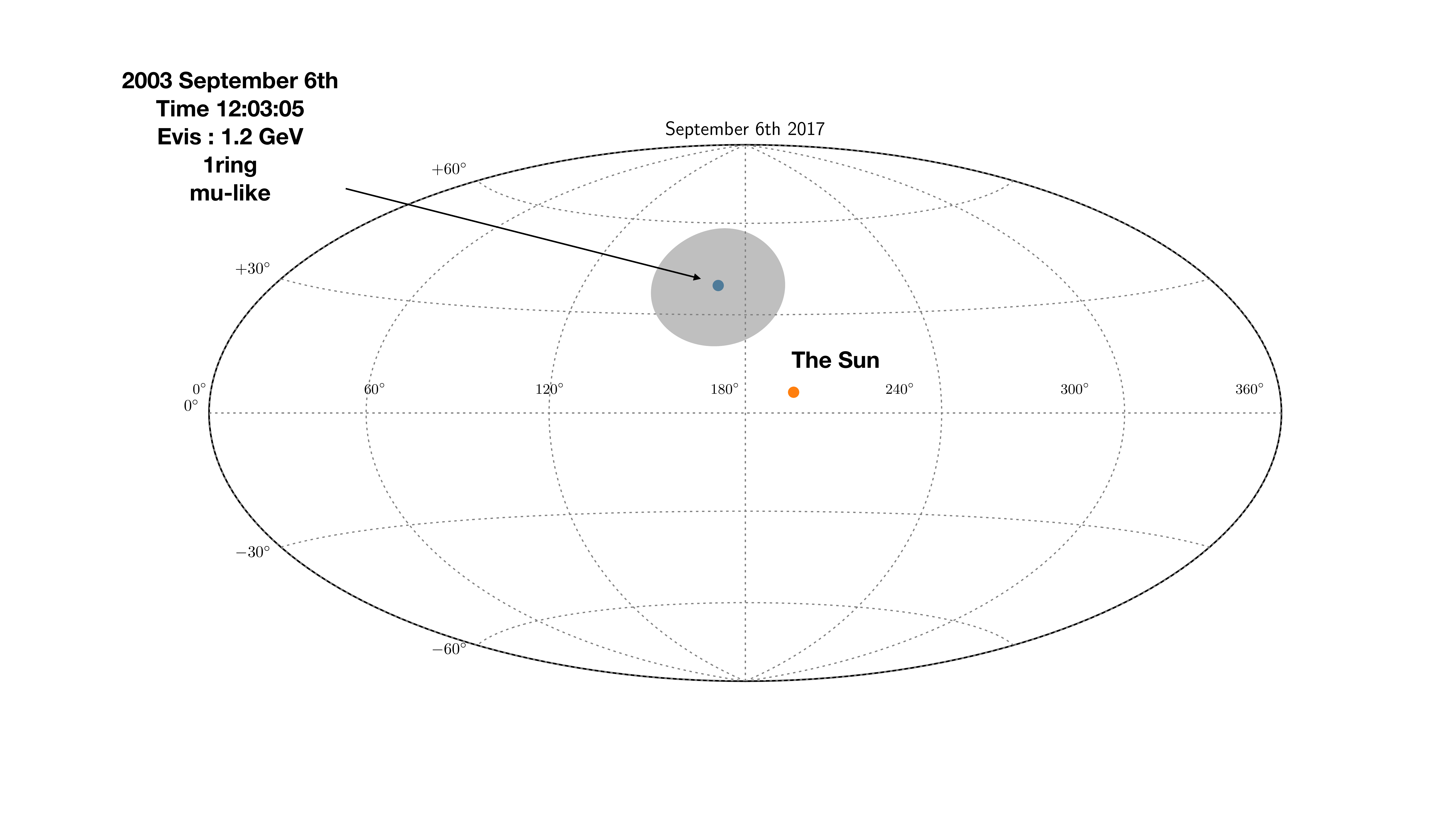}
    \end{minipage} 
   \caption{The event display and the sky-map of the observed event within the search window for the solar flare on September 6th, 2017 together with the location of the Sun at that time. The gray contour in the skymap represents the angular resolution for the observed event. \label{fig:skymap-visible2}}
\end{figure*}

The SK event displays of neutrino candidates from solar flares on the visible side of the Sun and sky maps together with the location of the Sun are shown in Figures~\ref{fig:skymap-visible1} and~\ref{fig:skymap-visible2}.

\section{Event display and skymap of the observed events from solar flares on the invisible side} \label{app:invisible}

\begin{figure*}[]
    \begin{minipage}{0.25\hsize}
        \centering\includegraphics[width=1.0\linewidth]{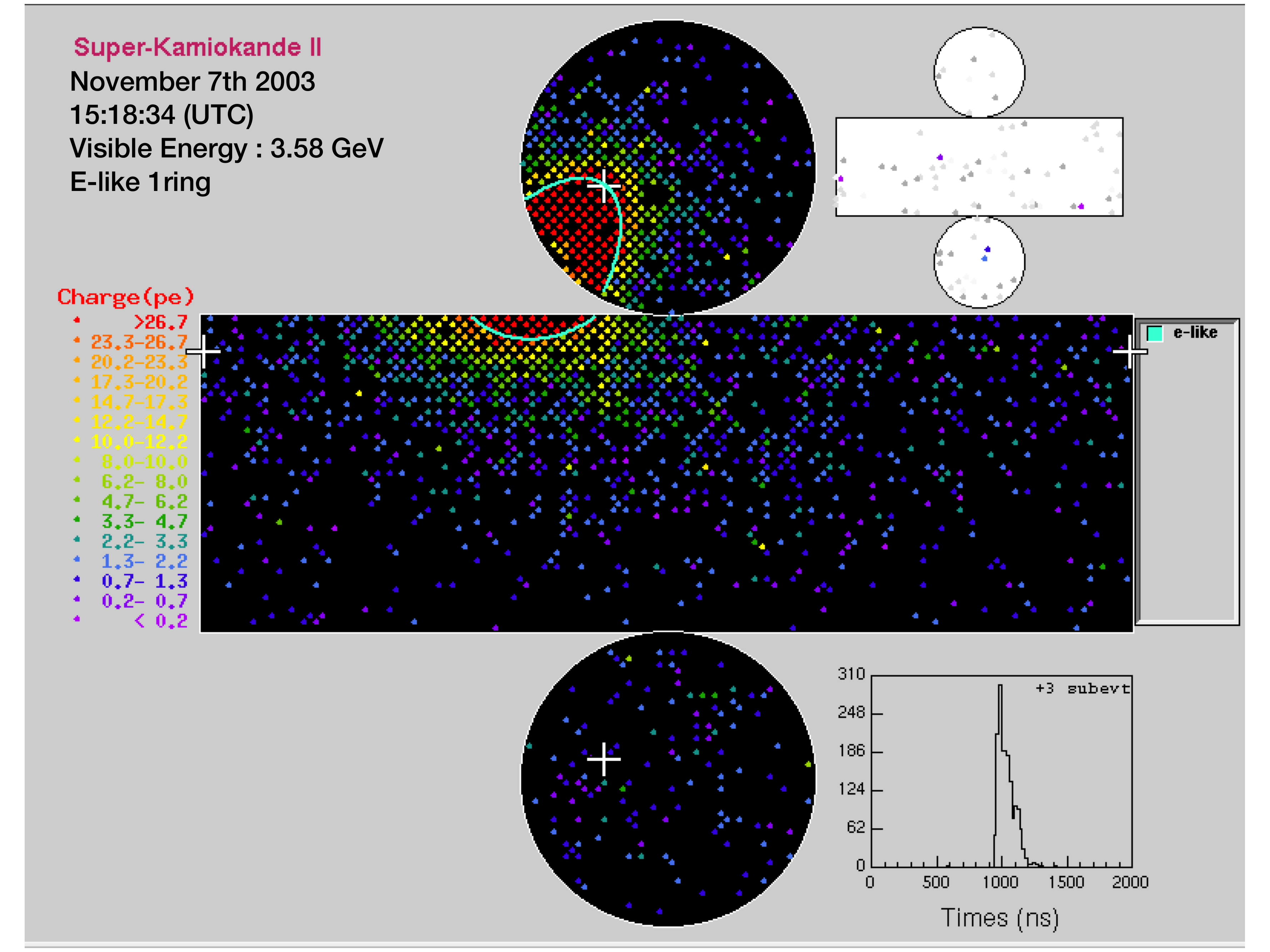}
    \end{minipage}
    \begin{minipage}{0.25\hsize}
        \centering\includegraphics[width=1.0\linewidth]{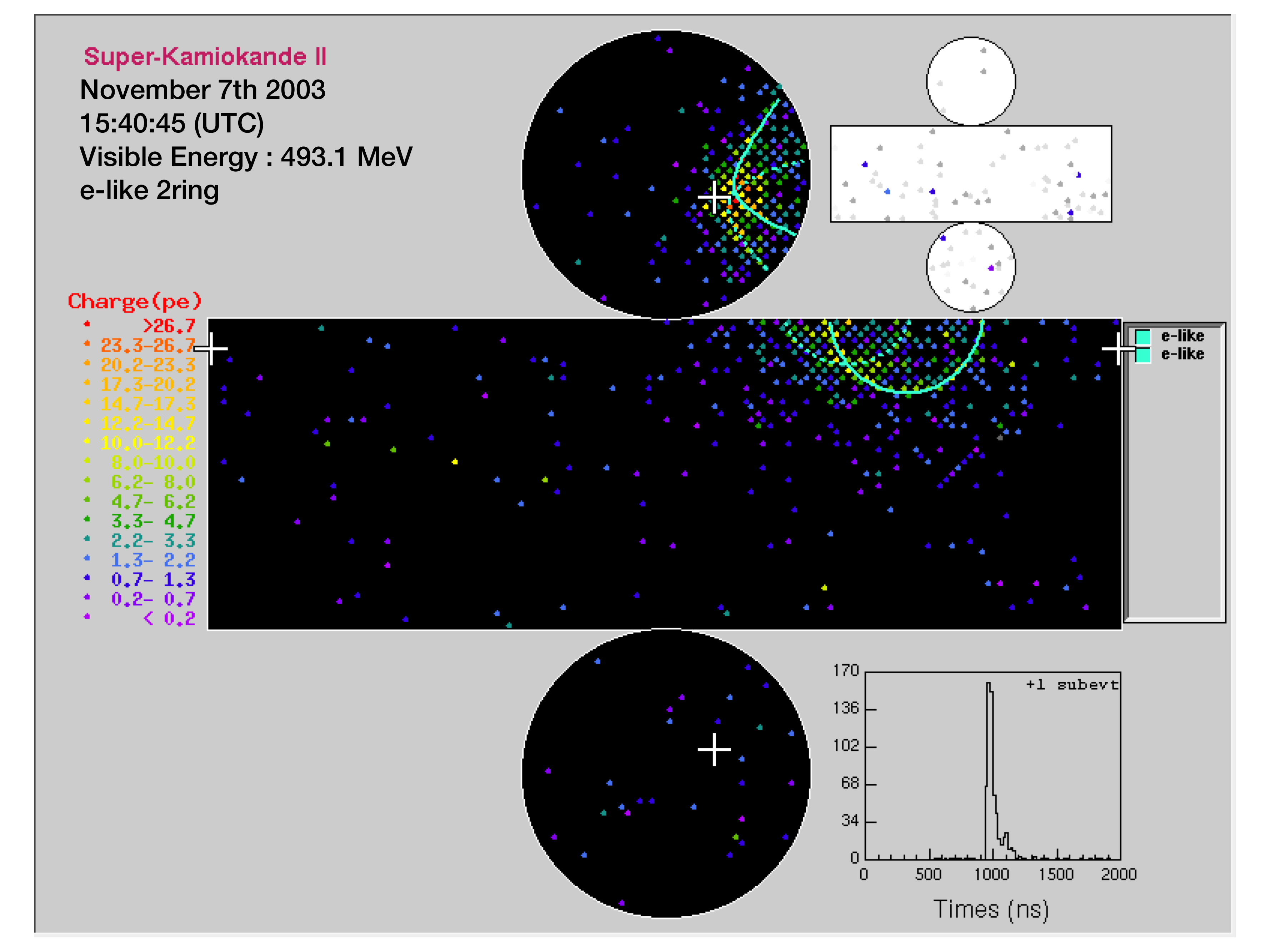}
    \end{minipage}
    \begin{minipage}{0.5\hsize}
        \centering\includegraphics[width=1.0\linewidth]{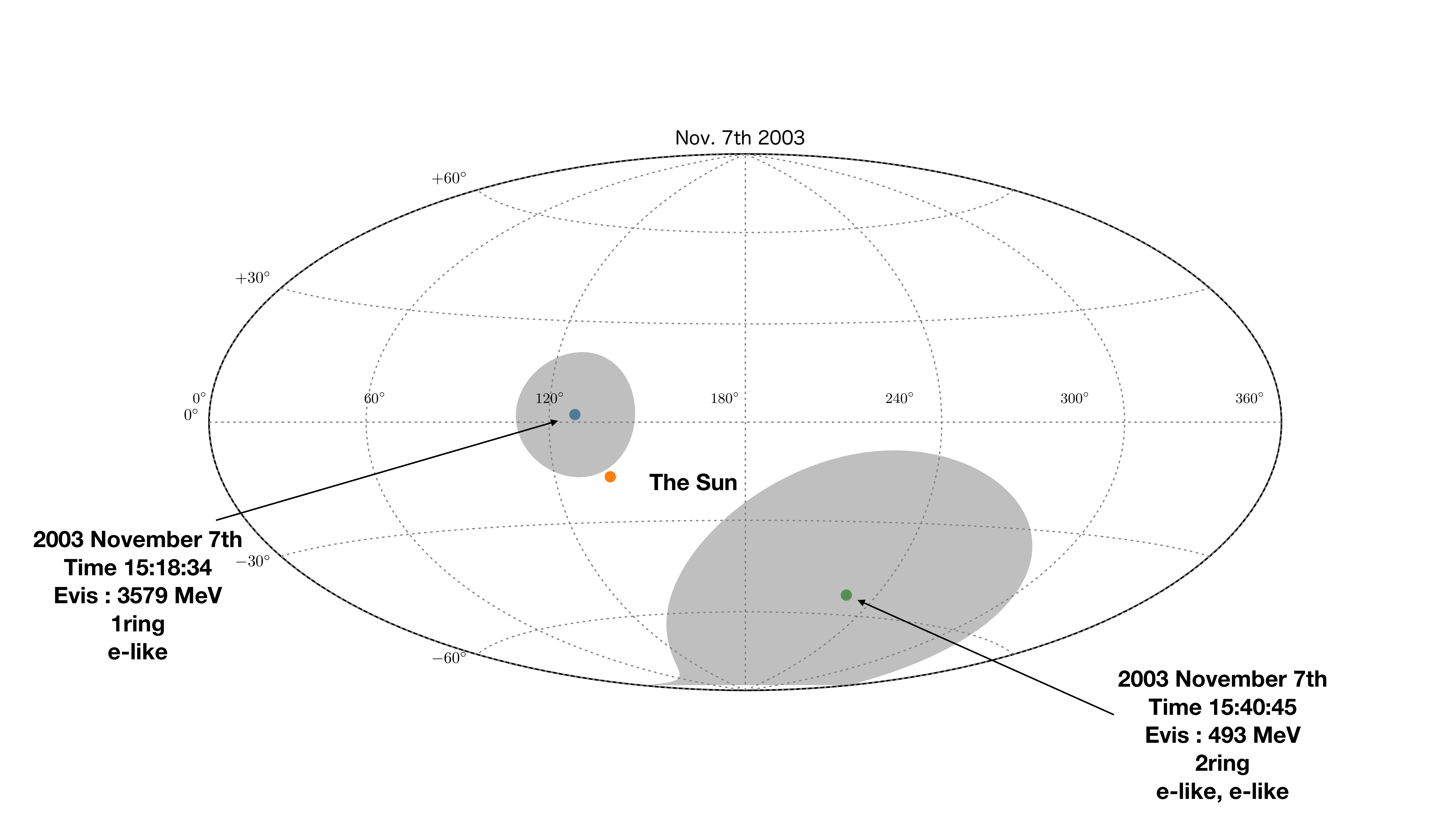}
    \end{minipage} 
   \caption{The event displays and the sky-map of the observed events within the search window for the CME occurring on the invisible side of the Sun on November 7th, 2003 together with the location of the Sun at that time. The gray contours in the skymap represent the angular resolutions for each observed event. \label{fig:skymap-invisible1}}
\end{figure*}

\begin{figure*}[]
    \begin{minipage}{0.25\hsize}
        \centering\includegraphics[width=1.0\linewidth]{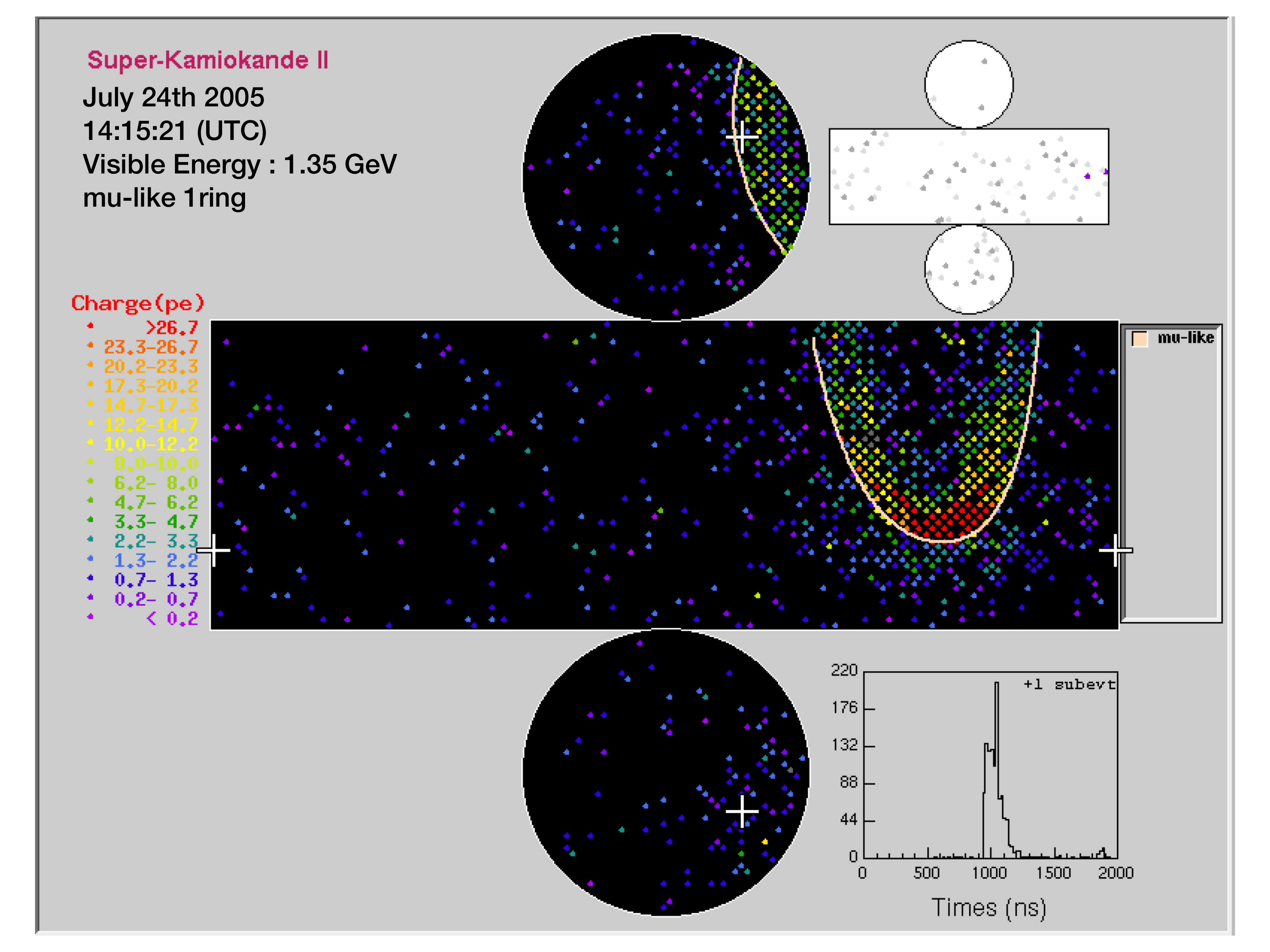}
    \end{minipage}
    \begin{minipage}{0.25\hsize}
        \centering\includegraphics[width=1.0\linewidth]{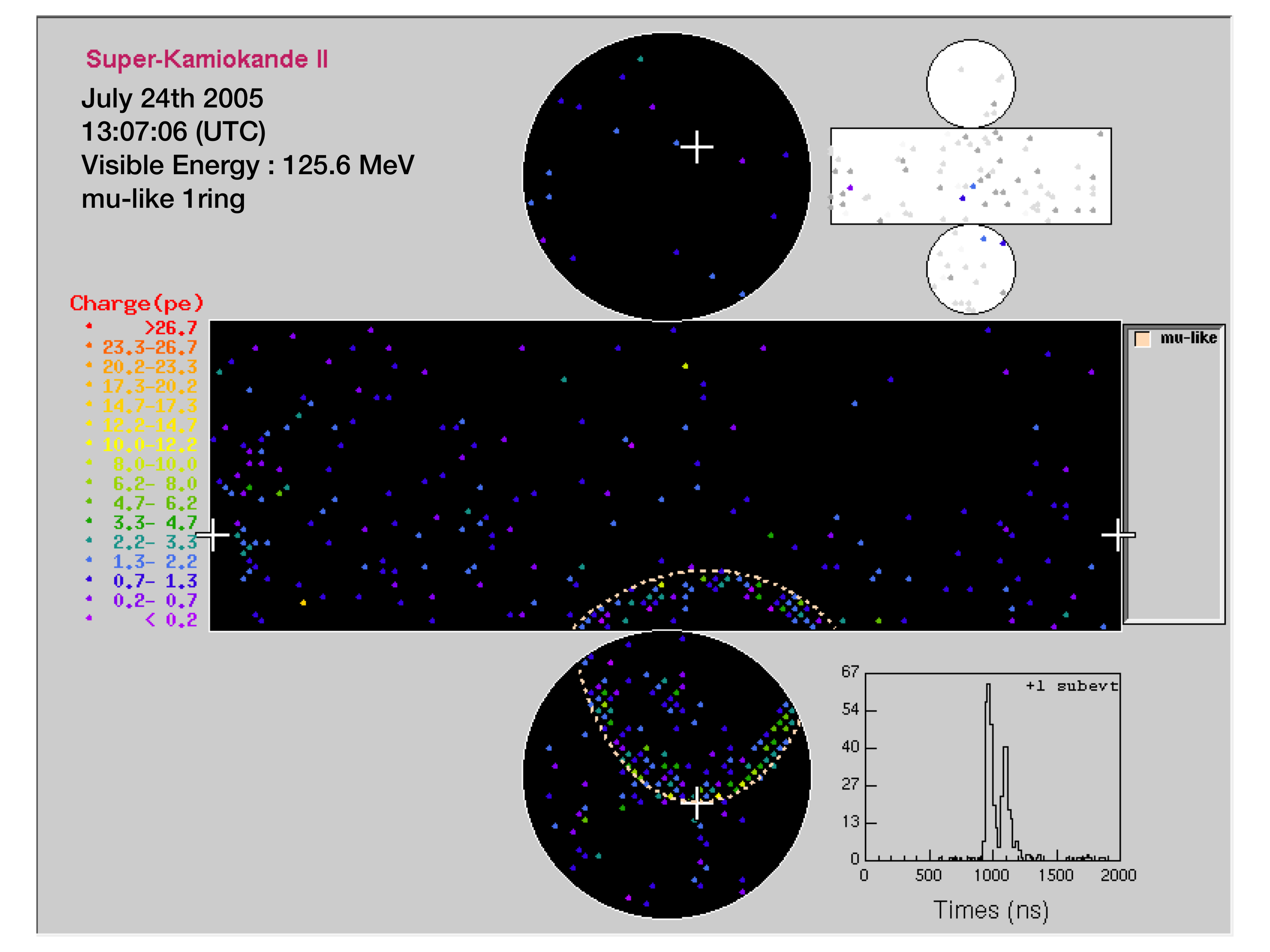}
    \end{minipage}
    \begin{minipage}{0.5\hsize}
        \centering\includegraphics[width=1.0\linewidth]{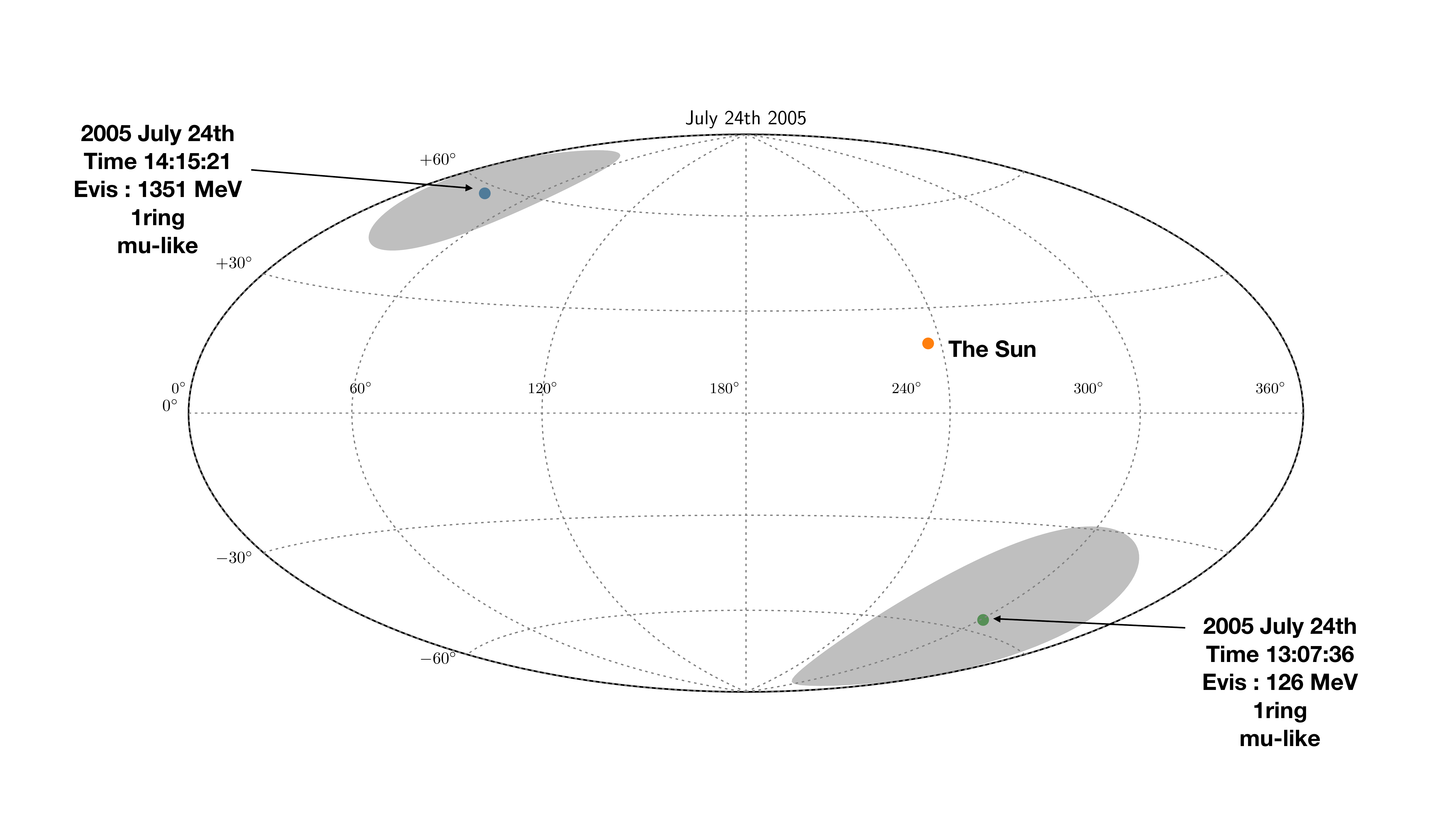}
    \end{minipage} 
   \caption{The event displays and the sky-map of the observed events within the search window for the CME occurring on the invisible side of the Sun on July 24th, 2005 together with the location of the Sun at that time. The gray contours in the skymap represent the angular resolutions for each observed event. \label{fig:skymap-invisible2}}
\end{figure*}

\begin{figure*}[]
    \begin{minipage}{0.5\hsize}
        \centering\includegraphics[width=1.0\linewidth]{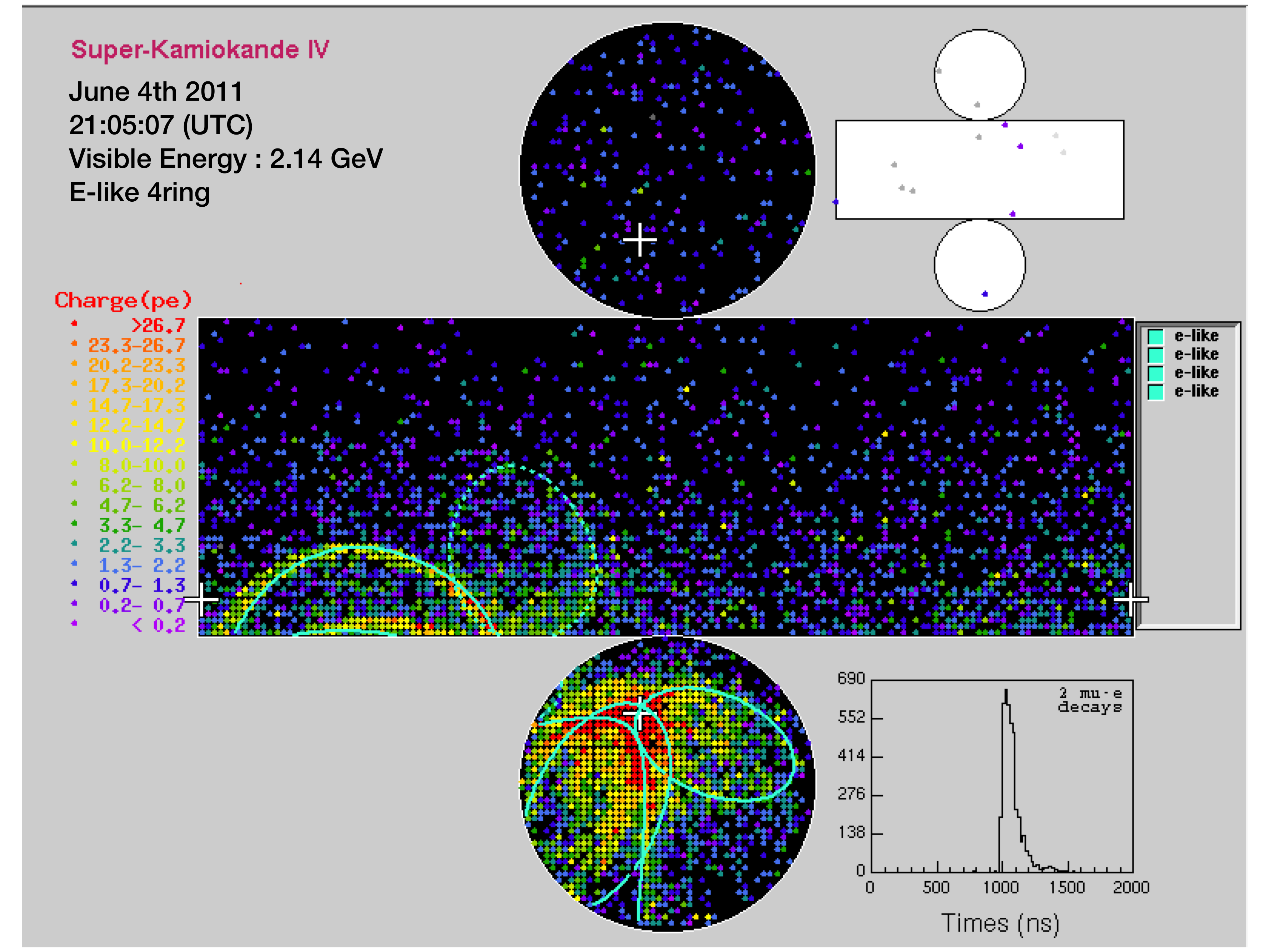}
    \end{minipage}
    \begin{minipage}{0.5\hsize}
        \centering\includegraphics[width=1.0\linewidth]{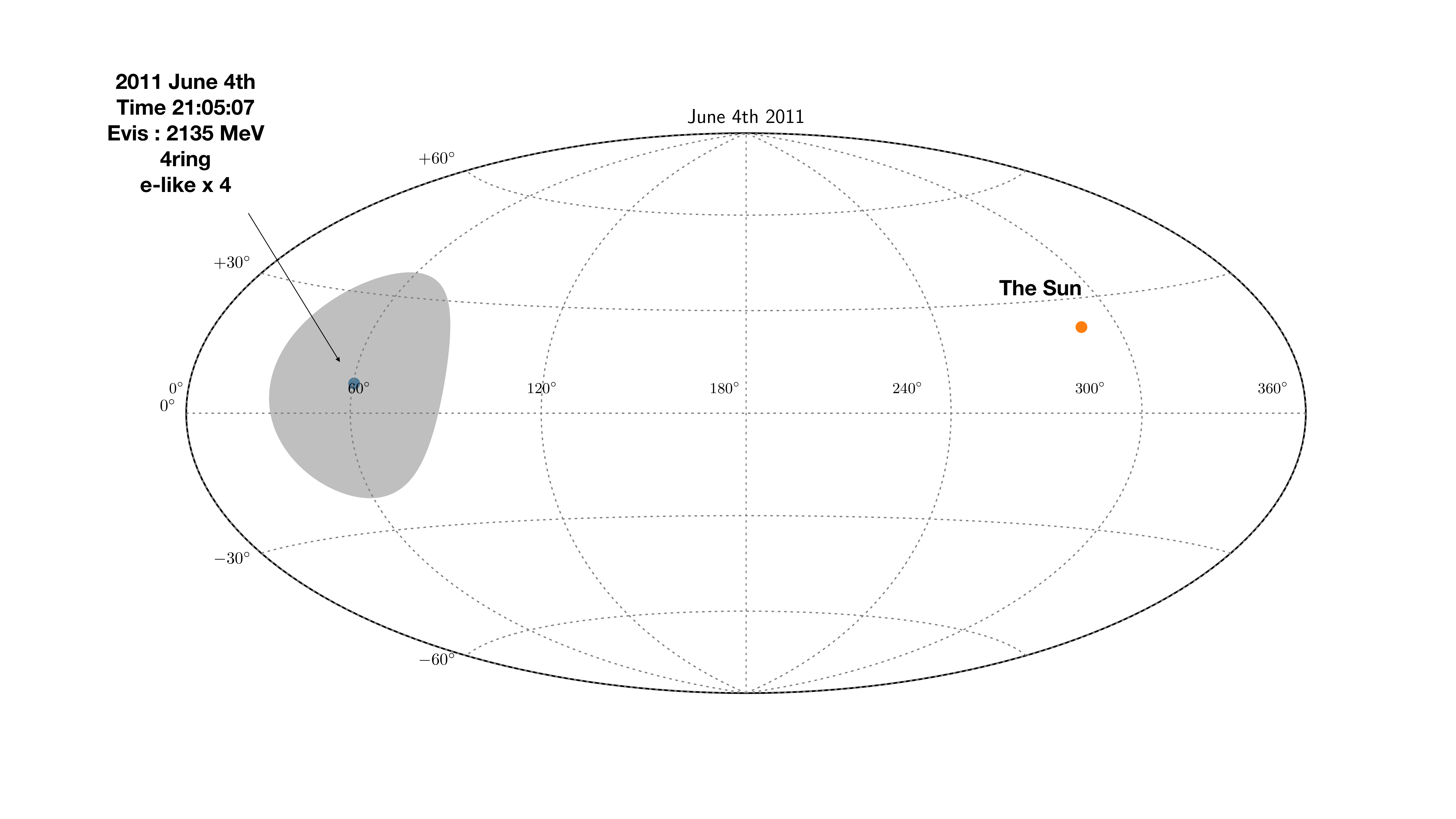}
    \end{minipage} 
   \caption{The event display and the sky-map of the observed event within the search window for the CME occurring on the invisible side of the Sun on June 4th, 2011 together with the location of the Sun at that time. The gray contour in the skymap represents the angular resolution for the observed event. \label{fig:skymap-invisible3}}
\end{figure*}

\begin{figure*}[]
    \begin{minipage}{0.5\hsize}
        \centering\includegraphics[width=1.0\linewidth]{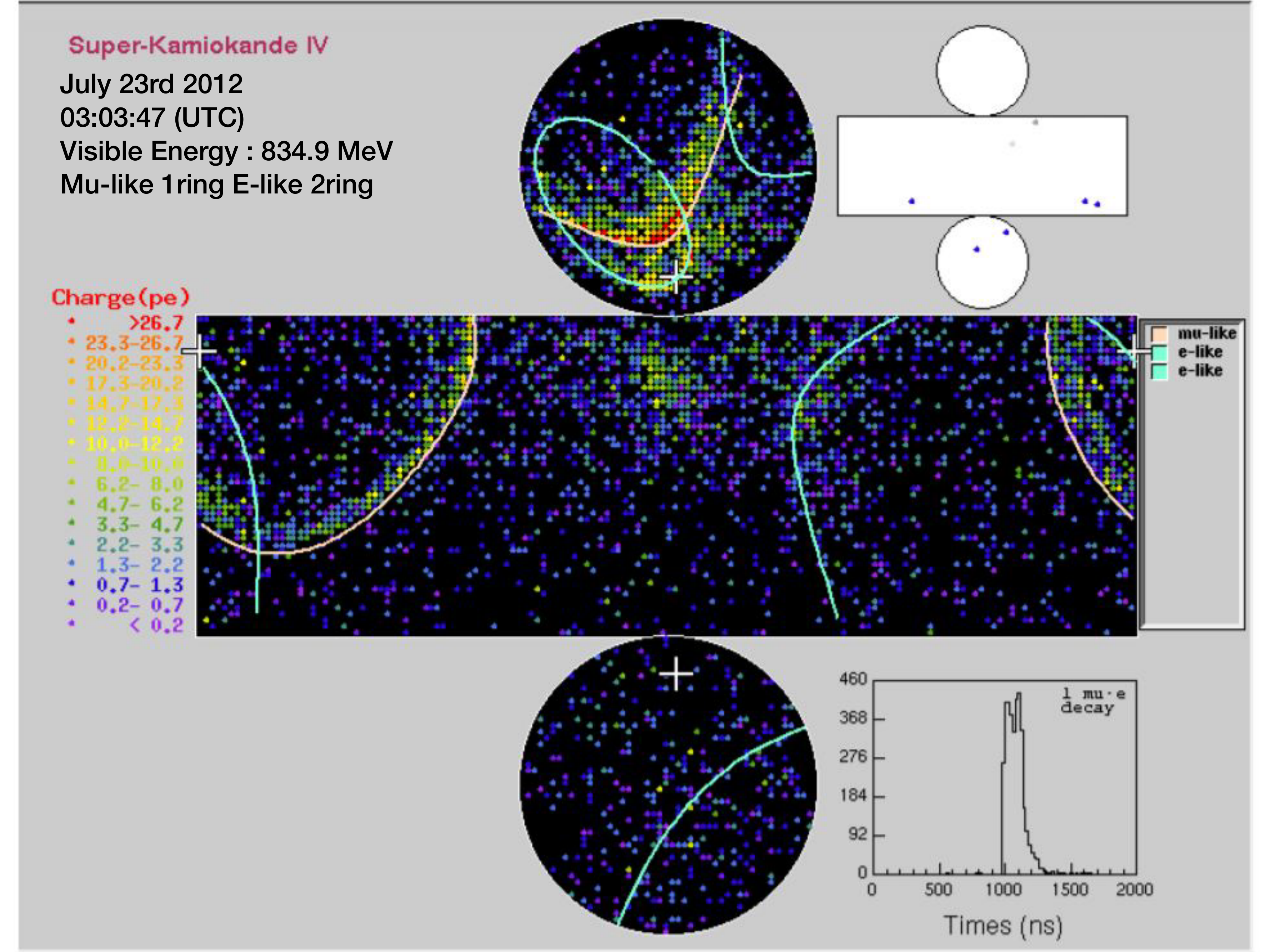}
    \end{minipage}
    \begin{minipage}{0.5\hsize}
        \centering\includegraphics[width=1.0\linewidth]{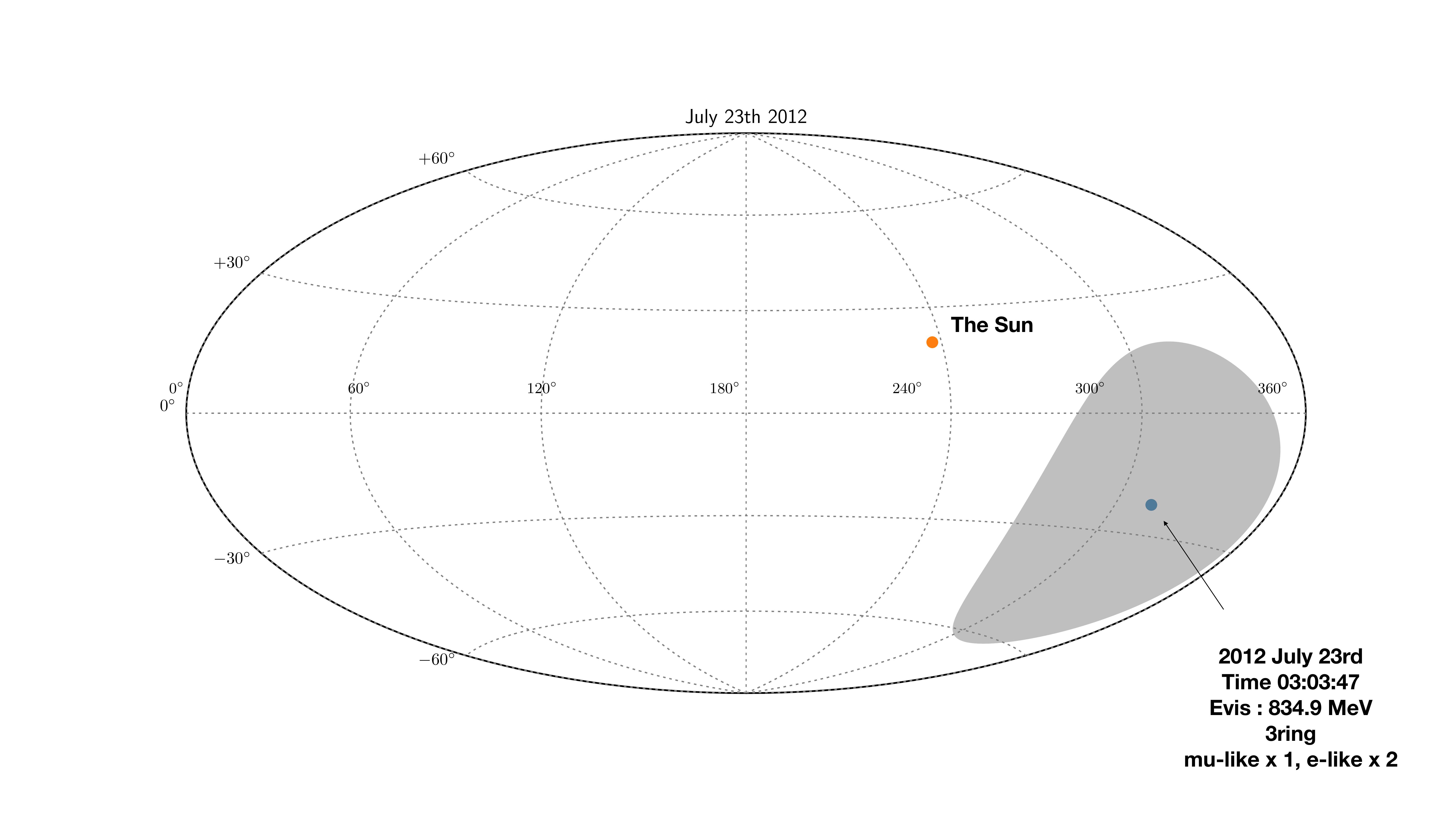}
    \end{minipage} 
   \caption{The event display and the sky-map of the observed event within the search window for the CME occurring on the invisible side of the Sun on July 23rd, 2012 together with the location of the Sun at that time. The gray contour in the skymap represents the angular resolution for the observed event. \label{fig:skymap-invisible4}}
\end{figure*}

The event displays of neutrino candidates from solar flares on the invisible side and sky maps together with the location of the Sun are shown from Figure~\ref{fig:skymap-invisible1} to Figure~\ref{fig:skymap-invisible4}.

\end{document}